\documentclass[twocolumn]{aastex631}
\usepackage{multirow}
\usepackage[graphicx]{realboxes}

\shorttitle{The effect of gravitational decoupling on constraining the mass and radius...}
\shortauthors{S. K. Maurya et al.}

\graphicspath{{./}{figures/}}

\begin{document}

\title{The effect of gravitational decoupling on constraining the mass and radius for the secondary component of GW190814 and other self-bound strange stars in $f(\mathcal{Q})$-gravity theory}

\author[0000-0003-4089-3651]{Sunil Kumar Maurya}
\affiliation{Department of Mathematical and Physical Sciences, College of Arts and Sciences, University of Nizwa, Nizwa 616, Sultanate of Oman, Email: sunil@unizwa.edu.om}

\author[0000-0001-9778-4101]{Ksh. Newton Singh}
\affiliation{Department of Physics, National Defence Academy, Khadakwasla, Pune 411023, India, Email: ntnphy@gmail.com}

\author[0000-0001-6110-9526]{Megandhren Govender}
\affiliation{Department of Mathematics, Durban University of Technology, Durban 4000, South Africa, Email: megandhreng@dut.ac.za}

\author[0000-0003-1409-2009]{Ghulam Mustafa}
\affiliation{Department of Physics, Zhejiang Normal University, Jinhua, 321004, People's Republic of China, Email:gmustafa3828@gmail.com}

\author[0000-0002-5909-0544]{Saibal Ray}
\affiliation{Centre for Cosmology, Astrophysics and Space Science (CCASS), GLA University, Mathura 281406, Uttar Pradesh, India, Email: saibal.ray@gla.ac.in}

\begin{abstract}
Inspired by the conundrum of the gravitational event, GW190814 which brings to light the coalescence of a 23 $ M_{\odot}$ black hole with a yet to be determined secondary component, we look to modelling compact objects within the framework of $f(\mathcal{Q})$ gravity by employing the method of gravitational decoupling. We impose a quadratic equation of state (EOS) for the interior matter distribution which in the appropriate limit reduces to the MIT bag model. The governing field equations arising from gravitational decoupling bifurcates into the $\rho=\theta^0_0$ and $p_r=\theta^1_1$ sectors leading to two distinct classes of solutions.  Both families of solutions are subjected to rigorous tests qualifying them to describe a plethora of compact objects including neutron stars, strange stars and the possible progenitor of the secondary component of GW190814. Using observational data of mass-radius relations for compact objects LMC X-4, Cen X-3, PSR J1614-2230 and PSR J0740+6620 we show that it is possible to generate stellar masses and radii beyond 2.0 $ M_{\odot}$ for neutron stars. Our findings reveal that the most { suitable and versatile model in this framework is the quadratic EOS}, which accounts for a range of low mass stars as well as typical stellar candidates describing the secondary component of GW190814.
\end{abstract}



\section{Introduction}  \label{sec1}
Over the years, compact objects such as neutron stars, pulsars and strange stars have served as cosmic laboratories for determining the nature of matter at  ultra-high densities. While Einstein's classical theory of general relativity can account for many observed features such as compactness, mass-radius relations, and surface redshifts of these objects, it has fallen short in accounting for peculiar observations of neutron stars with masses exceeding $M = 2 M_{\odot}$. The LIGO Scientific and Virgo collaborations (LVC) observations of gravitational waves such as GW190814 and GW170817 events have also cast light on the shortcomings of classical general relativity in accounting for supermassive black holes (BH) \citep{waves1,waves2,waves3} . In particular, the gravitational wave event GW190814 suggests that the source of the signals originated in a compact binary coalescence of a 22.2 to 24.3$ M_{\odot}$ black hole and a compact object having a mass in the range of 2.50 to 2.67$M_{\odot}$. The GW170817 event of August 17, 2017 is thought to be the merger of two neutron stars with masses in the range 0.86 - 2.26 $M_{\odot}$. 

There have been various proponents put forward to account for the observed signals of gravitational events including the nature of matter (equation of state) of stars making up the binary duo and modified gravity theories. The GW170817 event and its electromagnetic counterparts provided researchers with a new tool to propose more exotic equations of states for the neutron stars (NS) involved in this merger. The electromagnetic signal emanating from GW170817 was composed of two parts: a short gamma-ray burst GRB170817A with a delay of approximately 2 seconds with respect to the GW signal and a kilonova, AT2017gfo, peaking in luminosity after a few days after the merger. These observed delays have led researchers to speculate on the nature of the binary components. For example, the delay in the gamma-ray burst has prompted some to believe that remnant arising from the merger was most likely a hypermassive star that collapsed to a black hole within a few milliseconds \citep{ruiz}. Similarly, the kilonova signal points to a not too soft EOS \citep{radice}. In order to account for the possible ranges for the tidal deformability, $400 \leq  \Lambda \leq 800$ and the radius of the 1.5$M_{\odot}$ component which lies within $11.8\,\text{km} \leq R_{1.5} \leq 13.1\,,\text{km}$, the GW170817 event has been modeled as merger of an hadronic star with a strange star \citep{burgio}. Furthermore, to account for small radii and not too small $\Lambda$ it has been proposed that the stellar matter undergoes strong phase transitions at supranuclear densities giving rise to quark matter. 

The observed signal associated with GW190814 has led to speculation of the nature of the secondary component due to several factors including the mass of  2.59$_{-0.09}^{+0.08}M_{\odot}$, lack of significant tidal deformations and no accompanying electromagnetic signals. This points to the presence of either a neutron star (NS) or a BH. We are thus faced with a conundrum where the secondary component is either the heaviest NS or lightest BH ever observed in a double compact-object system. On the other hand, the GW190814 event has been modeled as a second-generation merger, i.e., a triple hierarchical system giving rise to a remnant from a primary bNS which was then captured by the 23 $ M_{\odot}$ black hole \citep{lu1}. Alternatively, the double merger scenario can be the result of a tight NS-NS scattering off a massive BH. These scenarios all call upon the very nature of matter (EOS) of the components making up the binary or triple hierarchical system. Using covariant density functional theory, \cite{fatt} explored the possible 2.6 $M_{\odot}$ stable bounded configuration while ensuring that existing constraints on the composition of neutron stars and the ground-state properties of finite
nuclei are preserved. The energy density functionals (EDF) used in their investigation predicted high pressures prevalent in stellar matter. A softening of the EOS accomplished by the addition of more interactions at high densities does not eradicate the problem with the internal pressure of the compact star. Their findings point to the secondary component of GW190814 event as most likely a BH. In a later study, \cite{tews} employed the NMMA framework to ascertain whether the GW190814 was the result of NS-BH or BH-BH merger. Their starting point was to use a set of 5000 EOSs that extended beyond 1.5 times the nuclear saturation density. Their conclusion was that GW190814 arose from a  binary black hole merger with a probability of $>99.9\%$. In an attempt to provide a theoretical basis for the existence of a 2.6 $M_{\odot}$ neutron star, \cite{god1} employed a Markov Chain Monte Carlo approach to generate about two million phenomenological equations of state with and without first-order phase transitions for the interior matter distribution. The only imposition in their study was the requirements of GR and causality. They observed that if the secondary component of GW190814 was indeed a nonrotating neutron star, then this constrains the EOS to densities of the order of $3\rho_{nuc}$. It is still possible to have a neutron star with a mass greater than 2.5 $M_{\odot}$ and an $R_{1.4}$ of 11.75 $\text{km}$. { In a more recent study,  \cite{mohanty1} employed a total of 60 EOS's to investigate the impact of anisotropy in neutron stars. Using a numerical approach they showed that it is possible to generate neutron star masses greater than 2.0 $M_{\odot}$ within the GR framework by varying the degree of anisotropy within the stellar core. }  

Researchers have also adopted alternative explorations within modified theories of gravity to account for the peculiar masses observed in gravitational events. Using a set of astronomical data and associated constraints on mass-radii relations, \cite{tang1} modeled the secondary component of the GW 190814 event as a quark star (QS) obeying a color-flavored locked-in (CFL) EOS within the framework of $f(R,T)$ gravity. They showed that the curvature coupling constant arising in $f(R,T)$ theory impacts on the maximum allowable masses and radii for stable compact objects to exist. By varying the bag constant and the color superconducting gap energy, masses up to 2.86 $M_{\odot}$ with a radius $R = 12.43 \text{km}$ were possible. Using Einstein-Gauss-Bonnet (EGB) gravity, Tangphati et. al modeled anisotropic quark stars satisfying a CFL EOS \citep{tang2}. They obtained masses up to 2.83 $M_{\odot}$ which exceeded the observed masses of pulsars. Such unconventional masses were possible through the variation of the EGB coupling constant. Further work in $R^2$-gravity coupled with a CFL EOS, has led to predicted masses of the order of 1.97 $M_{\odot}$ which falls in the observed range of the static neutron star PSR J1614-2230. In their attempt to model the mass of the secondary component inferred from the GW 190814 event, Astashenok et al. imposed a stiff EOS on the stellar fluid making up the compact object within the $f(R)$ framework\citep{Astash1}. They concluded that the secondary component could be an NS, BH or rapidly rotating NS and ruled out the possibility of it being a strange star. In a separate work, Astashenok et al. claimed that it was possible to have supermassive neutron stars with masses in the region of 3 $M_{\odot}$ provided that their spin was nonzero \citep{Astash01}. These stars were modeled within the context of $f(R)$ gravity where $f(R) = R + \alpha R^2$ and the quadratic contribution arises in the strong gravitation regime. 

In recent work on the modeling of compact objects the so-called $f(\mathcal{Q})$ gravity was employed with very interesting results, especially with regard to the upper bound on the mass limit of these bounded configurations. The study of anisotropic compact objects in which non-metricity, $\mathcal{Q}$ drives gravitational interactions in the presence of a quintessence field was carried out by \cite{Mandal1}. By choosing an exponential form for  $f(\mathcal{Q})$, they demonstrated that their models described physically realizable compact stellar objects such as HerX-1, SAXJ1808.4-3658, and 4U1820-30. In more recent work, Maurya et al. assumed that the interior matter distribution of a compact object obeyed an MIT bag EOS with the stellar fluid being composed of an anisotropic fluid originating from the superposition of two solutions via the Complete Geometric Deformation (CGD) \citep{sunil5d3}. They showed that contributions from the nonmetricity and the decoupling parameter tend to stabilise mass configurations beyond 2.5 $M_{\odot}$.

On the observational front, LIGO together with the advanced VIRGO observations of gravitational wave events have given us a glimpse into the possible nature of the sources of these signals. With the Laser Interferometer
Space Antenna (LISA) and the Einstein Telescope (ET) researchers hope to probe deeper for new physics during the merger of binary neutron stars. The increased resolution of these probes will enable researchers to gain a better understanding of the physics at play within the core of these massive stars. While the first observations from LISA is expected to be around 2030 and for ET a little later, researchers are currently generating simulated catalogs of standard sirens events for LIGO-Virgo, LISA and ET. Utilising these mock catalogues and a simple parametrization for the nonmetricity function, $f(\mathcal{Q})$ which replicates a $\Lambda$CDM cosmological background, \cite{ferr1} were able to calculate redshifts and $\Omega_m$ and compare their results to SNIa data. The idea is to find signatures or tell-tale signs of nonmetricity in observations of gravitational events. 

Motivated by the work of  \cite{god1} and \cite{tews}, amongst others, in which they utilized a wide spectrum of EOS's to simulate the matter configurations for the secondary components of GW190814 and GW170817, we employ a quadratic EOS within the $f(\mathcal{Q})$ gravity and MGD framework to model compact objects. We pay particular attention to the contributions from the quadratic term on observed mass-radii limits of neutron stars and pulsars which help constrain the free parameters in our model. 

The paper is organized as follows: In Section \ref{sec2}, model equations for $f(\mathcal{Q})$ gravity with extra source have been provided. In Section \ref{sec3}, extended gravitationally decoupled solution in $f(\mathcal{Q})$ gravity is studied along with mimicking of the density constraint (i.e., $\rho=\theta^0_0$ in \ref{solA}) and mimicking of the pressure constraint (i.e., $p_r=\theta^1_1$ in 
\ref{solB}). In Section \ref{sec4}, the matching conditions for the astrophysical system in connection to the exterior spacetime has been elaborated. The physical analysis of completely deformed strange star models and astrophysical implications the problem are done in Section \ref{sec5} where regular behavior of strange star models are discussed under the cases (i) for solution $\rho=\theta^0_0$ (in \ref{5.1.1}) and (ii) for solution $p_r=\theta^1_1$ (in \ref{5.1.2}) whereas we have provided the constraining upper limit of maximum mass for strange star via M-R diagrams under the cases (i) Linear EOS with constancy in bag constant, fixed decoupling parameter and varying EOS parameter (in \ref{5.2.1}), (ii) linear EOS with constancy in bag constant, fixed EOS parameter and varying decoupling constant (in \ref{5.2.2}), (iii)  Quadratic EOS with fixed bag constant, fixed decoupling constant with varying quadratic EOS parameter (in \ref{5.2.4}), and (iv)  Quadratic EOS with fixed bag constant, fixed quadratic EOS parameter with constant versus varying  decoupling constant for mimicking of radial pressure, $p_r=\theta^1_1$(in \ref{5.2.6}). We have discussed the energy exchange between the fluid distributions for $\hat{T}_{ij}$ and $\theta_{ij}$ under the cases (i) for solution $\rho=\theta^0_0$ (in \ref{6.0.1}) and (ii) for solution $p_r=\theta^1_1$ (in \ref{6.0.2}). In Section \ref{sec7} the comparative study of model's arising in GR, GR+CGD, $\lowercase{f}(\mathcal{Q})$ and $\lowercase{f}(\mathcal{Q})$+CGD gravity  have been presented while a few concluding remarks are discussed in Section \ref{sec8}.

\section{Model equations for $f(\mathcal{Q})$ gravity with extra source}\label{sec2}

Let us now have an exposure to the $f(\mathcal{Q})$ gravity and hence gravitationally decoupled systems in terms of symmetric teleparallel paradigm. The first essential point is that in $f(\mathcal{Q})$ theory the gravitational interaction is triggered by the non-metric scalar $\mathcal{Q}$. Therefore, it is always possible that using a second Lagrangian $\mathcal{L}_\theta$ for a different source $\theta_{ij}$, one may express the modified action of $f(\mathcal{Q})$ gravity for gravitationally decoupled systems in the following way:
\begin{eqnarray}
&& \hspace{-0.7cm}\mathcal{S}=\int{\left(\frac{1}{2}\,f(\mathcal{Q})+\lambda_{k}^{\varrho ij} R^{k}_{\varrho ij}+\lambda_{k}^{ ij} T^{k}_{ ij}+\mathcal{L}_m\right)\sqrt{-g}~d^4x}\nonumber\\&&\hspace{0.5cm}+\,\alpha \int{\mathcal{L}_\theta}\,\sqrt{-g}~d^4x. \label{eq1} 
\end{eqnarray}

In the above Eq. (\ref{eq1}), $\mathcal{L}_m$ is the matter Lagrangian density, $g$ is a determinant of the metric tensor (i.e., $g=|g_{ij}|$), $\alpha$ represents a decoupling constant, and $\lambda_{k}^{\varrho ij}$ defines the Lagrange multipliers. However, within the context of the $f(\mathcal{Q})$ gravity, the metric tensor $g_{ij}$ and the connection $\Gamma^k_{\,\,\,i j}$ are considered individually and the necessary connection for nonmetricity term can be expressed as follows:
\begin{equation}\label{eq2}
Q_{k i j}=\nabla_{_k}\,g_{i j}=\partial_i\, g_{jk}-\Gamma^l_{\,\,\,i j}\, g_{l k}-\Gamma^l_{\,\,\,i k} \,g_{j l}.
\end{equation}
where $\nabla_{k}$ defines the covariant derivative and $\Gamma^k_{\,\,\,i j}$ referred to as an affine connection. The final affine connection configuration is defined as
\begin{equation}\label{eq3}
\Gamma^k_{i j}=\lbrace^{\,\,\,k}_{\,i\,\,j} \rbrace+K^k_{\,\,\,ij}+ L^k_{\,\,ij},
\end{equation}
where $\lbrace^{\,\,\,k}_{\,i\,\,j} \rbrace$, $K^k_{\,\,\,ij}$ and $L^k_{\,\,ij}$ define the Levi-Civita connection, contortion tensor and disformation tensor respectively, which are described as:
\begin{eqnarray}
\label{eq4}
&&\hspace{-0.2cm} \lbrace^{\,\,\,k}_{\,i\,\,j} \rbrace=\frac{1}{2}g^{ kl}\left(\partial_i g_{l j}+\partial_ j g_{l i}-\partial_l g_{i j}\right),~~\nonumber\\
&&\hspace{-0.2cm}
L^k_{\,\,\,i j}=\frac{1}{2}Q^k_{\,\,\,i j}-Q_{(i \,\,\,\,\,\,j)}^{\,\,\,\,\,\,k},~~~ K^k_{\,\,\,ij}=\frac{1}{2} T^k_{\,\,\,i j}+T_{(i \,\,\,\,\,\,j)}^{\,\,\,\,\,\,k},
\end{eqnarray}
where $T^k_{\,\,i \,j}$ defines the anti-symmetric part of the affine connection, i.e., $T^k_{\,\,\,i \,j}=2\Gamma^l_{\,\,\,[i\,j]}$. 

In terms of the nonmetricity tensor, the superpotential is defined as:
\begin{eqnarray}
\label{eq5}
P^k_{\,\,\,\,ij}=\frac{1}{4}\left[-Q^k_{\,\,\,\,ij}+2 Q_{(i\,j)}^k+Q^k g_{ij}-\tilde{Q}^k g_{ij}-\delta^k_{(i}Q_{j)}\right].~~~
\end{eqnarray}

In particular, there are only two unique traces of the nonmetricity tensor $Q_{k i j}$ because of the symmetry of the metric tensor $g_{i j}$, which can be written as
\begin{eqnarray}
Q_{k}\equiv Q_{k\,\,\,~i}^{\,\,~i},~~~~~~~~\; \tilde{Q}^k \equiv Q^{\,\,\,\,ki}_{i}. \label{eq6}
\end{eqnarray}

Let us define the nonmetricity scalar in the background of connection as it will be useful in the current analysis and the calculated form of the nonmetricity scalar can be given as
\begin{equation}\label{eq7}
\mathcal{Q}=-\frac{1}{4} Q_{ijk} Q^{ijk}+\frac{1}{2} Q_{ijk} Q^{kji}+\frac{1}{4} Q_{i} Q^{i}-\frac{1}{2} Q_{i} \tilde{Q}^{i}.
\end{equation}
Furthermore, we define $T_{i j}$ and $\theta_{i j}$ for the current analysis as:
\begin{eqnarray}
\label{eq9}
\hat{T}_{ij}=\frac{2}{\sqrt{-g}}\frac{\delta\left(\sqrt{-g}\,\mathcal{L}_m\right)}{\delta g^{i j}}\,~~\hspace{-0.0cm}~\&~ \theta_{i j}=\frac{2}{\sqrt{-g}}\frac{\delta\left(\sqrt{-g}\,\mathcal{L}_\theta\right)}{\delta g^{i j}}.~~
\end{eqnarray}
The modified Einstein-Hilbert action within the purview of symmetric teleparallel gravity (i.e. Eq. (\ref{eq1})), can produce the following gravitational field equations by variation of action with respect to the metric tensor $g^{i j}$:
\begin{eqnarray}
\label{eq8}
\hspace{-0.2cm}\frac{2}{\sqrt{-g}}\nabla_k\left(\sqrt{-g}\,f_\mathcal{Q}\,P^k_{\,\,\,\,i j}\right)+\frac{1}{2}g_{i j}f 
+f_\mathcal{Q}\big(P_{i\,k l}\,Q_j^{\,\,\,\,k l} ~\nonumber\\-2\,Q_{k l i}\,P^{k l}_{\,\,\,\,\,j}\big)   =T_{i j},
~~~\text{where}~~T_{i j}=\big(\hat{T}_{i j}+\alpha \,\theta_{i j}\big),~~~~
\end{eqnarray}
where $f_\mathcal{Q}=\frac{\partial f}{\partial \mathcal{Q}}$, $T_{i j}$ is the stress energy-momentum tensor while $\theta_{i j}$ represents the involvement of extra source term. 

On the other hand, by varying the action with respect to the connection, one may obtain the following relation:
\begin{eqnarray}\label{eq10}
\nabla_{n} \,\lambda_{k}^{jin}+\lambda_{k}^{ij}=\sqrt{-g} \,f_{\mathcal{Q}} P_{ij}^{k}+H_{k}{ }^{ij},
\end{eqnarray}
where $H_{k}{ }^{ij}$ represents the density related to hyper-momentum tensor and can be defined as:
\begin{eqnarray}\label{eq11}
H_{k}{ }^{ij}=-\frac{1}{2} \frac{\delta L_{m}}{\delta \Gamma^{k}{ }_{ij}}.
\end{eqnarray}

Now, by using the antisymmetric property of $i$  and $j$ in the Lagrangian multiplier coefficients, the Eq. (\ref{eq10}) gives the following relation: 
\begin{eqnarray}
\nabla_i \nabla_j \left(\sqrt{-g}\,f_\mathcal{Q}\,P^k_{\,\,\,\,i j}+H^{k}_{\,\,\,\,i j}\right)=0.~~ \label{eq12}
\end{eqnarray}

Additionally, we can retrieve the constraint over the connection, $\bigtriangledown_i \bigtriangledown_j (H^k_{\,\,\,\,i j})=0$, according to the Eq. (\ref{eq12}) as follows:
\begin{eqnarray}\label{eq13}
\nabla_i \nabla_j \left(\sqrt{-g}\,f_\mathcal{Q}\,P^k_{\,\,\,\,i j}\right)=0.
\end{eqnarray}

Due to lacking of curvature and torsion, it can be more explicitly parameterized by a number of functions and finally, the affine connection gets the following form:
\begin{eqnarray}\label{eq14}
\Gamma^k_{\,\,\,i j}=\left(\frac{\partial x^k}{\partial\xi^l}\right)\partial_i \partial_j \xi^l.
\end{eqnarray}

Under the current scenario, an invertible relation is $\xi^k = \xi^k (x^i)$. As a consequence, there always does exist a possibility of discovering a coordinate system that eliminates the $\Gamma^k_{\,\,\,i j}$ connection, i.e. $\Gamma^k_{\,\,\,i j}=0$. It is to note that the coincident gauge is the covariant derivative $\nabla_i$ reduced to the partial one $\partial_i$. In any other coordinate system, where this affine relationship does not vanish, the metric development would be altered, resulting in a completely new theory \citep{Dimakis22}. Therefore, we find a coincident gauge coordinate and nonmetricity by Eq. (\ref{eq2}) which can be simplified as
\begin{eqnarray}\label{eq15}
Q_{k i j}=\partial_k \,g_{i j}.
\end{eqnarray}

One consequence of the operations stated above is that the metric makes the computation easier, as except for the standard General Relativity, here for the action diffeomorphism invariance no longer does exist. In principle, a covariant formulation of $f(\mathcal{Q})$ gravity can be used by determining the affine connection in the absence of gravity before choosing the affine connection in Eq. (\ref{eq14}). In this paper, we look into the gravitationally decoupled solutions for compact objects under $f(\mathcal{Q})$ gravity. 

For the current analysis, we consider the generic spherically symmetric metric, which is given as:
\begin{equation}
\label{eq16}
ds^2=e^{\Phi(r)}dt^2-e^{\mu(r)}dr^2-r^2 (d\theta^2+\sin^2{\theta} \,d\phi^2).
\end{equation}

In the present study, we consider that the spacetime is filled by the anisotropic matter distribution, then total energy-momentum tensor ($T_{ij}$) can be described as:
\begin{eqnarray}
T_{ij}=\epsilon\,u^i\,u_j-\mathcal{P}\,K^i_j+\Pi^i_j, \label{eq17}
\end{eqnarray}
where
\begin{eqnarray}
&&\hspace{-0.5cm} \mathcal{P}=\frac{P_r+2P_\perp}{3};~~~\Pi^i_j=\Pi \big(\xi^i \xi_j+\frac{1}{3} K^i_j\big);~~\nonumber\\&& \text{with}~~ \Pi=P_r-P_\perp; ~~~K^i_j=\delta^i_j-u^i u_j, \label{eq18}
\end{eqnarray}
and  fluid's four-velocity vector $u^i$ and unit space like vector  $\xi^i$ are given by $\{i=0,1,2,3\}$,
\begin{eqnarray}
u^i=(e^{-\Phi/2},~0,~0,~0)~~\text{and}~~\xi^i=(0,~e^{-\mu/2},~0,~0),~~ \label{eq19}
\end{eqnarray}
such that $\xi^i u_i=0$ and $\xi^i\xi_i=-1$. Moreover, $\epsilon$ denotes the total energy density while $P_r$ and $P_{\perp}$ represent the total radial and tangential pressure, respectively for the gravitationally decoupled system. In this regard, the components of the energy-momentum tensor for gravitationally decoupled system under the spherically symmetric line element (\ref{eq16}) are,
\begin{eqnarray}
&& T^0_0=\epsilon,~~~T^1_1=-P_r,~~~~T^2_2=T^3_3=-P_\perp, \label{eq20}
\end{eqnarray}

For the metric Equation (\ref{eq16}), we can calculate the nonmetricity scalar as follows:
\begin{equation}\label{eq21}
\mathcal{Q}=-\frac{2 e^{-\mu(r)} \left[1+r\, \Phi^\prime(r)\right]}{r^2}, 
\end{equation}
where the notation $'$ denotes the derivative over the radial coordinate $r$ only.

In the above expression, $\mathcal{Q}$ is based on the zero affine connections which can be absorbed via the equations of motion (\ref{eq8}) for the anisotropic fluid (\ref{eq18}) as follows:
\begin{eqnarray}
&& \hspace{-0.7cm}\epsilon =-\frac{f(\mathcal{Q})}{2}+f_{\mathcal{Q}} (\mathcal{Q}) \Big[\mathcal{Q}+\frac{1}{r^2}+\frac{e^{-\mu}}{r}(\Phi^\prime+\mu^\prime)\Big],\label{eq22}\\
&& \hspace{-0.7cm} P_r=\frac{f(\mathcal{Q})}{2}-f_{\mathcal{Q}}(\mathcal{Q})\Big[\mathcal{Q}+\frac{1}{r^2}\Big],\label{eq23}\\
&& \hspace{-0.7cm} P_{\perp}=\frac{f(\mathcal{Q})}{2}-f_{\mathcal{Q}}(\mathcal{Q})\Big[\frac{\mathcal{Q}}{2}-e^{-\mu} \Big\{\frac{\Phi^{\prime \prime}}{2}+\Big(\frac{\Phi^\prime}{4}+\frac{1}{2r}\Big) ~\nonumber\\ 
&&\hspace{0.15cm}\times (\Phi^\prime-\mu^\prime)\Big\}\Big],~~~~~~~~\label{eq24}\\
&& \hspace{-0.3cm}0=\frac{\text{cot} \,\theta}{2}\,\mathcal{Q}^\prime\,f_{\mathcal{Q}\mathcal{\mathcal{Q}}} (\mathcal{Q}), \label{eq25}
\end{eqnarray}
where $f_{\mathcal{Q}}(\mathcal{Q})=\frac{\partial f(\mathcal{Q})}{\partial \mathcal{\mathcal{Q}}}$ and $f_{\mathcal{Q}\mathcal{Q}}(Q)=\frac{\partial^2 f(\mathcal{Q})}{\partial \mathcal{Q}^2}$. 

It is to be noted that the nonzero off-diagonal metric components derived from the specific gauge choice for the field equations in the context of $f(T)$ theory put some constraints on the functional form of $f(T)$~\citep{rm7}. As a consequence, it imposes restrictions on the functional form of $f(\mathcal{Q})$ theory. In this connection,~\cite{Wang} derived the possible functional forms for $f(\mathcal{Q})$ gravity in the framework of the static and spherically symmetric spacetime by taking an anisotropic matter distribution. More specifically, they have shown that the exact Schwarzschild solution can exist only when $f_{\mathcal{Q}\mathcal{Q}}(\mathcal{Q})=0$, while the solution obtained by taking nonmetricity scalar $\mathcal{Q}^{\prime}=0$ or $\mathcal{Q}=\mathcal{Q}_0$, where $Q_0$ is constant, shows the deviation from the exact Schwarzschild solution (The detailed analysis of above derivation is given in section \ref{sec4}). Therefore, in order to solve the system of field equations in $f(\mathcal{Q})$-gravity theory for obtaining self-gravitating compact objects, we derive the functional form of $f(\mathcal{Q})$ by taking  only $f_{\mathcal{Q}\mathcal{Q}}$ to be zero as 
\begin{eqnarray}
f_{\mathcal{Q}\mathcal{Q}}(\mathcal{Q})=0~\Rightarrow~f_{\mathcal{Q}}(\mathcal{Q})=\beta_1~\Rightarrow~f(\mathcal{Q})=\beta_1\,\mathcal{Q}+\beta_2,~~~~  \label{eq26}
\end{eqnarray}
where $\beta_1$ and $\beta_2$ are constants. 

At this juncture, we would like to mention that the compatibility of static spherically symmetric spacetime with the coincident gauge, if one assumes the affine connection to be zero and $f(\mathcal{Q})$-gravity theory has vacuum solutions (i.e. $T_{ij}=0$), then the off-diagonal component can be given by 
\begin{eqnarray}
\frac{\text{cot}\,\theta}{2}\,\mathcal{Q}^\prime\,f_{\mathcal{Q}\mathcal{Q}}=0, \label{eq27}
\end{eqnarray}
where $\mathcal{Q}$ has been provided by Eq. (\ref{eq21}). 

As a consequence of the above Eq. \ref{eq26} it can be explored that $f(\mathcal{Q})$ to be linear, the equations of motion automatically turn into $f_{\mathcal{Q}\mathcal{Q}}=0$~\citep{Zhao}. This aspect is very essential as the non-linear function of $\mathcal{Q}$ would gives rise to inconsistent equations of motion. This implies that we need for a more generalized form of the spherically symmetric metric related to a fixed coincident gauge~\citep{Zhao}. Therefore, in the present study to make the spherically symmetric coordinate system compatible one with affine connection $\Gamma^k_{ij}=0$, we have opted for the linear functional form by considering $f_{\mathcal{Q}\mathcal{Q}}=0$ to derive the equations of motion. 

Now inserting Eqs. (\ref{eq21}) and (\ref{eq26}) into Eqs. (\ref{eq22})--(\ref{eq24}), the equations of motion can be obtained as follows:
\begin{eqnarray}
&& \hspace{-0.5cm}\epsilon = \frac{1}{2\,r^2} \Big[2\, \beta_1+2\, e^{-\mu}\, \beta_1  \left(r\, \mu^\prime-1\right)-r^2 \,\beta_2 \Big],\label{eq28}\\
&& \hspace{-0.5cm} P_r=\frac{1}{2\,r^2} \Big[-2\, \beta_1+2\, e^{-\mu} \,\beta_1  \left(r\, \Phi^\prime+1\right)+r^2\, \beta_2\Big],\label{eq29}\\
&& \hspace{-0.5cm} P_{\perp}=\frac{e^{-\mu}}{4\,r} \Big[2\, e^{\mu}\, r\, \beta_2 +\beta_1\,  \left(2+r \Phi^\prime\,\right) \left(\Phi^\prime-\mu^\prime\right) \nonumber\\&&\hspace{0.7cm} +2\, r\, \beta_1 \, \Phi^{\prime \prime}\Big],~~~~\label{eq30} 
\end{eqnarray}

The vanishing of the covariant derivative of the effective energy-momentum tensor is $\bigtriangledown_i T^i_j=0$, provides
\begin{eqnarray}
&&   -\frac{H^\prime}{2}(\epsilon+P_r)-(P_r)^{\prime}+\frac{2}{r}( P_{\perp}-P_r)=0.~~~\label{eq31}
\end{eqnarray}

It is to be noted that the above Eq. (\ref{eq31}) is nothing but the usual Tolman-Oppenheimer-Volkoff (TOV) equation~\citep{Tolman1939,OV1939}, with $f(\mathcal{Q})=\beta_1 \mathcal{Q}+\beta_2$. Therefore, in connection to the proposed compact stellar model we would like to employ gravitational decoupling under the CGD approach to get solution to the system of Eqs. (\ref{eq28})--(\ref{eq30}). For this specific purpose, the gravitational potentials $\Phi(r)$ and $\mu(r)$ are essential to modify as follows:
\begin{eqnarray}
&& \Phi(r) \longrightarrow H(r)+\alpha\, \eta(r)~ \label{eq32}\\
 && e^{-\mu(r)} \longrightarrow W(r)+\alpha\, \Psi(r),~~ ~~ \label{eq33}
\end{eqnarray}
where $\Phi(r)$ and $\mu(r)$ are two arbitrary deformation functions via the decoupling constant $\beta$ whereas in the resultant parts $\eta(r)$ and $\Psi(r)$ are the geometric deformation functions for the temporal and radial metric components, respectively. 

In the above Eqs. (\ref{eq32}) and (\ref{eq33}), for $\alpha = 0$, the standard $f(\mathcal{Q})$ gravity theory can be easily recovered. Essentially in the present work to continue, we must consider the non-zero value for both the deformation functions, i.e., $\eta(r) \ne 0$ and $\Psi(r) \ne 0$. The above transformations (\ref{eq32}) and (\ref{eq33}) can easily divide the decoupled system, viz. (\ref{eq28})--(\ref{eq30}), into two subsystems which are as follows: (i) the first system reflects the field equation in $f(\mathcal{Q})$ gravity under $T_{i\,j}$ and (ii) the second system represents the additional source $\theta_{i\,j}$. To include all these systems, we need to specify the form of energy-momentum tensor $T_{i\,j}$ in the following form: 
\begin{eqnarray}
\hat{T}^i_j=\rho\,\chi^i\,\chi_j-\mathcal{P}_\mathcal{Q}\,h^i_j+\hat{\Pi}^i_j~,\label{eq34}
\end{eqnarray}
where
\begin{eqnarray}
&&\hspace{-0.5cm} \mathcal{P}_\mathcal{Q}=\frac{p_r+2p_\perp}{3}~;~~~\hat{\Pi}^i_j=\Pi_\mathcal{Q} \big(\zeta^i \zeta_j+\frac{1}{3} h^i_j\big)~;\nonumber\\
&& ~~\text{with}~~~
 \Pi_\mathcal{Q}=p_r-p_t~; ~~h^i_j=\delta^i_j-\chi^i \chi_j~, \label{eq35}
\end{eqnarray}
and  $\chi^i$ (fluid's four-velocity vector) and $\zeta^i$ (unit space like vector) are given by,
\begin{eqnarray}
\chi^i=(e^{-H/2},~0,~0,~0)~~\text{and}~~\zeta^i=(0,~\sqrt{W},~0,~0), \label{eq36}
\end{eqnarray}
such that $\xi^i \zeta_i=0$ and $\zeta^i\zeta_i=-1$. The $\epsilon$, $P_r$ and $P_\perp$ can be written as,
\begin{eqnarray}
 \epsilon=\rho+\alpha \,\theta^0_0,~~P_r=p_{_r}-\alpha\,\theta^1_1,~~P_\perp=p_{_t}-\alpha\,\theta^2_2,~~~~  \label{eq37}
\end{eqnarray}
and the corresponding total anisotropy,
\begin{eqnarray} 
\Delta=P_\perp-P_r= \Delta_{\mathcal{Q}}+\Delta_{\theta}, \label{eq38}
\end{eqnarray}
where $~\Delta_{\mathcal{Q}}= p_{_t}-p_{_r}~~\text{and}~~\Delta_\theta= \alpha (\theta^1_1-\theta^2_2)\nonumber.$

One may note that in the present anisotropic compact stellar system there are two types of anisotropies, viz. $T_{i\,j}$ and $\theta_{i\,j}$. Another anisotropy $\Delta_{\theta}$ comes in the picture due to gravitational decoupling which has a definite role in the transformation processes. 

By putting Eqs. (\ref{eq32}) and (\ref{eq33}) in the system (\ref{eq28})--(\ref{eq30}), the set of equations of motion dependent on the gravitational potentials ($H(r)$ and $W(r)$, or when $\alpha = 0$), are produced:
\begin{eqnarray}
&&\hspace{-0.5cm}\rho= \frac{\beta_1 (1-W) }{r^2}-\frac{W^{\prime} \beta_1 }{r}-\frac{\beta_2 }{2},\label{eq39} \\
&&\hspace{-0.5cm} p_{_r}=\frac{\beta_1 (W-1) }{r^2}+\frac{H^{\prime} W \beta_1 }{r}+\frac{\beta_2 }{2}, \label{eq40}\\
&&\hspace{-0.5cm} p_{_t}=\frac{\beta_1(W^{\prime} H^{\prime} + 2 H^{\prime \prime} W +H^{\prime 2} W)  }{4} +\frac{\beta_1\,(W^{\prime} +H^{\prime} W) }{2 r}\nonumber\\&&\hspace{0.5cm} +\frac{\beta_2}{2},~~~ \label{eq41}
\end{eqnarray}
and according to the TOV Eq. (\ref{eq31}),
\begin{eqnarray}
-\frac{H^\prime}{2}(\rho+p_{_r})-(p_{_r})^{\prime}+\frac{2}{r}( p_{_t}-p_{_r})=0.~~\label{eq42}
\end{eqnarray}

Consequently, the spacetime that follows can provide the corresponding solution:
\begin{equation}
ds^2=-e^{H(r)}dt^2+\frac{dr^2}{W(r)}+r^2d\theta^2+r^2\text{sin}^2\theta d\phi^2. \label{eq43}
\end{equation}

Moreover, the system of field equations for $\theta$-sector is derived by turning on $\alpha$ as,
\begin{eqnarray}
&&\hspace{-0.15cm}\theta^{0}_0=-\beta_1 \Big(\frac{\Psi   }{r^2}+\frac{\Psi^\prime }{r}\Big), \label{eq44}\\
&&\hspace{-0.15cm}\theta^1_1=-\beta_1\Big[\frac{\Psi  }{r^2}+\frac{(\Phi^{\prime} \Psi +W\,\eta^{\prime})  }{r}\Big], \label{eq45}\\
&&\hspace{-0.15cm}\theta^2_2=-\beta_1 \Big[\frac{(\Psi^\prime \Phi^{\prime} + 2 \Phi^{\prime \prime} \Psi+\Phi^{\prime 2} \Psi +W^{\prime}\,\eta^{\prime}) }{4}    +\frac{(\Psi^\prime +\Phi^{\prime} \Psi) }{2 r}\Big]\nonumber\\&&\hspace{0.5cm} -\beta_1\Big[\frac{W}{4}\,\big(2\,\eta^{\prime\prime}+\alpha\,\eta^{{\prime}\,2}+\frac{2\,\eta^{\prime}}{r}+2\,H^{\prime}\,\eta^{\prime}\big)\Big],~~~~~~ \label{eq46}
\end{eqnarray}
and the associated conservation is
\begin{eqnarray}\label{eq47}
-\frac{\Phi^{\prime}}{2} (\theta^0_0-\theta^1_1)+ (\theta^1_1)^\prime+\frac{2}{r} (\theta^1_1-\theta^2_2)=\frac{\eta^{\prime}}{2}\,({p_r}+{\rho}).
\end{eqnarray}

However, the mass function for both systems is given by
\begin{eqnarray}\label{eq48}
 m_{\mathcal{Q}}=\frac{1}{2} \int^r_0 \rho(x)\, x^2 dx~~\text{and}~~m_{\theta}= \frac{1}{2}\,\int_0^r \theta^0_0 (x)\, x^2 dx, ~~~
\end{eqnarray}
where the relevant mass functions for the sources $T_{ij}$ and $\theta_{ij}$ are $m_{\mathcal{Q}}(r)$ and $m_{\theta}(r)$, respectively. Then, in the context of $f(\mathcal{Q})$ gravity, the interior mass function of the minimally deformed space-time (\ref{eq16}) may be expressed as
\begin{eqnarray} \label{eq49}
\tilde{m}(r)=m_\mathcal{Q}(r)-\frac{\beta_1\,\alpha}{2}\,r\,\Psi(r).
\end{eqnarray}

Therefore, from all the previous steps towards the generation of the mass functions, the need as well as advantage of CGD-decoupling becomes very straightforward as such it can be pointed out that one can extend any known solutions associated with the action $\mathcal{S}_{\mathcal{Q}} $ with solution $\{T_{ ij}, W, H\}$ of the system (e.g. Eqs. (\ref{eq39})--(\ref{eq41})) into the domain beyond of $f(\mathcal{Q})$ gravity theory associated with the action $\mathcal{S}$. It is to be noted that in the process equation of motion are displayed in Eqs. (\ref{eq28})--(\ref{eq30}) and the unconventional gravitational system of equations are Eqs. (\ref{eq44})--(\ref{eq48}) to determine $\{\theta_{ij}$, $\Psi$, $\eta\}$. 

Let us now generate the {\it $\theta$-version} of any $\{T_{ ij}, W,H\}$-solution as 
\begin{eqnarray}
\{T_{ ij},~ H(r),~ W(r)\} \Longrightarrow \{T^{\text{tot}}_{ ij}, ~\Phi(r),~ \mu(r)\}. \label{eq50}
\end{eqnarray}
which describes a definite way to investigate the results that is beyond the symmetric teleparallel gravity.

\section{Extended gravitationally decoupled solution in $ \lowercase{f}(\mathcal{Q})$ gravity} \label{sec3}

In this Section, we will solve both systems of equations (\ref{eq36})--(\ref{eq39}) and (\ref{eq41})--(\ref{eq44}) related to the sources ${T}_{\mu\nu}$ and $\theta_{\mu\nu}$. It is mentioned that the energy-momentum tensor ${T}_{ij}$ describes an anisotropic fluid matter distribution, therefore,  the $\theta_{ij}$ may enhance the total anisotropy of the system which helps in preventing the gravitational collapse of the system. Furthermore, if we look at the second system, it shows clearly that the solution of the second system depends on the first system.  Then it is mandatory to solve the first system initially. For solving the first system (\ref{eq36})--(\ref{eq39}) in $f(\mathcal{Q})$-gravity, we use a generalized polytropic equation of state (EOS) of the form,
\begin{eqnarray}
p_r &=& a\, \rho^{1+1/n}+b \rho + c, \label{eq51}
\end{eqnarray}
where $a,~ b$ and $c$ are constant parameters with proper dimensions and $n$ denotes a polytropic index. 

Let us trace back the generic history of the polytropic equation of state $p_r=a\rho^{1+\frac{1}{n}}$ which has been extensively used to analyse the physical attributes of the compact stellar objects in the context of various requirements~\citep{EOS1,EOS2,EOS3,EOS4,EOS5,EOS6,EOS7,EOS8,EOS9,EOS14}. In connection to cosmological scenario~\citet{EOS1c} first tried to generalize the polytropic EOS in the form $p_r=\gamma \rho^{1+\frac{1}{n}}+\beta \rho$. In the context of the late universe~\citet{EOS2c} further studied the issues by considering negative indices which was immediately received attention to the study of quantum fluctuations and constant-density cosmology~\citep{EOS3c}. But, this EOS is mostly applicable for a stellar system with vanishing pressure when the density goes to zero. Therefore, for the self-bound compact objects, a further modified EOS $p_r=\gamma \rho^{1+\frac{1}{n}}+\beta \rho +\chi$ was extensively considered by several investigators~\citep{EOS10,EOS13,EOS12,EOS11}.   

It is interesting to note that the polytropic EOS (\ref{eq51}) may represent a MIT bag EOS for the specifications: $a=0,~b=\frac{1}{4}$ and $c=-\frac{4}{3}\mathcal{B}_g$, where $\mathcal{B}_g$ is a bag constant. Due to highly non-linearity of the field equations, as a simple case we assume the value for the polytropic index as $n=1$. In this context one may note that the contribution for quadratic term (i.e. $a \rho^2$) appears in the EOS to express the neutron liquid in Bose-Einstein condensate form whereas the linear terms (i.e. $b \rho+c$) come from the free quarks model of the MIT bag model, with $b=1/3$ and $c=-4\mathcal{B}_g/3$. 

{In addition to the above studies, there are various works employing more realistic  EOSs. Analytical representations of more realistic EOS based on parameters arising in nuclear physics such as the FPS EOS due to Pandharipande \& Ravenhall \citep{Pandhar}, SLy EOS of Douchin \& Haensel \citep{Douchin} and the unified EOSs BSK19, BSk20 and BSk21 \citep{Potekhin}, amongst others have been previously employed using numerical techniques to generate neutron star models. The quadratic EOS can be viewed as a truncation of these more realistic EOSs. The models derived using the quadratic EOS can be used as first approximation to test more complicated numerical codes. Furthermore, Haensel and Potekhin \citep{Haensel2004} pointed out that unified EOSs (presented in the form of tables) introduce ambiguities in the calculations of various parameters in neutron star modeling. These pathologies arise in the interpolation between the tabulated points as well as the calculation of derivatives regarding thermodynamical quantities. They further highlight the point that analytical EOSs ensure that shortcomings are circumnavigated and allow for high-precision neutron star modeling.}

Now, by using Eqs. (\ref{eq39}) and (\ref{eq40}) in Eq. (\ref{eq51}), one can find the differential equation
\begin{eqnarray}
&& \hspace{-0.2cm}  4 \beta_1 r^2 \left(a \beta_2-a \beta_1 W^{\prime 2}-a \beta_2 W+b W-b+W-1\right)+4 \beta_1 r^3 \nonumber\\&& \times (-a \beta_2 W^{\prime}+b W^{\prime}+H^{\prime} W)+\beta_2 r^4 (-a \beta_2+2 b+2)-8 a \beta_1^2 \nonumber\\&& \times W^{\prime} (W-1) r-4 a \beta_1^2 (W-1)^2-4 c r^4=0. \label{eq52}
\end{eqnarray}
The Eq.(\ref{eq52}) leads directly to,
\begin{equation} \label{EOS1}
H(r) = \int{G(W, W',{\bar r)\,{ d\bar r}}}\end{equation}  
where
\begin{eqnarray} 
\label{EOS2}
G(W, W', r)  &=& \frac{1}{4\beta_1r^3W}\Big[ -4 \beta_1 r^2 \Big(a \beta_2-a \beta_1 W^{\prime 2}-a \beta_2 W \nonumber\\&&\hspace{-0.3cm} +b W-b+W-1\Big)-4 \beta_1 r^3 (-a \beta_2 W^{\prime}+b W^{\prime}) \nonumber\\&&\hspace{-0.3cm} -\beta_2 r^4 (-a \beta_2+2 b+2)+8 a \beta_1^2 W^{\prime} (W-1) r \nonumber\\&&\hspace{-0.3cm}-4 a \beta_1^2 (W-1)^2+4 c r^4 \Big]. \nonumber
\end{eqnarray}

The above differential equation (\ref{EOS1}) depends on one undermined unknown $W$. Therefore, we assume a well-known ansatz for $W$ for the Buchdahl model to solve the above differential equation,
\begin{eqnarray}
&& W(r)=\frac{1+Lr^2}{1+Nr^2},  \label{eq53} 
\end{eqnarray}
where $L$ and $N$ are constants with dimensions $\text{length}^{-2}$ and $\text{length}^{-4}$, respectively.  Now plugging of Eq. (\ref{eq53}) into Eq. (\ref{EOS1}) and integrate, we obtain the solution for $H(r)$ of the form, 
\begin{small}
\begin{eqnarray} \label{eq54}
&&\hspace{-0.2cm} H(r)=\frac{1}{8} \Bigg[\frac{1}{\beta_1 L^2 (L-N)}\Big(\log \left(L r^2+1\right) \Big\{4 \beta_1 L^3 (3 a \beta_2-6 a \beta_1\nonumber\\&&\hspace{0.2cm} \times N-3 b-1)+L^2 [\beta_2  (a \beta_2-2 b-2)+8 \beta_1 N (1-2 a \beta_2+2 b)\nonumber\\&&\hspace{0.2cm} +4 a \beta_1^2 N^2]-2 L N (\beta_2 (a \beta_2-2 b-2)+2 \beta_1 N  (-a \beta_2+b+1)) \nonumber\\&&\hspace{0.2cm} +\beta_2 N^2 (a \beta_2-2 b-2)+36 a \beta_1^2 L^4+4 c (L-N)^2\Big\}\Big)+\frac{1}{\beta_1 L} \beta_1\nonumber\\&&\hspace{0.2cm} \times \Big[N r^2 (\beta_2 (a \beta_2-2 b-2)+4 c)\Big]+\frac{\beta_1}{N-L}\big[4 \log \left(N r^2+1\right) \beta_1\nonumber\\&&\hspace{0.2cm} \times \big\{a (9 \beta_1 L^2+2 \beta_2 L-6 \beta_1 L N +\beta_1 N^2-2 \beta_2 N) \nonumber\\&&\hspace{0.2cm} -2 b (L-N)\big\}\big] +\frac{16 a \beta_1 (2 L-N)}{N r^2+1}+\frac{8 a \beta_1 (L-N)}{\left(N r^2+1\right)^2}\Bigg]+F,
\end{eqnarray}
\end{small}
where $F$ is an arbitrary constant of integration. Using the expression for $W(r)$ and $H(r)$, we find the expressions for $\rho$, $p_r$, and $p_t$ as,
\begin{small}
\begin{eqnarray} \label{eq55}
&&\hspace{-0.15cm}\rho(r)= \frac{1}{2 \left(N r^2+1\right)^2}\Big[N^2 \left(2 \beta_1 r^2-\beta_2 r^4\right)-\beta_2-2 \beta_1 L \nonumber\\&& \hspace{0.5cm} \times \left(N r^2+3\right)+N \left(6 \beta_1-2 \beta_2 r^2\right)\Big] ,~~~~~~\\ \label{eq56}
&&\hspace{-0.15cm} p_r(r) =\frac{1}{4 \left(N r^2+1\right)^4}\Big[\Big(\beta_2+2 \beta_1 L \left(N r^2+3\right)+N^2 \nonumber\\&& \hspace{0.5cm} \times \left(\beta_2 r^4-2 \beta_1 r^2\right)  +N \left(2 \beta_2 r^2-6 \beta_1\right)\Big) \Big\{a \Big(\beta_2+2 \beta_1 L \nonumber\\&& \hspace{0.5cm} \times \left(N r^2+3\right)+\beta_2 N^2 r^4-2 \beta_1 N^2 r^2-6 \beta_1 N \nonumber\\&& \hspace{0.5cm} +2 \beta_2 N r^2\Big)-2 b \left(N r^2+1\right)^2\Big\}\Big]+c,\\ 
&&\hspace{-0.15cm} p_{_t}(r)=\frac{1}{64} \Bigg[\frac{f_1^2(r) \left(L r^2+1\right) \beta_1 r^2}{N r^2+1}+\frac{8 f_1(r) \left(L r^2+1\right) \beta_1}{N r^2+1}\nonumber\\&& \hspace{0.6cm} +32 \beta_2+\frac{1}{\left(L r^2+1\right) \left(N r^2+1\right)^5}\Big[8 (L-N) \big(4 c r^2 \nonumber\\&& \hspace{0.6cm} \times \big(N r^2+1\big)^4+r^2 \beta_2 (-2 b +a \beta_2-2) \big(N r^2+1\big)^4 \nonumber\\&& \hspace{0.6cm} +4 \beta_1 \big(N^2 (b-a \beta_2+1) r^4+3 N  (b-a \beta_2 +1) r^2+L \big(N \nonumber\\&& \hspace{0.6cm} \times  (a \beta_2 +1) r^2-b \big(N r^2+3\big)+3 a \beta_2+1\big) r^2+2\big)  \nonumber\\&& \hspace{0.6cm} \times \big(N r^2+1\big)^2+4 a (L-N)^2 r^2  \big(N r^2+3\big)^2 \beta_1^2\big)\Big] \nonumber\\&& \hspace{0.6cm} +\frac{1}{N r^2+1} \Bigg(\frac{8 f_2(r)\big(L r^2+1\big) \beta_1}{\big(L r^2+1\big)^2 \big(N r^2+1\big)^4 \beta_1} \nonumber\\&& \hspace{0.6cm} +\frac{8 f_3(r) \big(L r^2+1\big) \beta_1}{\big(N r^2+1\big)^3 \big(L \beta_1 r^2+\beta_1\big)}\Bigg)\Bigg], \label{eq57}
\end{eqnarray}
\end{small}
where the used coefficients in the above expressions are mentioned in the Appendix. 

In the above set of Eqs. (\ref{eq55})--(\ref{eq57}), we have the complete spacetime geometry for the seed solution. However, corresponding to the  $\theta$-sector, we need to find the solution of the second system of Eqs. (\ref{eq44})--(\ref{eq46}). However, note that there are three independent equations with five unknowns. This situation, therefore, demands for a need of two additional information to close the $\theta$-system, e.g. $\Psi$ and $\eta(r)$. It is known that physical viability $\Phi (r)$ should have a monotonic increasing feature towards the boundary and so $H(r)+\alpha \eta (r)$ must be an increasing function of $r$. Hence for simplicity, we assume $\eta(r)=H(r)$, which provides eventually $\Phi(r)=(1+\alpha) H (r)$.

Related to the system of equations (\ref{eq44}) and (\ref{eq45}) for the source $\theta_{ij}$, after imposing the constraint $\beta_2\ne 0$, we opt for the following preferences:
\begin{eqnarray}
&&\hspace{-0.5cm} \Psi(r)=W(r)-1+\frac{\beta_2 r^2}{6 \beta_1}~~\label{eq58}\\
&&\hspace{-0.5cm} \Psi(r)=\frac{1}{ 1+r\,\Phi'(r)}- \frac{W(r) \left[1+r \{H'(r)+\eta'(r)\big\}\right]}{1+r\,\Phi'(r)} \nonumber\\&& \hspace{0.6cm} -\frac{\beta_2 r^2}{2 \beta_1}.~~~~~~~\label{eq59}
\end{eqnarray}

One can note that $\Psi(r)$ is free from any singularity and also $\Psi(0)=0$. As a result, these allow us to mimic (i) $\theta^0_0$ with the energy density $\rho$, i.e. $\rho=\theta^0_0$ from Eq. \ref{eq58} and (ii) $\theta^1_1$ with the radial pressure $p_r$, i.e. $p_r=\theta^1_1$ from Eq. (\ref{eq59}). Several authors ~\citep{sharif1,sharif2,sharif3,sunil5d1,sunil5d2,sunil5d3,sunil5d4} have successfully applied this technique in modeling compact objects in GR as well as modified gravity theories and its gravitational cracking concept under gravitational decoupling \citep{Contreras}. Motivated by these works, we use the mimic approach in the present study for the system of equations (\ref{eq39})-(\ref{eq41}). Therefore, we adopt the following approaches: (i) Mimicking of the density constraint (i.e. $\rho=\theta^0_0$) and (ii) Mimicking of the radial pressure constraint (i.e. $p_r=\theta^1_1$) [vide for details the following Ref.~\citep{OvallePRD2017}].

\subsection{\textbf{Mimicking of the density constraint (i.e., $\rho=\theta^0_0$)}} \label{solA}

To solve the $\theta$-sector, here we mimic the seed energy density $\rho$ to $\theta^0_0$ from Eqs. (\ref{eq39}) and (\ref{eq44}), we find first order linear differential equation in $\Psi(r)$ as, 
\begin{eqnarray}\label{eq60}
\Psi^{\prime}+\frac{\Psi}{r}+\frac{1}{2\beta_1 r}\Big[2\beta_1(1-W-r\,W^{\prime})  - r^2\,\beta_2\Big] =0.
\end{eqnarray}

Now we obtain the expression of deformation function $\Psi(r)$ after integrating the above equation by using the known potential $H(r)$ as
\begin{eqnarray} \label{eq61}
\Psi(r)=\frac{r^2 \left(\beta_2+6 \beta_1 L-6 \beta_1 N+\beta_2 N r^2\right)}{6 \beta_1 \left(1+ N r^2\right)}.
\end{eqnarray}

The arbitrary constant of integration has been taken to be zero to ensure the non-singular nature of $\Psi(r)$ at the center. Furthermore, we take the deformation function $\eta(r)=H(r)$ as mentioned above in order to find the expression for $\theta$-sector.  Hence, the $\theta$-sector components are obtained as
\begin{small}
\begin{eqnarray}\label{eq62}
&&\hspace{-0.2cm}\theta^0_0(r)= \frac{1}{2 \left(N r^2+1\right)^2}\Big[N^2 \left(2 \beta_1 r^2-\beta_2 r^4\right)-\beta_2-2 \beta_1 L \nonumber\\&& \hspace{0.6cm} \times \left(N r^2+3\right)+N \left(6 \beta_1-2 \beta_2 r^2\right)\Big] ,~\\
&&\hspace{-0.2cm}\theta^1_1(r)=\frac{-1}{6(1+ N r^2)}\Big[\beta_2+6 \beta_1 \theta_{11}(r) \big((\alpha +2) L r^2-(\alpha +1) \nonumber\\&& \hspace{0.6cm} \times N r^2+1\big)+\beta_2 (\alpha +1) \theta_{11}(r) r^2 \left(N r^2+1\right)+6 \beta_1 L \nonumber\\&& \hspace{0.6cm} -6 \beta_1 N+\beta_2 N r^2\Big], ~~~\label{eq63}\\
&&\hspace{-0.2cm} \theta^2_2(r)= \frac{-1}{24 \left(N r^2+1\right)^2}\Big[2 \theta_{22}(r) \left(N r^2+1\right) \big[6 \beta_1 \big(1+(\alpha +2) L r^2 \nonumber\\&& \hspace{0.4cm} -(\alpha +1)  N r^2\big)+(\alpha +1) \beta_2 r^2  \left(N r^2+1\right)\big] + \theta_{23} (r)\Big], \label{eq64}
\end{eqnarray}
\end{small}
where the explicit expressions for $\theta_{22}(r)$ and $\theta_{23}(r)$ are given in Appendix. 

\subsection{\textbf{Mimicking of the pressure constraint (i.e., $p_r=\theta^1_1$)}} \label{solB}

In this mimic constraints approach, we are mimicking the  seed radial pressure $p_r$ with the $\theta^1_1(r)$ from Eqs. (\ref{eq40}) and (\ref{eq45}),  we obtain the expression for deformation function $\Psi(r)$ as
\begin{small}
\begin{eqnarray} \label{eq65}
&& \hspace{-0.2cm}\Psi(r)=-\frac{1}{\left(N r^2+1\right) \Psi_{21}(r)}\Big[2 r^2 \big(L r^2+1\big) \Big\{a \beta_2^2-2 \beta_1 L \big[-6 a \beta_2  \nonumber\\&&\hspace{0.2cm} +N^3 \big(r^6 (1-2 a \beta_2)+4 a \beta_1 r^4\big) +N^2 \big(r^4 (3-10 a \beta_2)+24 a \nonumber\\&&\hspace{0.2cm} \times \beta_1 r^2\big) +\Psi_{22} (r)\Big\}\Big], ~~~~~
\end{eqnarray}
\end{small}
Finally using the same deformation function $\eta(r)= H(r)$, we find the $\theta$-components for this solution as
\begin{small}
\begin{eqnarray}\label{eq66}
&&\hspace{-0.2cm} \theta^0_0(r)= \frac{2 \beta_1 \big[4 N r^2 \big(L r^2+1 \big) \big(N r^2+1\big) \Psi_{00} (r) +f_4(r)\big]}{3 f_3(r) \left(N r^2+1\right)^2}  ,~~~~~\\
&&\hspace{-0.15cm} \theta_{1}^{1}(r)= \frac{1}{4 \left(N r^2+1\right)^4}\Big[\Big(\beta_2+2 \beta_1 L \left(N r^2+3\right)+N^2 \big(\beta_2 r^4 \nonumber\\&&\hspace{0.3cm} -2 \beta_1 r^2\big)  +N \left(2 \beta_2 r^2-6 \beta_1\right)\Big) \Big\{a \Big(\beta_2+2 \beta_1 L \left(N r^2+3\right)\nonumber\\&&\hspace{0.3cm}+\beta_2 N^2 r^4-2 \beta_1  N^2 r^2-6 \beta_1 N+2 \beta_2 N r^2\Big) \nonumber\\&&\hspace{0.3cm} -2 b (N r^2+1)^2\Big\}\Big]+c, \label{eq67}
\end{eqnarray}
\end{small}
where the explicit expressions for $\Psi_{21}(r)$ and $\Psi_{22}$, and $\Psi_{00}(r)$ are given in Appendix. 

\section{Exterior spacetime and matching conditions} \label{sec4}

At this juncture, let us apply the boundary condition to have the expressions for the constants as well as the physical parameters to study the features of the compact star. For this purpose, we need to match the interior spacetime smoothly with a suitable exterior vacuum solution at the pressureless bounding surface (i.e. at $r=R$) for the functional form of $f(\mathcal{Q})=\beta_1 \mathcal{Q}+ \beta_2$.  

Following the analysis provided by~\citet{Wang} for the off-diagonal component as appeared in Eq. (\ref{eq20}), the solutions of $f(Q)$ gravity are restricted to the following two cases: 
\begin{eqnarray}
 && f_{{\mathcal{Q}\mathcal{Q}}} =0\,~~~\Rightarrow~~f(\mathcal{Q})=\beta_1\,\mathcal{Q}+\beta_2, \label{Ap1} \\
  && \mathcal{Q}^{\prime}=0~~ \Rightarrow ~~ \mathcal{Q}=\mathcal{Q}_0. \label{Ap2}
\end{eqnarray}
where $\beta_1$, $\beta_2$ and $\mathcal{Q}_0$ constants. 

One can note that this result is similar to the one in $f(T)$ gravity theory-related work in Ref.~\citep{Boehmer:2011gw}, where they have considered the constraint case $f_{TT}=0$ or $T^{\prime}=0$ for the spherically symmetric static distributions with diagonal tetrad as background formalism. One can also note that for the cosmological constant assumed to be $\beta_2/\beta_1$, the first solution Eq. (\ref{Ap1}) is equivalent to GR as it reduces to STGR. We shall discuss this point later on in a conversant way.

We have the physical $\epsilon=P_r=P_\perp=0$ in the case of vacuum. Therefore, the equation of motions (\ref{eq21})--(\ref{eq23}), due to Eq. (\ref{Ap1}), take the forms 
\begin{eqnarray}
 && \Phi^{\prime}+\mu^{\prime} =0, \label{Ap3}\\
 && \frac{\beta_2}{\beta_1}-\frac{2}{r^2}=\mathcal{Q}, \label{Ap4}\\
 && \frac{\beta_2}{2}+\beta_1 e^{-\mu}\Bigg[\frac{\Phi^{\prime \prime}}{2}+\Big(\frac{\Phi^{\prime}}{4}+\frac{1}{2r}\Big)(\Phi^{\prime}-\mu^{\prime})\Bigg]=0. \label{Ap5}
\end{eqnarray}

Now, from Eq. (\ref{Ap3}), one can get in a straightforward way the following one
\begin{eqnarray}
  \Phi(r)=-\mu(r)+\mu_0,  \label{Ap6}
\end{eqnarray} 
where $\mu_0$ is a constant of integration.

It is to note that, for the sake of convenience, $\mu_0$ can be absorbed via the rescaling of the time coordinate $t$ to $e^{-\mu_0/2}$. As a result, the $rr$-component in Eq. (\ref{eq14}) becomes the inverse of the $tt$ component, so that one may obtain 
\begin{eqnarray}
    \Phi(r)=-\mu(r). \label{Ap7}
\end{eqnarray}

In Eq. (\ref{Ap4}), we consider the term $\beta_2/\beta_1$ as the cosmological constant of Einstein $\Lambda$ with a reversed sign due to the convention of nonmetricity of Eq. (\ref{eq7}).

Now, from the relation (\ref{eq16}) along with Eqs. (\ref{Ap3}) and (\ref{Ap4}), we get the expression for ${\mu}$ as
\begin{eqnarray}
    && e^{\mu}=\Big(1+\frac{\mu_1}{r}-\frac{\beta_2}{6\beta_1} r^2\Big)^{-1}, \label{Ap8}
\end{eqnarray}
where $\mu_1$ is an integration constant.

In a similar way $\Phi(r)$ can be found from Eq. (\ref{Ap7}) and (\ref{Ap8}) as follows:
\begin{eqnarray}
   && e^{\Phi}=\Big(1+\frac{\mu_1}{r}-\frac{\beta_2}{6\beta_1} r^2\Big). \label{Ap8a}  
\end{eqnarray}

Hence the explicit line element can be provided as
\begin{eqnarray}
&&\hspace{-0.5cm} ds^2=-\Big(1+\frac{\mu_1}{r}-\frac{\beta_2}{6\beta_1} r^2\Big) dt^2+\Big(1+\frac{\mu_1}{r}-\frac{\beta_2}{6\beta_1} r^2\Big)^{-1}\nonumber\\
&&\hspace{0.2cm} dr^2+r^2(d\theta^2 +\text{sin}^2\theta \,d\phi^2). \label{Ap9} 
\end{eqnarray}
A few interesting observations from the above metric (\ref{Ap9}) are as follows: straightforwardly it represents the Schwarzschild (anti–)de Sitter solution with (i) the cosmological constant $\Lambda=\beta_2/\beta_1$, (ii) the mass of the stellar object $\mu_1=2 \mathcal{M}$. This comparison and resemblance indicate that there exists the Schwarzschild (anti–)de Sitter solution only for the linear $f(\mathcal{Q})$ gravity which is equivalent to GR.

Let us now treat the following situations: (i) solution for $\mathcal{Q}=\mathcal{Q}_0$ as obtained in Eq. (\ref{Ap2}), which acts as a constraint on the functional form of $f(\mathcal{Q})$, and (ii) the nonmetricity scalar constant $\mathcal{Q}_0$ which can be considered as equivalent to the cosmological constant $\Lambda$, as shown in Eq. (\ref{Ap4}). Hence, in the present case the non-metricity scalar $\mathcal{Q}$ can be expressed as 
\begin{eqnarray}
    \mathcal{Q}_0=-\frac{2e^{-\mu}}{r}\Big(\Phi^{\prime}+\frac{1}{r}\Big).\label{Ap10a}
\end{eqnarray}

Now, under the constant non-metricity scalar (i.e. $\mathcal{Q}=\mathcal{Q}_0$) and vacuum case (i.e. $\epsilon=P_r=P_\perp=0$), the equation of motions (\ref{eq22})--(\ref{eq24}) become 
\begin{eqnarray}
  &&\hspace{-0.5cm}  f_{\mathcal{Q}}(\mathcal{Q}_0) \frac{e^{-\mu}}{r} (\Phi^{\prime}+\mu^\prime)=0,\label{Ap10} \\
  &&\hspace{-0.5cm}  - \frac{f_{\mathcal{Q}}(\mathcal{Q}_0)}{2}+f_{\mathcal{Q}}(\mathcal{Q}_0) \Big(\mathcal{Q}_0+\frac{1}{r^2}\Big)=0, \label{Ap11} \\
  &&\hspace{-0.5cm} f_{\mathcal{Q}}(\mathcal{Q}_0) \Bigg[\frac{\mathcal{Q}_0}{2}+\frac{1}{r^2}+e^{-\mu}\Bigg\{\frac{\Phi^{\prime \prime}}{2}+\Big(\frac{\Phi^{\prime}}{4}+\frac{1}{2r}\Big) \nonumber\\&&\hspace{3.5cm} \times (\Phi^{\prime}-\mu^{\prime})\Bigg\} \Bigg]=0.~~~~~~~~~ \label{Ap12} 
  \end{eqnarray}
  
The above set of equations (\ref{Ap10}) and (\ref{Ap11}) readily provide
\begin{eqnarray}
  f(\mathcal{Q}_0)=0,~~~\text{or}~~~f_{\mathcal{Q}}(\mathcal{Q}_0)=0.  \label{Ap13}
\end{eqnarray}

These two restrictions immediately give clue that the functional form of $f(\mathcal{Q})$ can be expressed in terms of power series expansion of $f(\mathcal{Q})$ around $\mathcal{Q}=\mathcal{Q}_0$ which may be as follows: 
\begin{eqnarray}
 && \hspace{-0.2cm}f(\mathcal{Q})=\mathcal{Q}_1 (\mathcal{Q}-\mathcal{Q}_0)^1+\mathcal{Q}_2 (\mathcal{Q}-\mathcal{Q}_0)^2+\mathcal{Q}_3 (\mathcal{Q}-\mathcal{Q}_0)^3\nonumber\\
 &&\hspace{0.8cm} +\mathcal{Q}_4 (\mathcal{Q}-\mathcal{Q}_0)^4+.......... ~~~~ \label{Ap14}
\end{eqnarray}
and which can be written in the general form as
\begin{eqnarray}
  f(\mathcal{Q})=  \sum_{n=1} \mathcal{Q}_n (\mathcal{Q}-\mathcal{Q}_0)^n, \label{Ap14a}
\end{eqnarray}
$\mathcal{Q}_1$,~ $\mathcal{Q}_2$, ~$\mathcal{Q}_3$, ..... being the constant coefficients. 

In the present situation of $f(\mathcal{Q})$ gravity-related non-trivial solution one should keep in mind that Eq. (\ref{Ap13}) should be satisfied by the functional form of $f(\mathcal{Q})$. This is essential to obtain new solutions which are distinctly different than GR. 

Now, from Eq. (\ref{Ap10}), one may get another situation which is
\begin{eqnarray}
   &&  \Phi^\prime+\mu^\prime=0. \label{eq83}
\end{eqnarray}

Then, substituting Eq. (\ref{eq83}) in Eq. (\ref{Ap10a}) we get
\begin{eqnarray}
    e^{\Phi(r)}=\frac{\mu_1}{r}-\frac{Q_0}{6}r^2~~\text{and}~~e^{-\mu(r)}=\frac{\mu_1}{r}-\frac{Q_0}{6}r^2. \label{Ap15}
\end{eqnarray}

Therefore, the metric (\ref{eq14}) eventually takes the form as follows: 
\begin{eqnarray}
 ds^2=-\Big(\frac{\mu_1}{r}-\frac{Q_0}{6}r^2\Big) dt^2+\Big(\frac{\mu_1}{r}-\frac{Q_0}{6}r^2\Big)^{-1}dr^2 \nonumber\\~+r^2d\theta^2
 +r^2\text{sin}^2\theta \,d\phi^2.~~~~~~ \label{Ap16}
\end{eqnarray}
It is noticeable that the above line element is not the same as the Schwarzschild solution which implies that the exact Schwarzschild solution does not at all exist for the non-trivial functional form of $f(\mathcal{Q})$.

By taking into account the above discussion, we opt the Schwarzschild anti-de Sitter spacetime in $f(\mathcal{Q})$ gravity under the functional form (\ref{Ap1}) which can be provided  as, 
\begin{eqnarray}\label{eq4.1}
ds^2_+ =-\bigg(1-\frac{2{\mathcal{M}}}{r}-{\Lambda \over 3}~r^2\bigg)\,dt^2+\frac{dr^2}{\bigg(1-\frac{2{\mathcal{M}}}{r}-{\Lambda \over 3}~r^2\bigg)} \nonumber\\ +r^2 \Big(d\theta^2  +\sin^2\theta\,d\phi^2 \Big),~~~~~~
\end{eqnarray} 
where $\mathcal{M}=\hat{M}_\mathcal{Q}/\beta_1$ and $\Lambda=\beta_2/2\beta_1$, where $\hat{m}_\mathcal{Q}(R)=\hat{M}_\mathcal{Q}$. Therefore, it is clearly observed that when $\beta_1=1$ and $\beta_2=0$, the Schwarzschild anti-de Sitter spacetime (\ref{eq4.1}) reduces to the Schwarzschild exterior solution. 

On the other hand, the minimally deformed interior spacetime for the region ($0 \le r \le R$)  is given by
\begin{eqnarray}\label{eq4.2}
 ds^2_{-}= -e^{H(r)+\alpha\,\eta(r)}\,dt^2+ \big[W(r)+\alpha\,\Psi(r)\big]^{-1} dr^2 \nonumber\\+r^2(d\theta^2  +\sin^2\theta\,d\phi^2). \label{metric2}~~~~~ 
\end{eqnarray} 

As usual, here we employ the Israel-Darmois matching conditions~\citep{Israel1966,Darmois1927}, i.e. to satisfy the first and second form at the interface and mathematically, this can be given as 
\begin{small}
\begin{eqnarray}
 \label{eq4.3}
&& \hspace{-0.2cm}  e^{\Phi^{-}(r)}|_{r=R}=e^{\Phi^{+}(r)}|_{r=R}~\mbox{and}~e^{\lambda^{-}(r)}|_{r=R}=e^{\lambda^{+}(r)}(r)|_{r=R},~~~~~\\
&& \hspace{-0.15cm}\big[G_{i\,\varepsilon}\,r^{\varepsilon}\big]_{\Sigma}\equiv \lim_{r \rightarrow R^{+}} (G_{\epsilon\,\varepsilon})-\lim_{r \rightarrow R^{-}} (G_{\epsilon\,\varepsilon})=0~~\nonumber\\&& \hspace{0.5cm}  \Longrightarrow~~~ \big[T^{\text{eff}}_{\epsilon\,\varepsilon}\,r^{\varepsilon}\big]_{\Sigma}=\big[(T_{\epsilon\,\varepsilon}+\alpha\,\theta_{\epsilon\,\varepsilon})\,r^{\varepsilon}\big]_{\Sigma}=0,~~~\label{eq4.4}
\end{eqnarray}
The conditions (\ref{eq4.3}) and (\ref{eq4.4}) yields,
\begin{eqnarray} 
&&\hspace{-0.2cm} e^{H(R)+\alpha \eta(R)} = \bigg(1-\frac{2{\mathcal{M}}}{R}-{\Lambda \over 3}~R^2\bigg)~~\nonumber\\&&\hspace{-0.2cm}\text{and}~~
W(R)+\alpha\,\psi(R) = \bigg(1-\frac{2{\mathcal{M}}}{R}-{\Lambda \over 3}~R^2\bigg), \label{eq92}\\
&& \hspace{-0.2cm} P_r(R) = p_r(R)+\alpha\,\beta_1\, \,\Big[\psi_{_\Sigma}\Big(\frac{1 }{R^2}+\frac{H^{\prime}_{_\Sigma}   }{R}\Big)+\frac{W_{_\Sigma}\,\eta^{\prime}_{_\Sigma}}{R} \Big]=0. ~~~~~\label{eq93}
\end{eqnarray}
\end{small}
In order to calculate the numerical values we employ Eqs. (\ref{eq92}) and (\ref{eq93}) and thus determine the unknown parameters such as the constant ($F$), mass ($\mathcal{M}$) and arbitrary constant ($c$).

\begin{figure*}
     \centering
     \rotatebox{90}{%
   \begin{minipage}{1\textheight}
      \includegraphics[width=4.1cm,height=4.1cm]{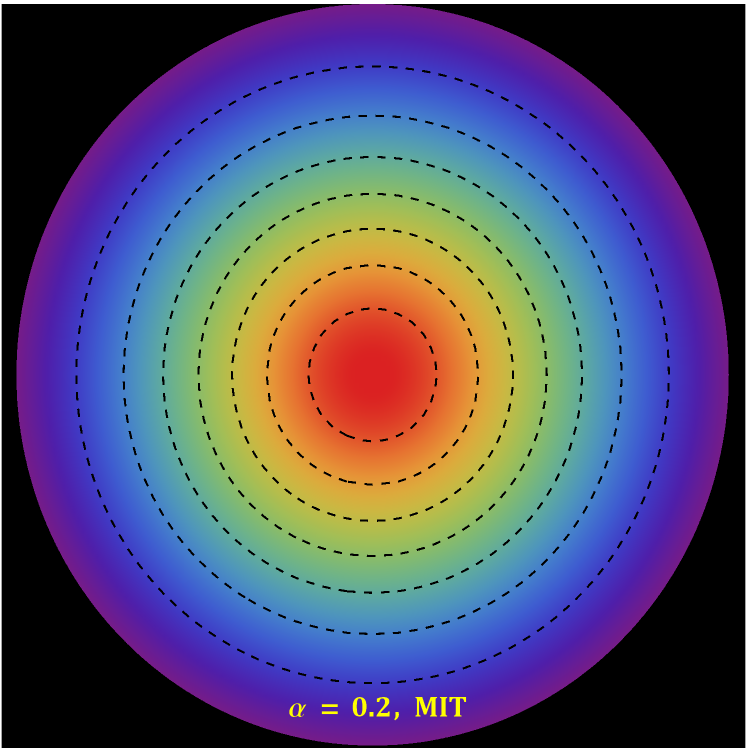}\,\includegraphics[width=1.2cm,height=4.1cm]{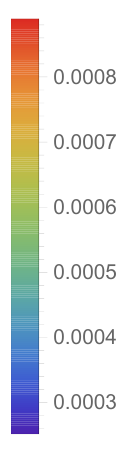} ~~~\includegraphics[width=4.1cm,height=4.1cm]{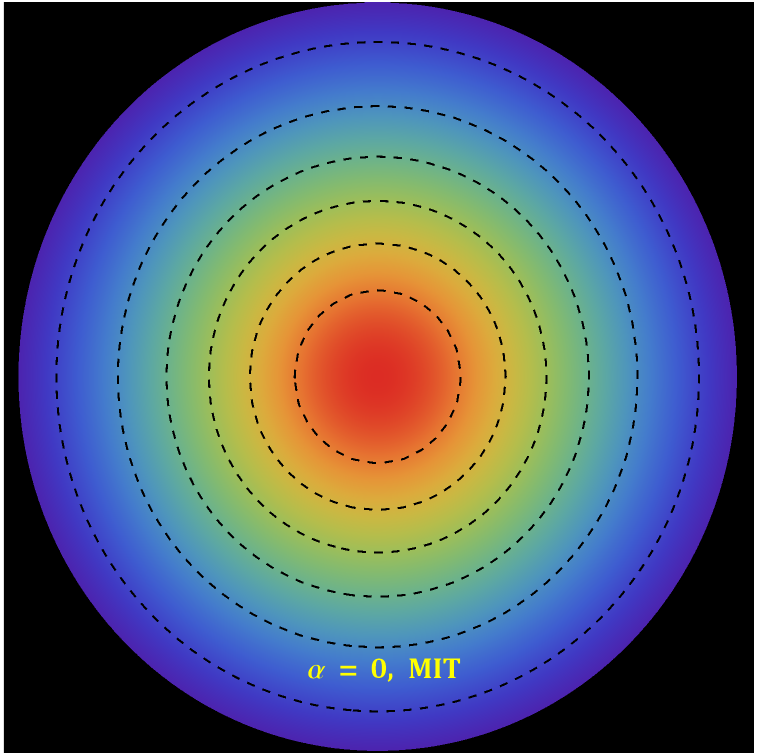}\includegraphics[width=1.2cm,height=4.1cm]{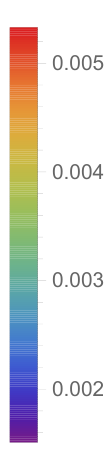}\,~~~
      \includegraphics[width=4.1cm,height=4.1cm]{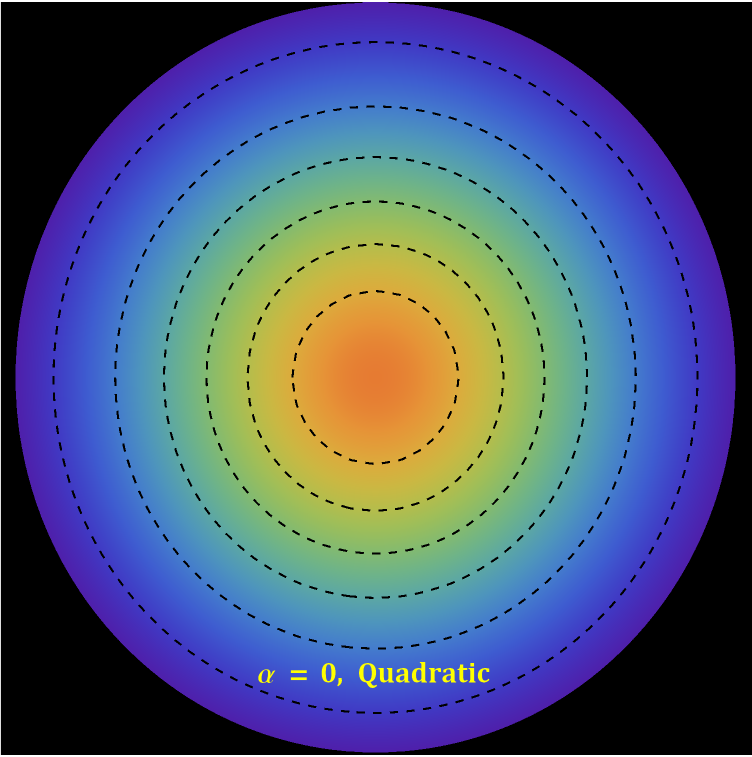}\,\includegraphics[width=1.2cm,height=4.1cm]{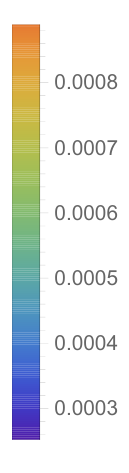} ~~~\,\includegraphics[width=4.1cm,height=4.1cm]{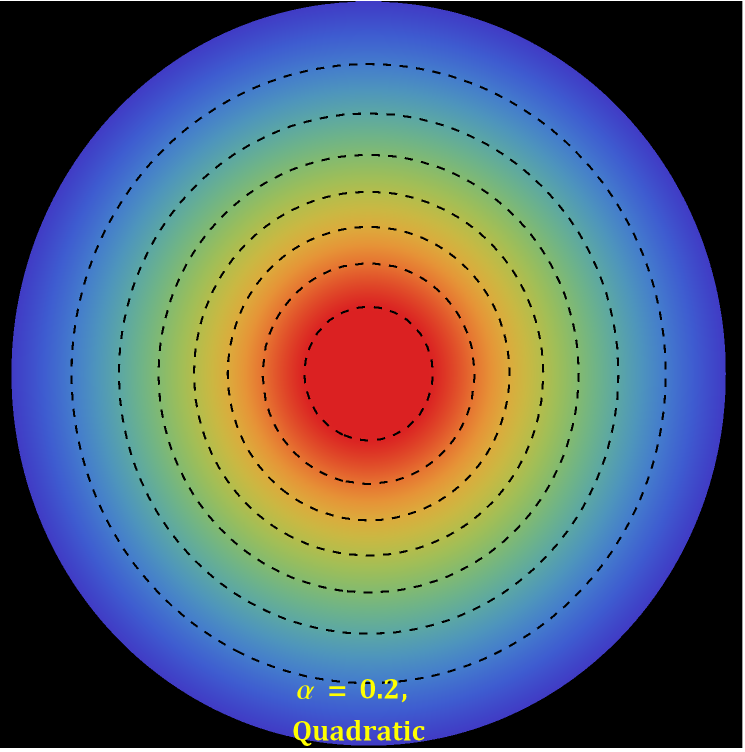}\,\includegraphics[width=1.2cm,height=4.1cm]{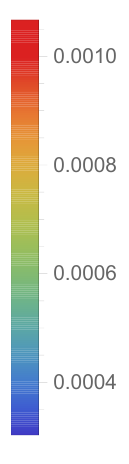}
      \caption{The distribution of energy density {[$\epsilon(r)$ in $\text{km}^{-2}]$ versus radial distance $r$ from center to boundary of star}  for MIT Bag model by taking $a=0$ and $\mathcal{B}_g=60\,MeV/fm^3$ (first two panels) and quadratic model by taking $a=5\,\text{km}^2$ and $\mathcal{B}_g=60\,MeV/fm^3$ {[$1\,MeV/fm^3=1.3234 \times 10^{-6}~ \text{km}^{-2}$]} (right two panels) with two different values of $\alpha$ for solution \ref{solA} ($\rho=\theta^0_0)$). We have chosen the following fixed values of constants $L=0.0011/\text{km}^2$, $N= 0.00787/\text{km}^2$, radius $R=12.5\,\text{km}$, $\beta_1= 1.1$, and $\beta_2=10^{-46}/\text{km}^{2}$ to plot these curves. {The color bar shows the amount of energy density [$\epsilon(r)$ in $\text{km}^{-2}]$ for $r=0$ to $r=12.5~\text{km}$. }} \label{fig1}
\includegraphics[width=4.1cm,height=4.1cm]{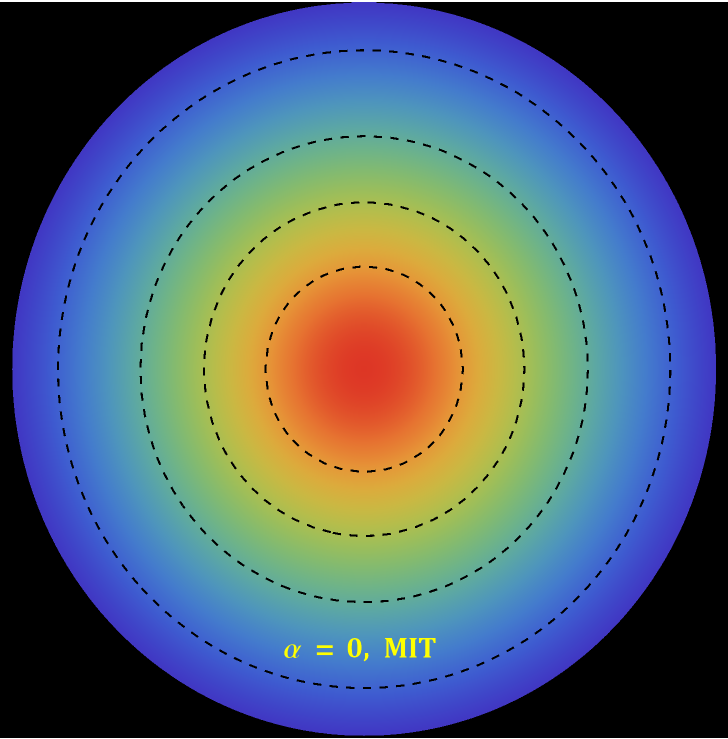}\,\includegraphics[width=1.2cm,height=4.1cm]{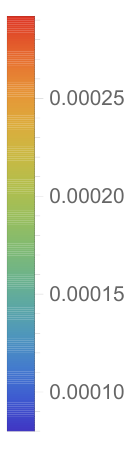}~~~\includegraphics[width=4.1cm,height=4.1cm]{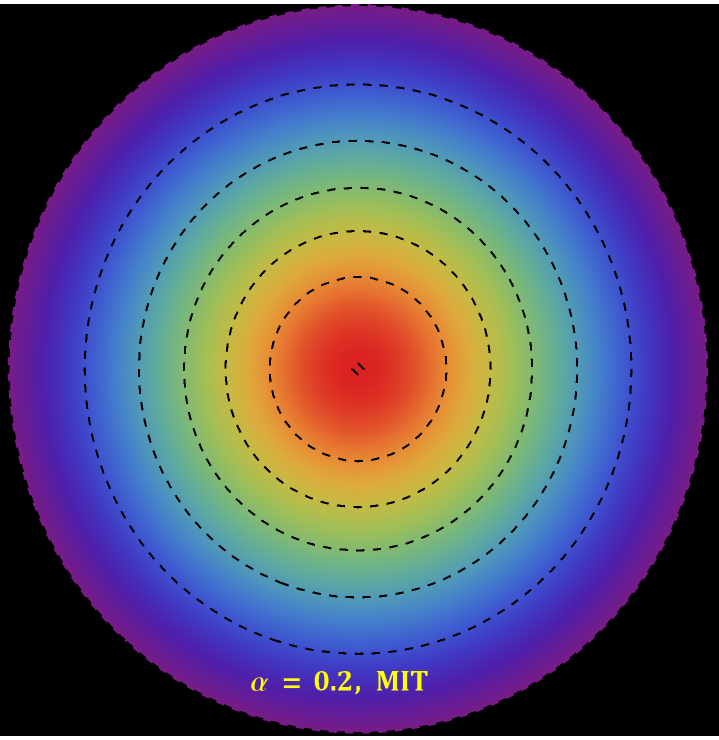}\,\includegraphics[width=1.2cm,height=4.1cm]{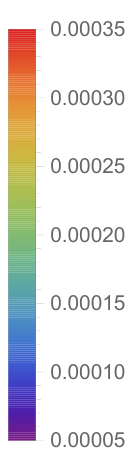}\,~~~
      \includegraphics[width=4.1cm,height=4.1cm]{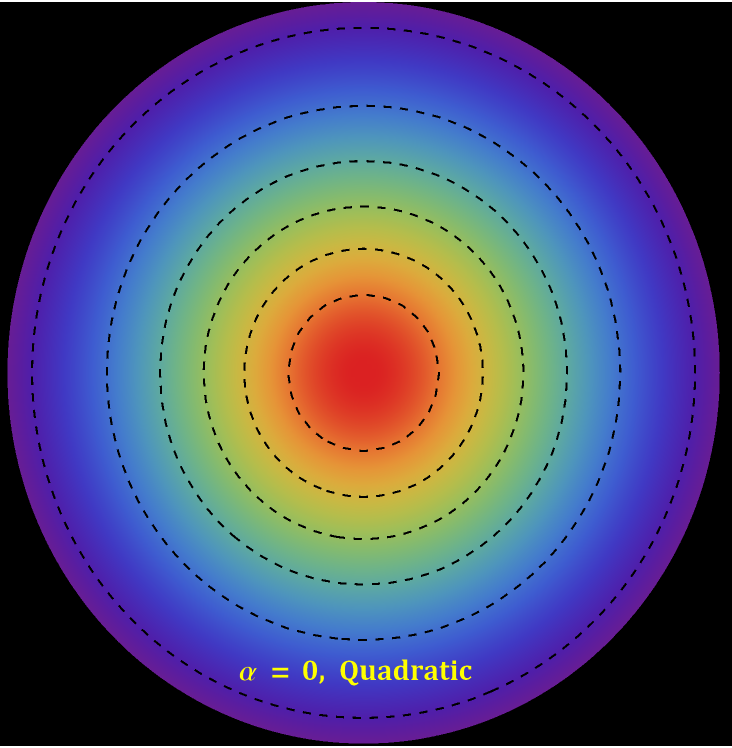}\,\includegraphics[width=1.2cm,height=4.1cm]{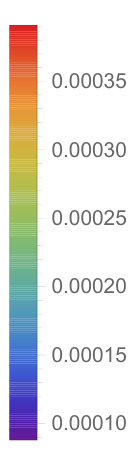} ~~~\,\includegraphics[width=4.1cm,height=4.1cm]{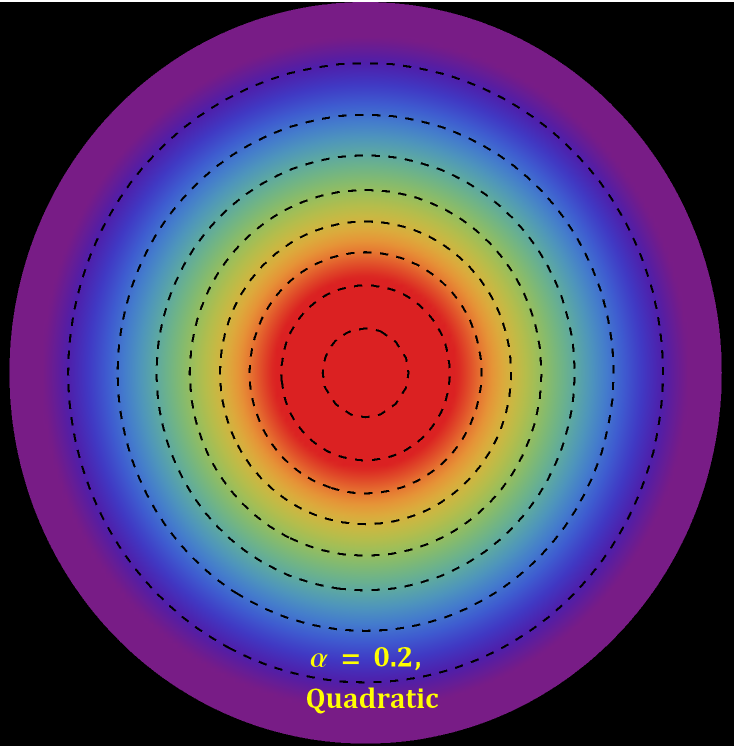}\,\includegraphics[width=1.2cm,height=4cm]{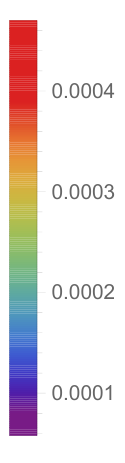}
    \caption{The distribution of radial pressure {[$P_r(r)$ in $\text{km}^{-2}]$ versus radial distance $r$ from center to boundary of star}  for MIT Bag model by taking $a=0$ and $\mathcal{B}_g=60\,MeV/fm^3$ (first two panels) and quadratic model by taking $a=5\,\text{km}^2$ and $\mathcal{B}_g=60\,MeV/fm^3$ (right two panels) with two different values of $\alpha$ for solution \ref{solA} ($\rho=\theta^0_0)$). The same values of constants are employed here as used in Fig.\ref{fig1}. {The color bar shows the amount of radial pressure [$P_r(r)$ in $\text{km}^{-2}]$ for $r=0$ to $r=12.5~\text{km}$. }}
    \label{fig2}
       \includegraphics[width=4cm,height=4.1cm]{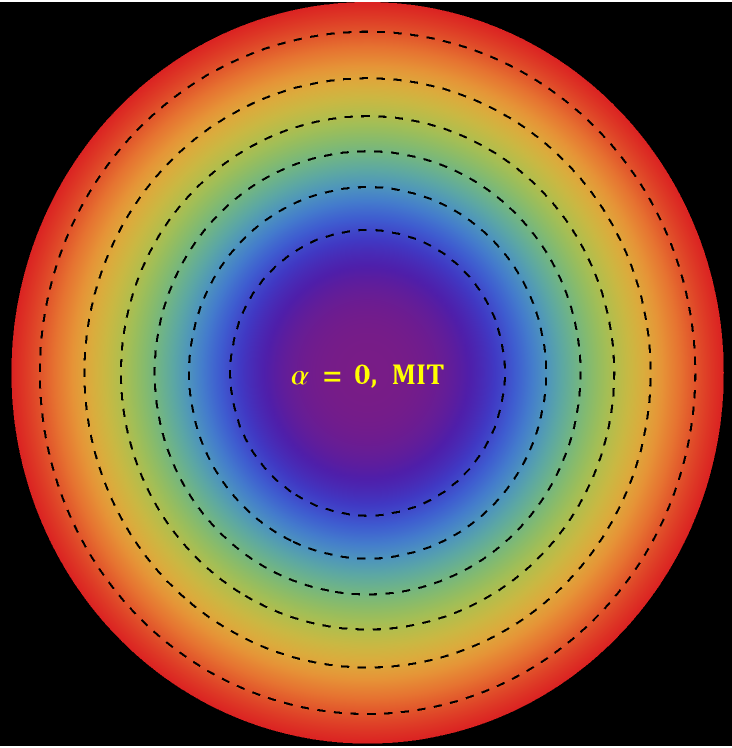}\,\includegraphics[width=1.2cm,height=4.1cm]{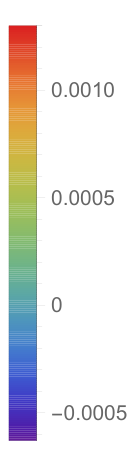}~~~\includegraphics[width=4.1cm,height=4.1cm]{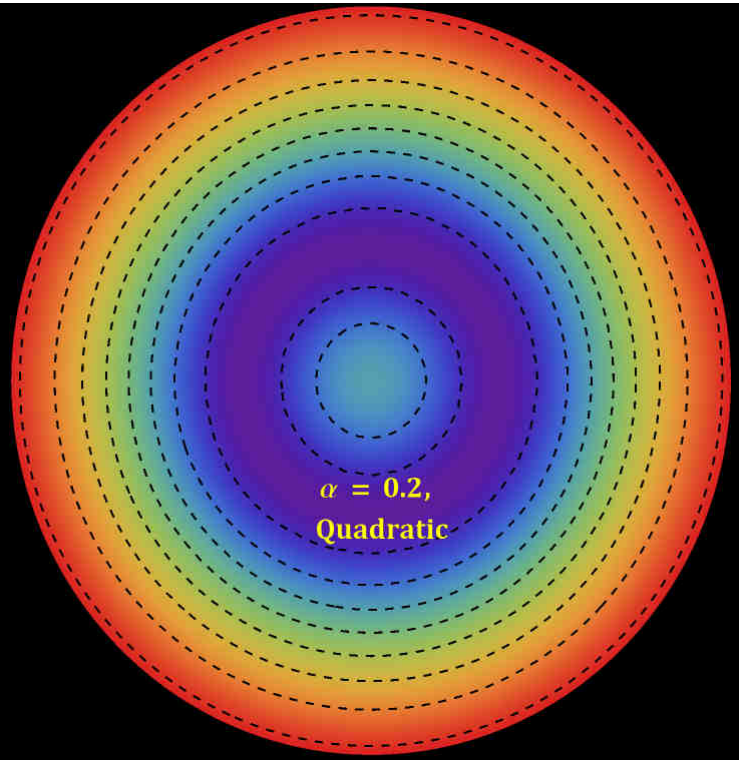}~~~\includegraphics[width=1.2cm,height=4.1cm]{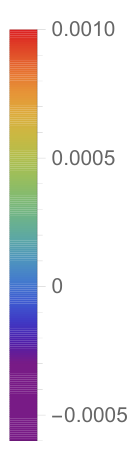}\,
      \includegraphics[width=4.1cm,height=4.1cm]{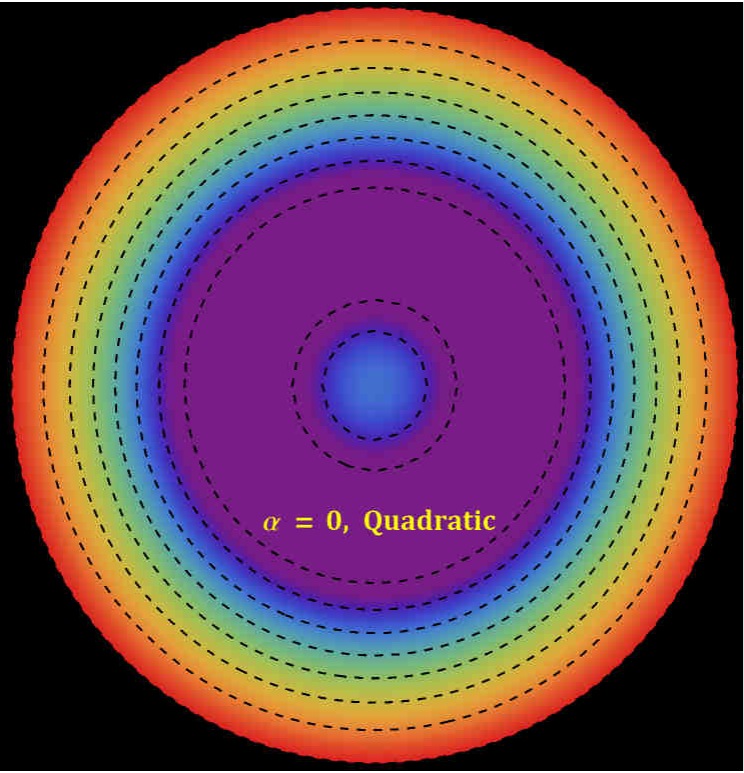}\,\includegraphics[width=1.2cm,height=4.1cm]{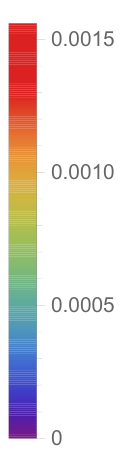} ~~~\,\includegraphics[width=4.1cm,height=4cm]{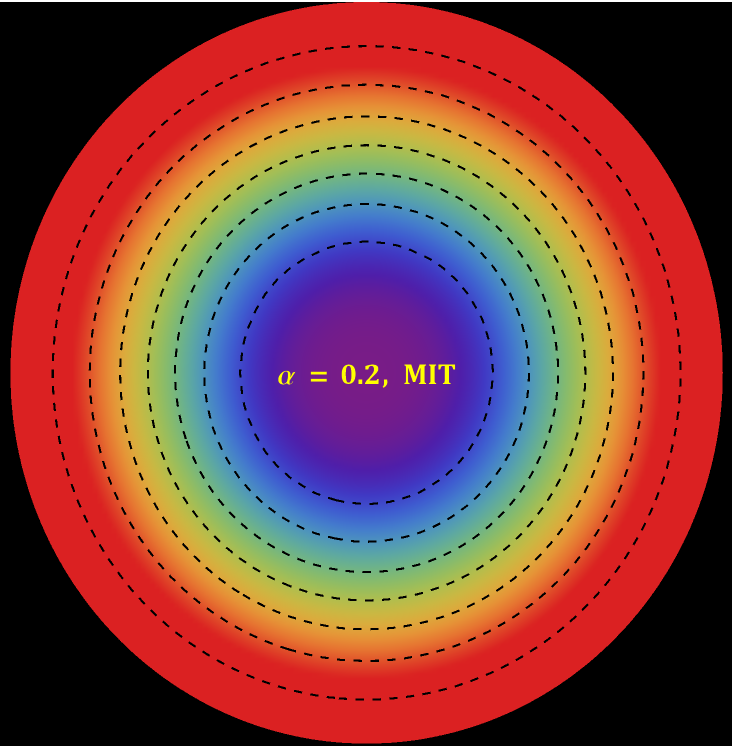}\,\includegraphics[width=1.2cm,height=4.1cm]{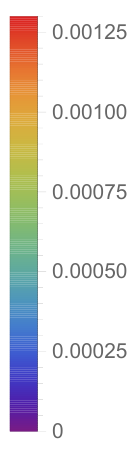}
    \caption{The distribution of anisotropy {[$\Delta(r)$ in $\text{km}^{-2}]$ versus radial distance $r$ from center to boundary of star} for MIT Bag model by taking $a=0$ and $\mathcal{B}_g=60\,MeV/fm^3$ (first two panels) and quadratic model by taking $a=5\,\text{km}^2$ and $\mathcal{B}_g=60\,MeV/fm^3$ (right two panels) with two different values of $\alpha$ for solution \ref{solA} ($\rho=\theta^0_0)$). The same values of constants are employed here as used in Fig.\ref{fig1}. {The color bar shows the amount of anisotropy [$\Delta(r)$ in $\text{km}^{-2}]$ for $r=0$ to $r=12.5~\text{km}$. }}
    \label{fig3}
      \end{minipage}}
\end{figure*}


\section{Physical analysis of completely deformed SS models and astrophysical implications } \label{sec5}

\subsection{Regular behavior of  strange star (SS) models} \label{5.1}

\subsubsection{For solution \ref{solA} ($\rho=\theta^0_0$)} \label{5.1.1}
We now turn our attention to the physical analysis of the models obtained for the $\rho=\theta^0_0$ sector. The energy density is plotted in Fig 1. Keep in mind that for $\alpha = 0$ we recover the standard $f(\mathcal{Q})$ gravity. Starting from the left of Fig. \ref{fig1}. the first and second plots reveal the behavior of the energy density as a function of the radial coordinate for a fluid obeying the MIT bag model EOS. We observe that the energy density is regular at all interior points of the fluid configuration, attaining a maximum at the center. It is clear that contributions from the decoupling constant, $\alpha$ leads to higher core densities in the linear regime as compared to the standard $f(\mathcal{Q})$ gravity models. The third and fourth panels reveal the trend in the energy density for a fluid obeying a quadratic EOS. We note that for the quadratic EOS, contributions from $\alpha$ lead to a significant increase in the energy density, particularly in the central regions of the star. More ever, the closeness of the contours reveals that these higher densities lead to more compact configurations. If we compare the linear EOS to the quadratic EOS for nonzero decoupling constant, we observe that stars with quadratic EOS have higher core densities and are more compact than their linear EOS counterparts. Panels one and three reveal that in the absence of the decoupling constant $(\alpha = 0)$, the energy density in the linear regime dominates its quadratic counterpart. The linear model predicts high core densities but as one moves to the surface layers of the star, the densities are of similar magnitude. A comparison of panels 2 and 4 (MIT versus quadratic EOS for nonvanishing decoupling parameter) reveals the density for the quadratic model is higher than the linear model at each interior point of the bounded configuration. 

We now consider the trends in the behaviour of the radial pressure (plotted in Fig. \ref{fig2}) throughout the interior of the compact object for the $\rho=\theta^0_0$ sector. A comparison of panels one and three show that the pressure in the quadratic models dominate over the linear models when the decoupling parameter vanishes. This increased pressure in the quadratic models leads to greater stability against the inwardly driven gravitational force, with this effect being enhanced in the central regions of the star. The surface pressure in the quadratic models dominates their linear counterparts thus leading to stable surface layers. It is clear from the second and fourth panels that the pressure is enhanced in the presence of the decoupling parameter. In addition, the radial pressure in the quadratic models dominates their linear counterparts at each interior point from the center through to the stellar surface with the highest pressures achieved in models with quadratic EOS and nonzero $\alpha$.  

Before we embark on a discussion of the trends in anisotropy in our models, we highlight some work which motivates the inclusion of pressure anisotropy in stellar configurations. In our models the radial and transverse stresses at each interior point of the compact object are unequal. Obviously, equality of the transverse component of pressure does ensure the spherical symmetry of the model~\citep{Gleiser2002}. However, there are various reasons for consideration of the origin of anisotropy inside the compact stars~\citep{Ruderman1972,Sawyer1972,Jones1975,Sokolov1980,Kippen1990,Weber1999,Liebling2012} and review work on the local anisotropy by~\cite{Herrera1997} may be informative in this regard. Very recently,~\citet{Hererra2020} has investigated the conditions for the (in)stability of the isotropic pressure condition in collapsing spherically symmetric, dissipative fluid distributions. Herrera demonstrated that an initially isotropic configuration upon leaving hydrostatic equilibrium evolves into an anisotropic regime.

The anisotropy parameter is displayed in Fig. \ref{fig3}. We observe that the anisotropy changes sign within the compact object. We recall that a positive anisotropy factor signifies a repulsive anisotropic force which is directed outwards \citep{hansraj1}. This helps stabilise the stellar configuration against gravity.  In Fig. \ref{fig3}, the extreme left panel shows that the anisotropy parameter is minimum at the centre of the fluid and is negative up to a finite radius, $r_0$. This negative anisotropic factor is accompanied by an inwardly driven force which sums with the gravitational force, leading to an unstable interior. As one moves beyond $r = r_0$, the anisotropy factor becomes positive. The repulsive force associated with $\Delta > 0$ stabilises the surface layers of the star. In direct comparison, panel 3 reveals a peculiar behaviour of the anisotropy factor. It starts off negative in the central regions and remains negative up to a large radius, $r_1$ where $r_1 > r_0$. It appears that the quadratic EOS model is more unstable than its linear counterpart in the central regions of the compact object. As one moves further out, the anisotropy becomes positive. The first and third panels indicate that the quadratic EOS model has a narrower band of stable surface layers as compared to the linear EOS model. Its quadratic counterpart with a vanishing decoupling parameter (third panel) shows a marked difference between the two models. In the quadratic model, we see an interesting variation in the anisotropy parameter, i.e., starting from the centre of the star, the anisotropy is negative for some central region, $0 < r < r_0$ then becomes more negative as one moves outwards towards the boundary. The anisotropy remains negative for $r_0 < r < r_1$ where $r_1 > r_0$. Beyond $r_1$, the anisotropy becomes positive, rendering the surface layers stable due to the repulsive anisotropic force. The change in sign of $\Delta$ within the quadratic EOS model can be attributed to phase transitions in different regions within the stellar fluid. In the second and fourth panels, we observe the effect of a nonvanishing decoupling constant in both the linear and quadratic EOS models, respectively. In the second panel, we observe that $\Delta < 0$ from the centre to some finite radius. Thereafter, $\Delta$ remains positive up to the boundary of the star. The fourth panel shows that anisotropy is negative within the central regions of the configuration. The anisotropy factor starts off with a finite negative value from $r = 0$ up to some radius $r = r_0$. Thereafter $\Delta$ becomes more negative as one move outwards. After some radius, $r_1$ the anisotropy factor changes sign and becomes positive. A comparison of the second and fourth panels clearly shows that the surface layers of the linear model are more stable than their quadratic counterpart. It is interesting to observe the behavior of the anisotropy in the quadratic models, particularly the change of sign of $\Delta$ in different regions within the stellar fluid. The change in the nature of the anisotropic force (from attraction to repulsion) leads to an unstable core but stable surface layers. This interesting behavior in anisotropy was also observed \citep{sunil5d4}. In this work, they modeled compact objects in $f(\mathcal{Q})$ gravity in which the stellar fluid obeyed the MIT bag model EOS. 
\begin{figure*}
     \centering
     \rotatebox{90}{%
   \begin{minipage}{1\textheight}
  \centering
  \includegraphics[width=4.15cm,height=4.1cm]{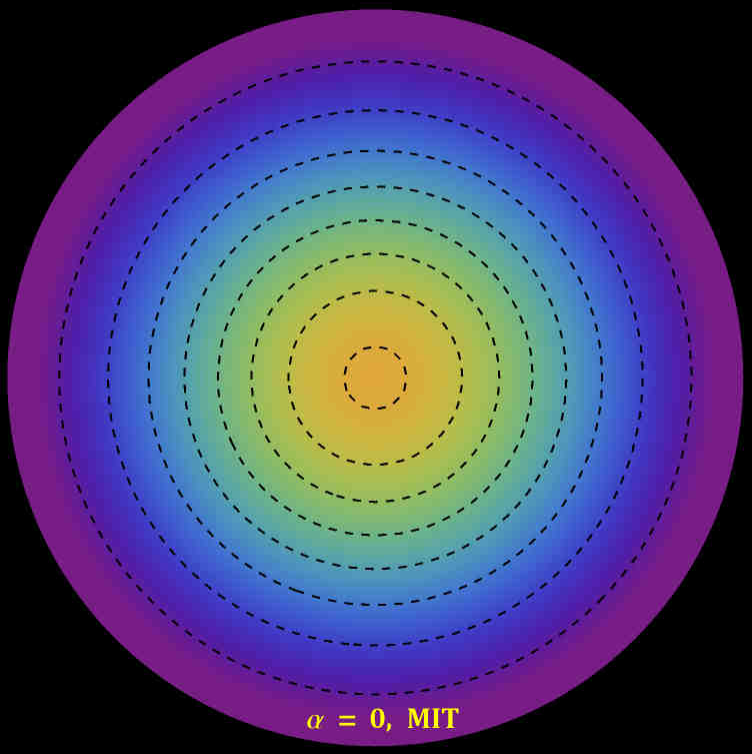}\,\includegraphics[width=1.2cm,height=4.1cm]{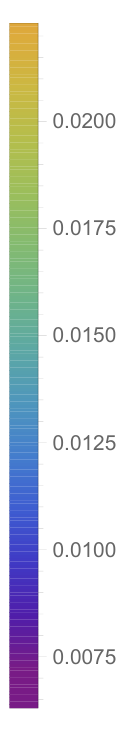} ~~~\includegraphics[width=4.15cm,height=4.1cm]{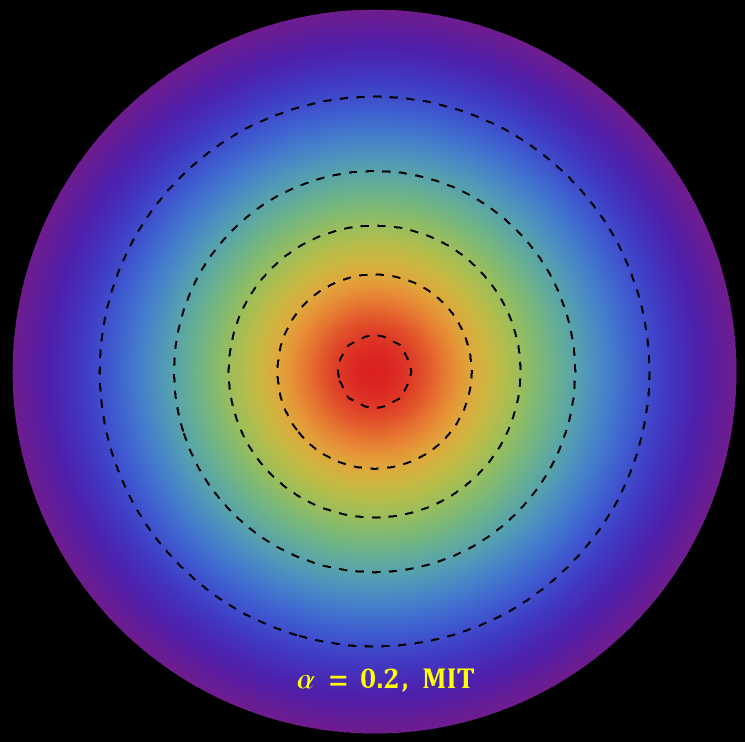}\includegraphics[width=1.2cm,height=4.1cm]{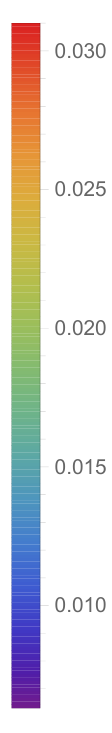}\,~~~
      \includegraphics[width=4.15cm,height=4.1cm]{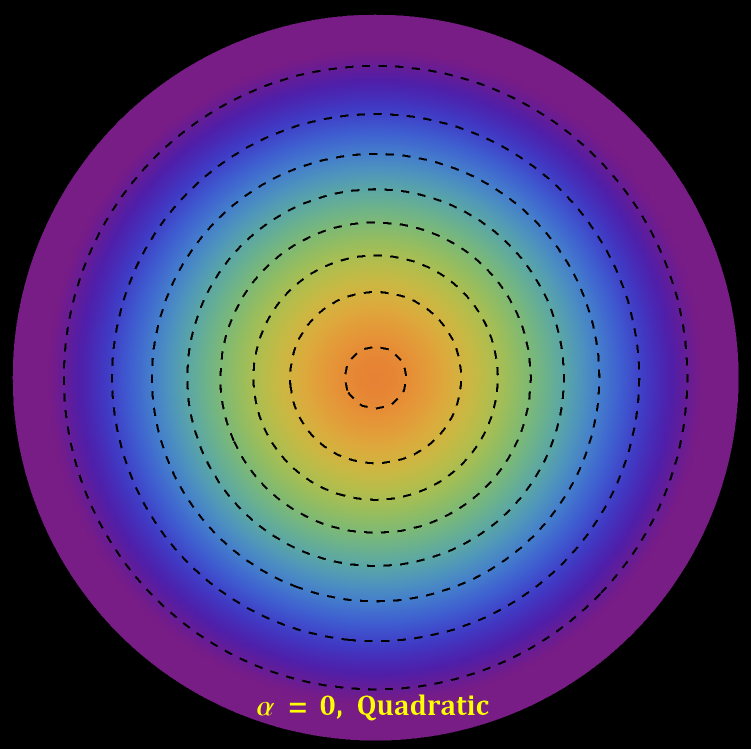}\,\includegraphics[width=1.2cm,height=4.1cm]{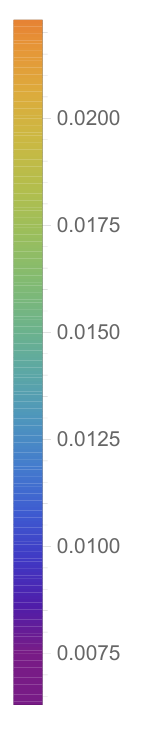} ~~~\includegraphics[width=4.15cm,height=4.1cm]{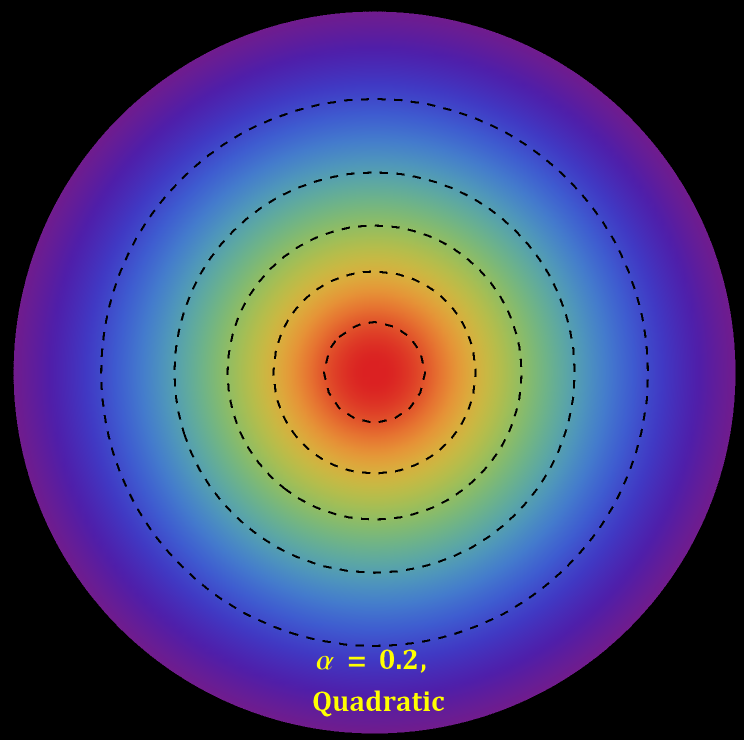}\includegraphics[width=1.2cm,height=4.1cm]{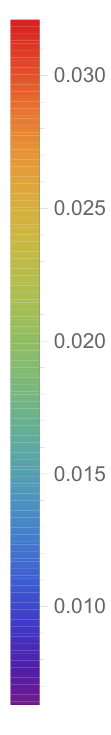}
    \caption{The distribution of energy density {[$\epsilon(r)$ in $\text{km}^{-2}]$ versus radial distance $r$ from center to boundary of star} for MIT Bag model by taking $a=0$ and $\mathcal{B}_g=60\,MeV/fm^3$ (first two panels) and quadratic model by taking $a=2\,\text{km}^2$ and $\mathcal{B}_g=60\,MeV/fm^3$ (right two panels) with two different values of $\alpha$ for solution~ \ref{solB}($p_r=\theta^1_1)$). We have chosen the following fixed values of constants $L=0.0011/\text{km}^2$, $N= 0.00787/\text{km}^2$, radius $R=12.5\,\text{km}$, $\beta_1= 1.1$, and $\beta_2=10^{-46}/\text{km}^{2}$ to plot these curves. {The color bar shows the amount of energy density [$\epsilon(r)$ in $\text{km}^{-2}]$ for $r=0$ to $r=12.5~\text{km}$. } }
    \label{fig4}
   \includegraphics[width=4.15cm,height=4.1cm]{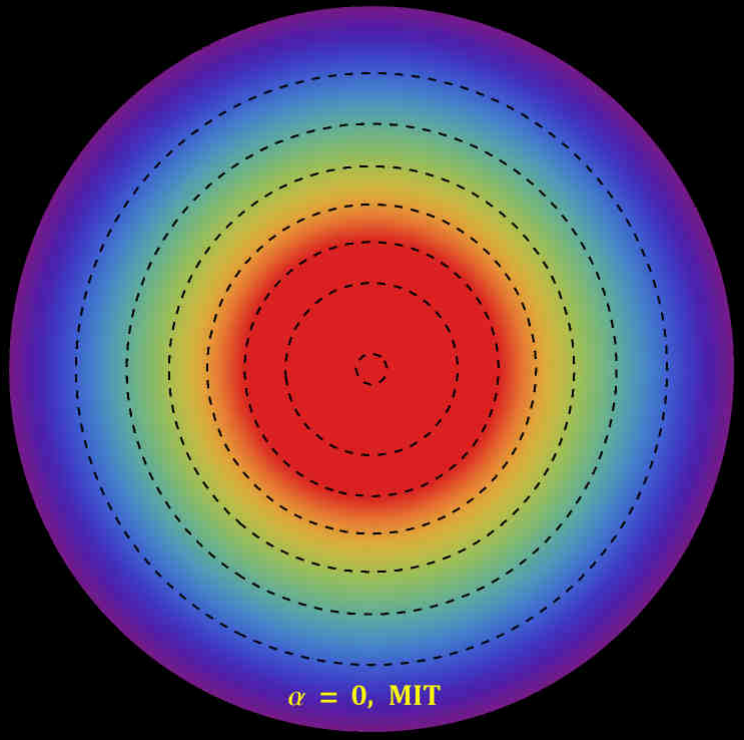}\includegraphics[width=1.2cm,height=4.1cm]{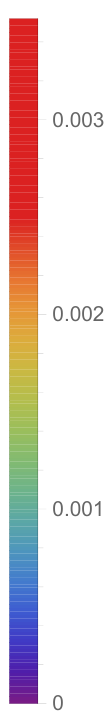} ~~~\includegraphics[width=4.1cm,height=4.1cm]{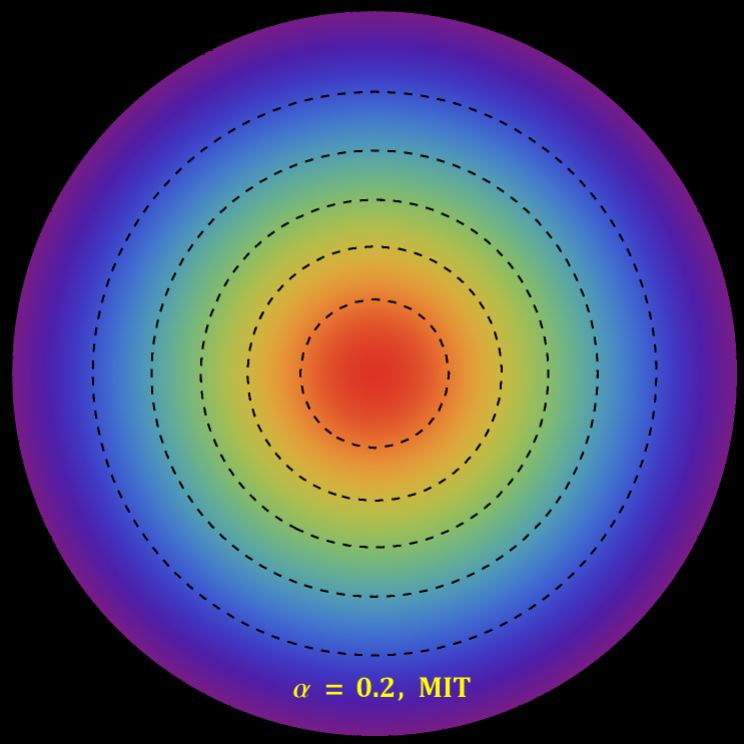}\includegraphics[width=1.2cm,height=4.1cm]{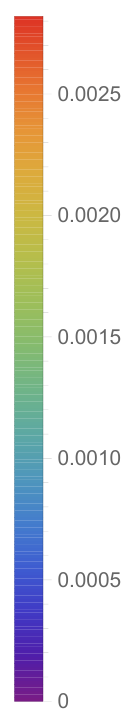}\,~~~
      \includegraphics[width=4.15cm,height=4.1cm]{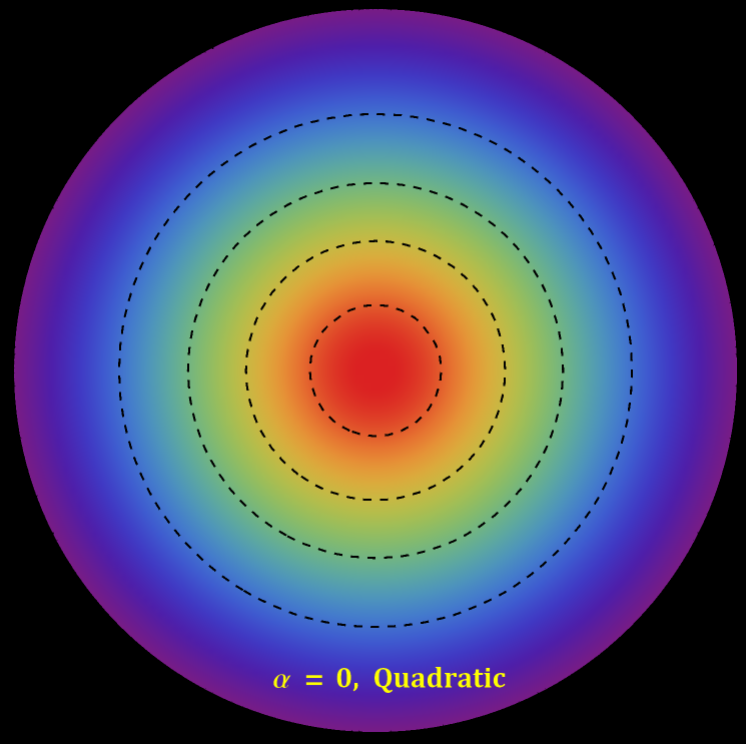}\includegraphics[width=1.2cm,height=4.1cm]{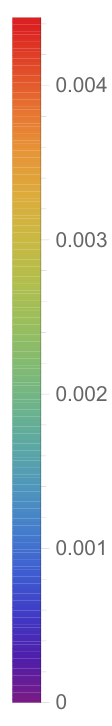} ~~~\includegraphics[width=4.15cm,height=4.1cm]{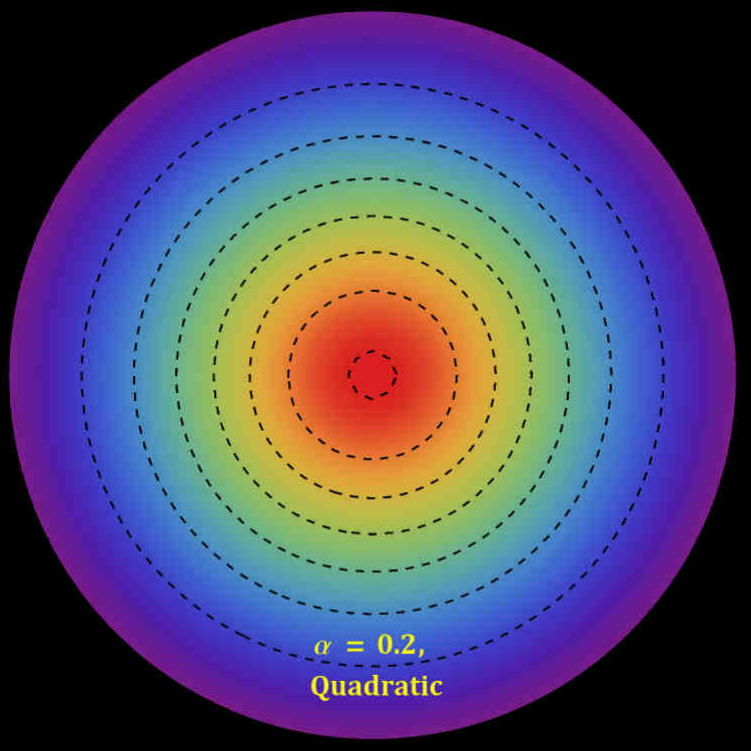}\includegraphics[width=1.2cm,height=4.1cm]{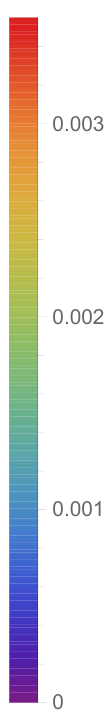}
    \caption{The distribution of radial pressure {[$P_r(r)$ in $\text{km}^{-2}]$ versus radial distance $r$ from center to boundary of star} for MIT Bag model by taking $a=0$ and $\mathcal{B}_g=60\,MeV/fm^3$ (first two panels) and quadratic model by taking $a=2\,\text{km}^2$ and $\mathcal{B}_g=60\,MeV/fm^3$ (right two panels) with two different values of $\alpha$ for solution \ref{solB} ($p_r=\theta^1_1)$). The same values of constants are employed here as used in Fig.\ref{fig4}. {The color bar shows the amount of radial pressure [$P_r(r)$ in $\text{km}^{-2}]$ for $r=0$ to $r=12.5~\text{km}$. }}
    \label{fig5}
   \includegraphics[width=4.15cm,height=4.1cm]{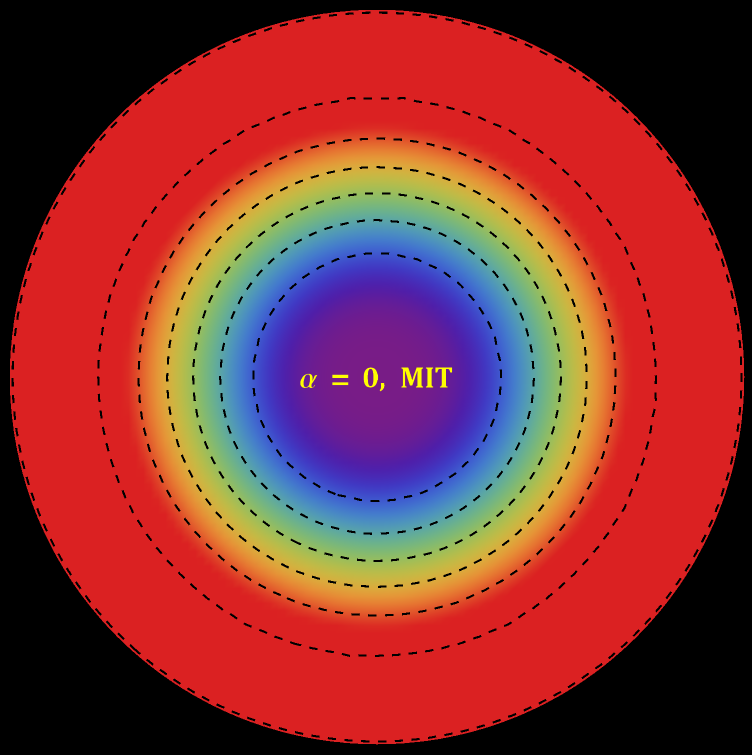}\includegraphics[width=1.2cm,height=4.1cm]{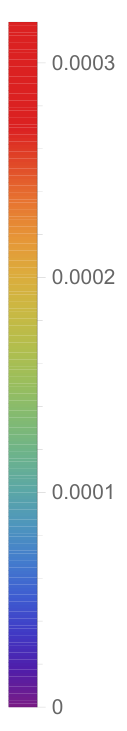} ~~~\includegraphics[width=4.15cm,height=4.1cm]{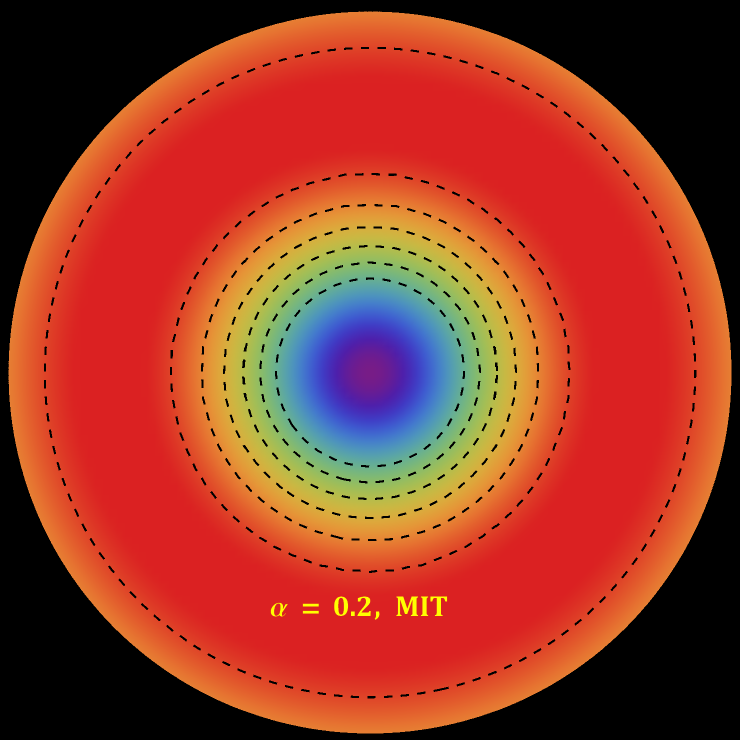}\includegraphics[width=1.2cm,height=4.1cm]{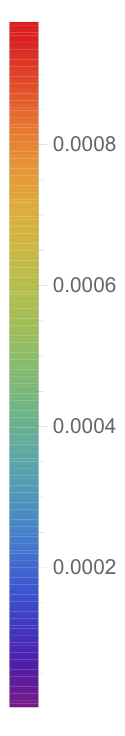}~~~
      \includegraphics[width=4.15cm,height=4.1cm]{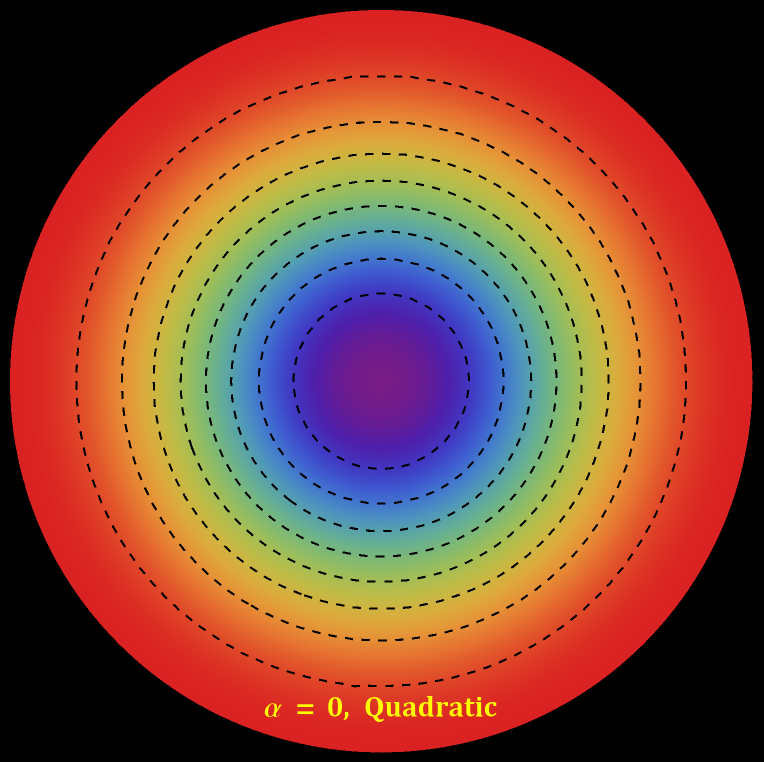}\includegraphics[width=1.2cm,height=4.1cm]{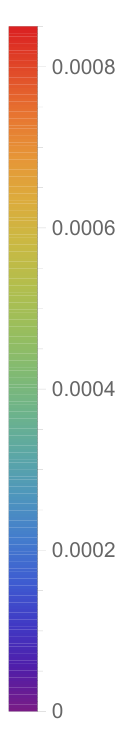}~~~\includegraphics[width=4.15cm,height=4.1cm]{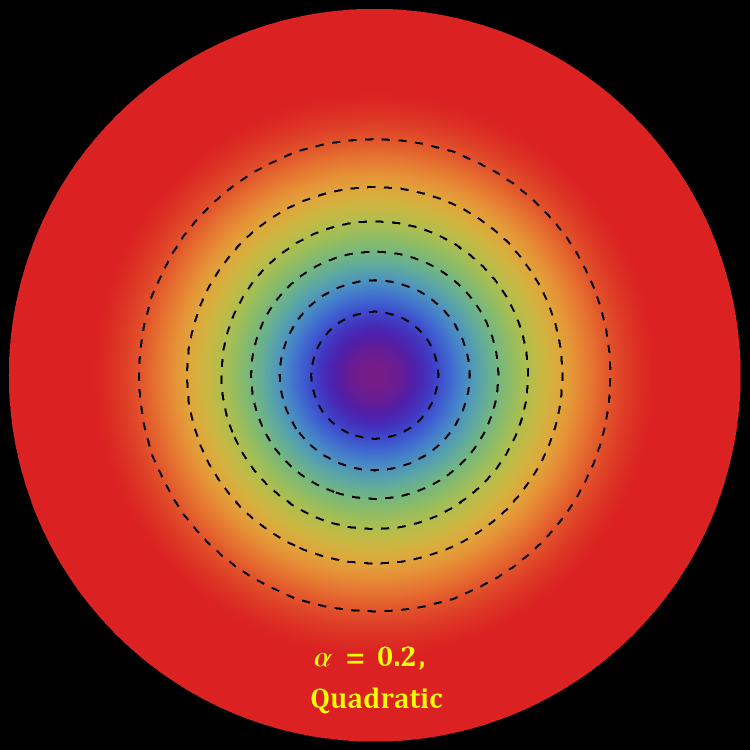}\includegraphics[width=1.2cm,height=4.1cm]{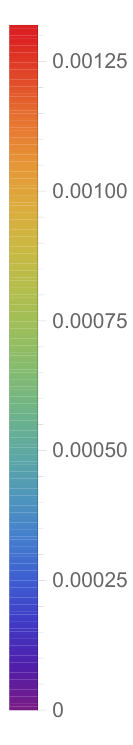}
    \caption{The distribution of anisotropy {[$\Delta(r)$ in $\text{km}^{-2}]$ versus radial distance $r$ from center to boundary of star }for MIT Bag model by taking $a=0$ and $\mathcal{B}_g=60\,MeV/fm^3$ (first two panels) and quadratic model by taking $a=2\,\text{km}^2$ and $\mathcal{B}_g=60\,MeV/fm^3$ (right two panels) with two different values of $\alpha$ for solution \ref{solB} ($p_r=\theta^1_1)$). The same values of constants are employed here as used in Fig.\ref{fig4}. {The color bar shows the amount of anisotropy [$\Delta(r)$ in $\text{km}^{-2}]$ for $r=0$ to $r=12.5~\text{km}$. }} \label{fig6}
    \end{minipage}}
\end{figure*}


 \subsubsection{For solution \ref{solB} ($p_r=\theta^1_1$)} \label{5.1.2}

In this subsection, we move our attention to the physical analysis of the models discovered for the solution $p_r=\theta^1_1$. Figure \ref{fig4} exhibits the energy density plot. As usual $\alpha = 0$ leads to the standard $f(\mathcal{Q})$ gravity. Initially, let us look at the left two panels of Fig. \ref{fig4}, the first and second plots show the behavior of the energy density against the radial coordinate $r$ for a fluid distribution following the MIT bag model EOS. It is evident that the energy density obtained here is regular at all internal points of the stellar structure and attaining a maximum at the core of the object. From the first and second panels, it is clear that the contribution of the decoupling constant ($\alpha$) allows a higher core densities in the MIT bag model EOS as compared to the pure $f(\mathcal{Q})$ gravity stellar models.  Furthermore, the third and fourth panels indicate the trend in the energy density for the fluid distribution in the quadratic EOS.  We notice that the contributions from $\alpha$ in the context of quadratic EOS gives to a substantial growth in the energy density, particularly in the central regions of the star which lead to more compact configurations. On the other hand, we also detect that stars obeying quadratic EOS have higher central densities and are more compact than their linear EOS counterparts as happened in the first solution. 

Now we move to the Fig. \ref{fig5} to check the behavior of the radial pressure within the stellar object for the $p_r=\theta^1_1$ sector. On comparing of the panels to the first and third in absence of the gravitational decoupling, we observe that the radial pressure of models under the quadratic EOS dominates over models in a linear regime. Furthermore, this increment in the pressure for quadratic models leads to greater stability in the central regions of the star against the inwardly directed gravitational force. From the second and fourth panels, it is clear that the pressure decreases in the presence of the gravitational decoupling parameter. Apart from this, the radial pressure of the model in the context of quadratic EOS dominates their linear counterparts everywhere inside the star from the center to the surface as well as the highest pressure is achieved in models obeying quadratic EOS under vanishing $\alpha$. 

A scrutiny of Fig. 6 reveals that the anisotropy is regular at all interior points of each of the models displayed in the four panels. In addition, the anisotropy is positive and increases from the center of the star toward the boundary. This is in direct contrast to our models $\rho=\theta^0_0$ sector. A comparison of panels 1 and 3, shows the trend in anisotropy in the absence of the decoupling constant for the linear and quadratic EOS respectively. The central anisotropy is lower in the linear model compared to its quadratic counterpart. As one approaches the surface layers, we observe that anisotropy increases, with the increase more significant over a larger portion of the surface layers in the linear model. Since $\Delta > 0$, the repulsive anisotropic renders the surface layers more stable than the central core regions. Comparatively, the relative magnitudes of the anisotropy show that the linear model is more stable than the quadratic model for the vanishing decoupling parameter. We now turn our attention to the second and third panels of Fig. 6.  In these models, the anisotropy increases from the center of the configuration toward the boundary. The increase in $\Delta$ is more profound at each interior point in the linear model. This indicates that the contribution from the repulsive nature of the anisotropic force renders the linear model more stable. A peculiar observation in the behavior of the anisotropy is observed in the linear model. While the contributions from anisotropy increase steadily from the center, $r_0$ to some finite radius $r_1$, we observe a decrease in $\Delta$ for $r_1 < r \leq b$. This trend is not observed in the quadratic model. If we now compare panels 1 and 2, ie., the linear models for $\alpha = 0$ and nonvanishing$\alpha$, respectively, we observe that the decoupling parameter stabilizes the central region by enhancing the anisotropy. A comparison of panels 3 and 4 clearly shows the effect of the decoupling parameter on $\Delta$ in the quadratic models. We note that the contributions from $\alpha$ lead to enhanced anisotropy throughout the stellar configuration thus leading to greater stability of concentric matter shells centered about the origin.  .

\begin{figure*}
    \centering    
 \includegraphics[width=8cm,height=6.5cm]{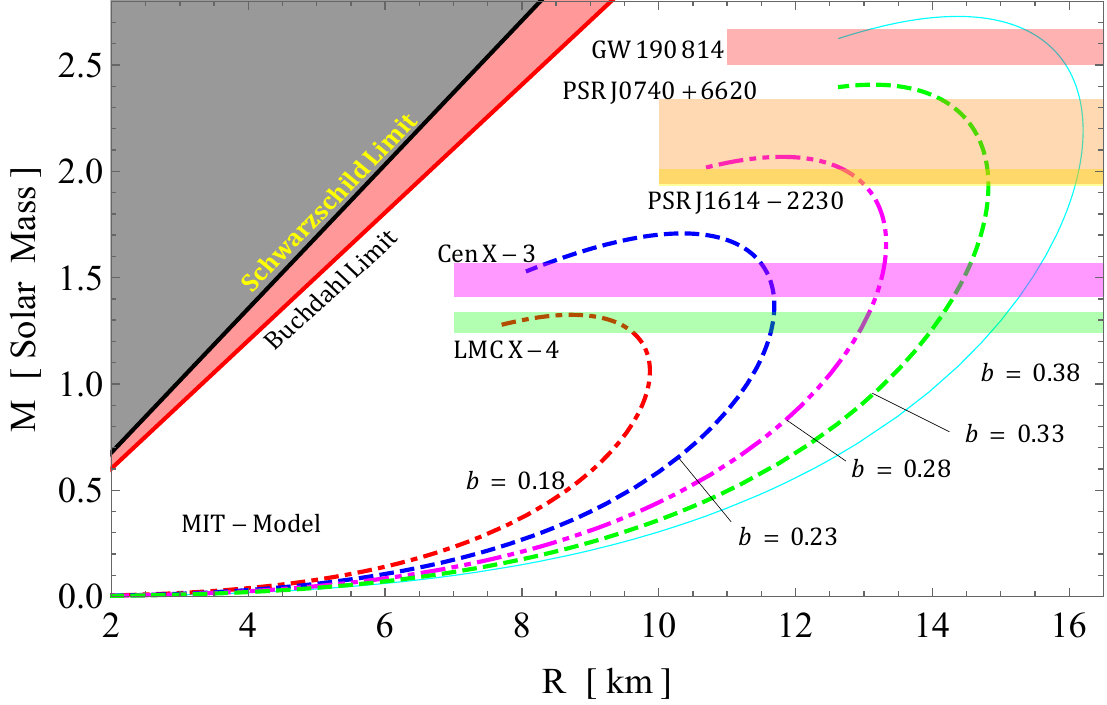}~~~~~~ \includegraphics[width=8cm,height=6.7cm]{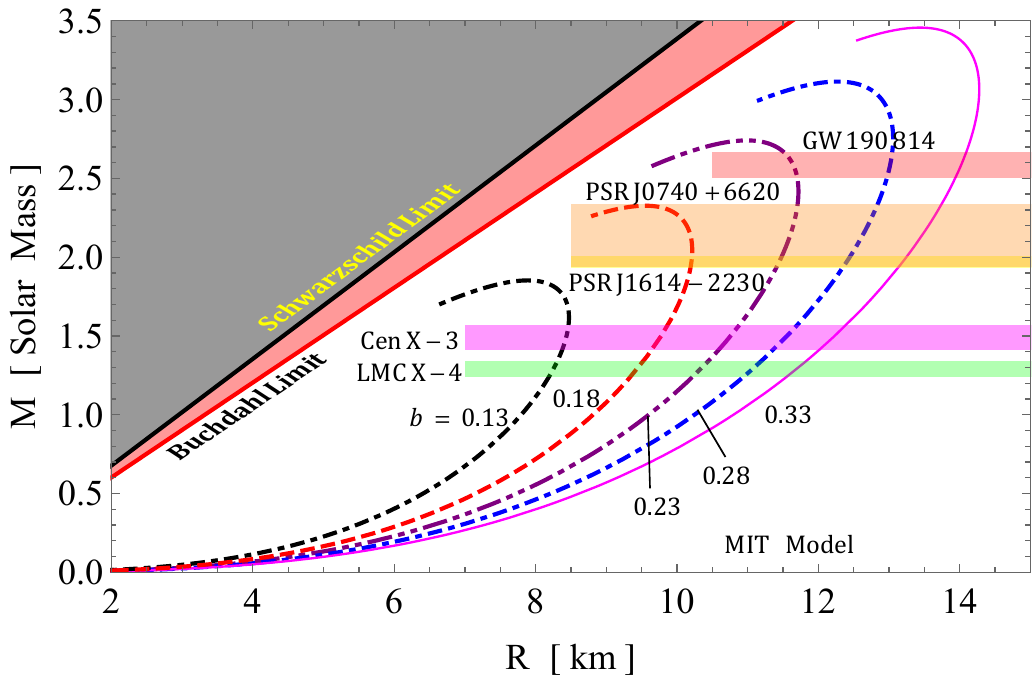}
    \caption{The above $M-R$ curves are plotted for describing the upper limit of mass-radius relationship in a purely MIT bag model ($a=0$) for different $b$ with fixed $a=0,~\alpha= 0.3\,\text{km}^2$ and $\mathcal{B}_g=60\,MeV/fm^3$ for solution \ref{solA} ($\theta^0_0=\rho$)~-~(left panel) and with fixed $a=0,~\alpha= 0.2\,\text{km}^2$ and $\mathcal{B}_g=60\,MeV/fm^3$ solution \ref{solB} ($\theta^1_1=p_r$)~~~(right panel), respectively.}
    \label{fig7a}
\end{figure*}

\begin{figure*}
    \centering    
 \includegraphics[width=8cm,height=7.5cm]{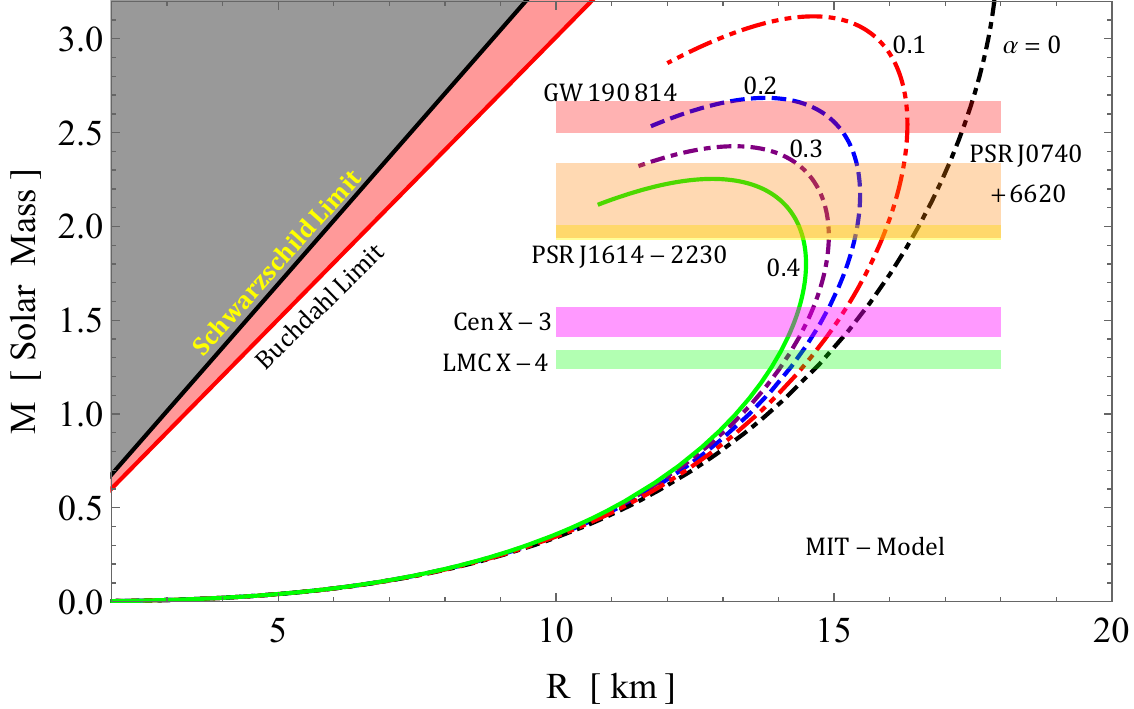}~~~~~~  \includegraphics[width=8cm,height=7.5cm]{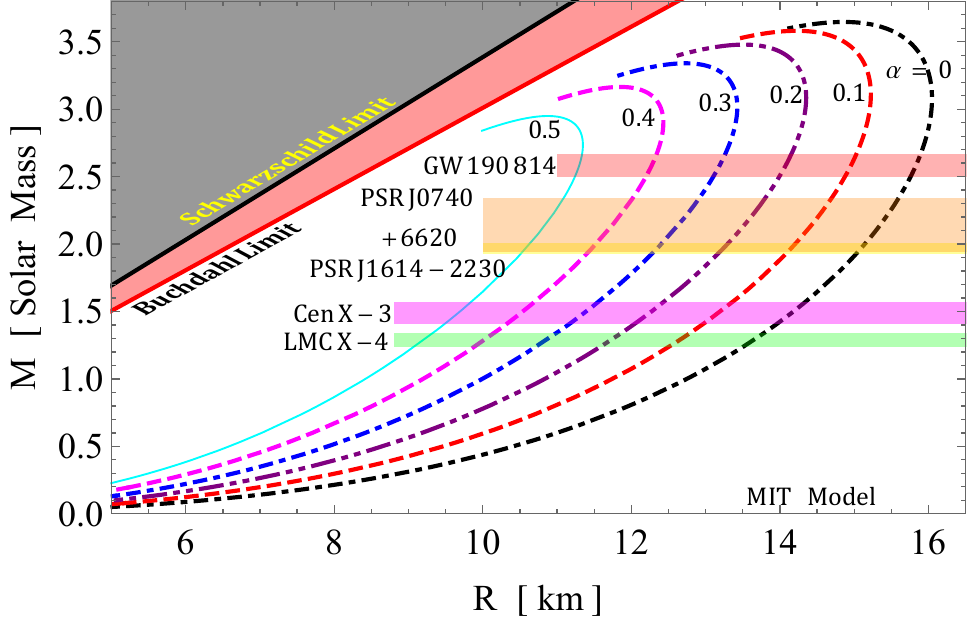}
    \caption{The above $M-R$ curves are plotted for describing the upper limit of mass-radius relationship in MIT-Model for different $\alpha$ with fixed $a=0,~b= 1/3$ and $\mathcal{B}_g=60\,MeV/fm^3$ for solution \ref{solA} in ($\theta^0_0=\rho$)~-~(left panel) and solution \ref{solB} in ($\theta^1_1=p_r$)~~~(right panel), respectively.}
    \label{fig8a}
\end{figure*}


\begin{figure*}
    \centering    
     \includegraphics[width=8cm,height=6.5cm]{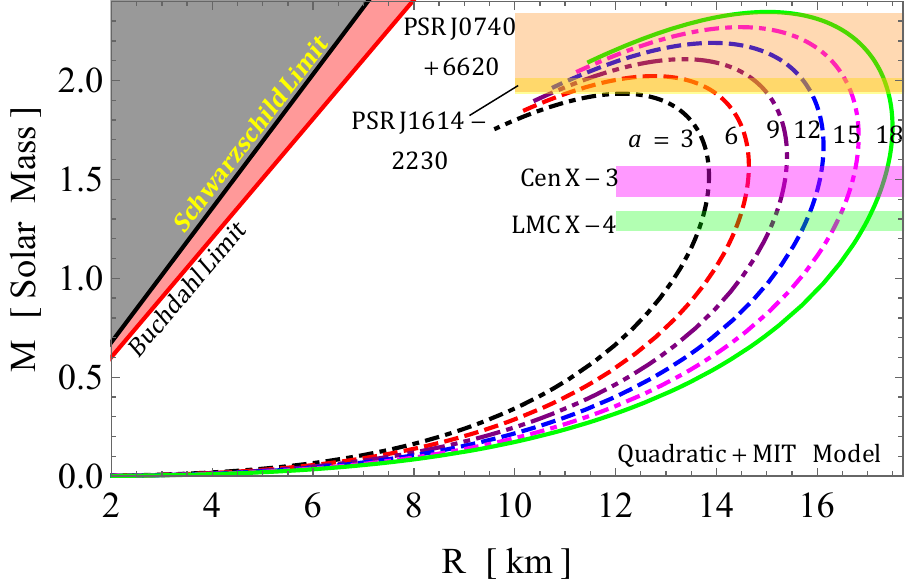}~~~~~~\includegraphics[width=8cm,height=6.5cm]{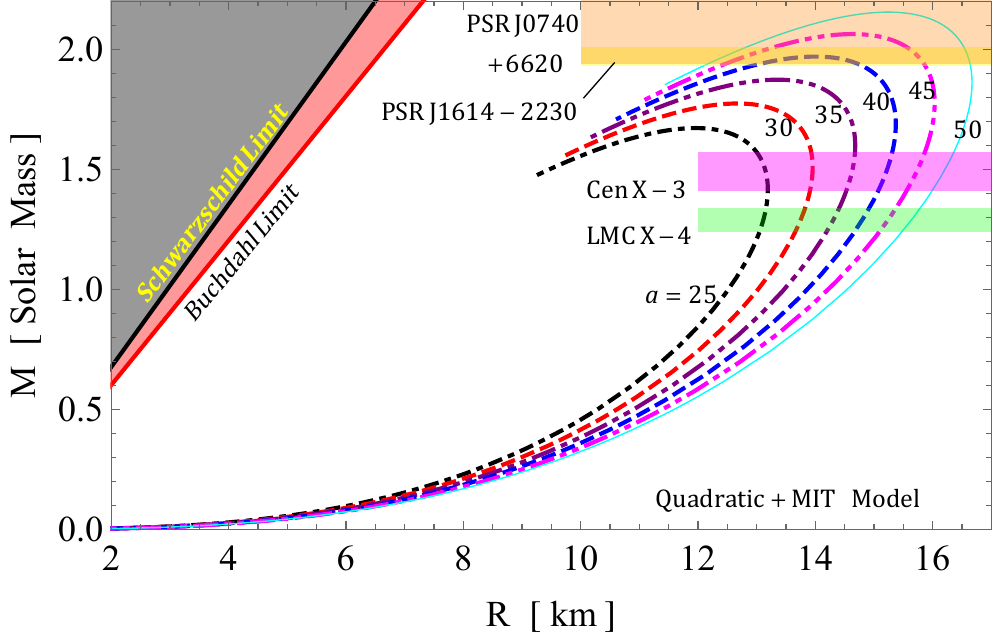}
      \caption{The above $M-R$ curves are plotted for describing the upper limit of mass-radius relationship in Quadratic+MIT bag model for different $a$ with fixed $\alpha=0.3\,\text{km}^2$,  $b=1/3$ and $\mathcal{B}_g=60\,MeV/fm^3$ for solution \ref{solA} ($\theta^0_0=\rho$)~-~(left panel) and $\alpha=0.2\,\text{km}^2$,  $b=1/3$ and $\mathcal{B}_g=60\,MeV/fm^3$ solution \ref{solB} ($\theta^1_1=p_r$) - (right panel), respectively.}
     \label{fig10a}
\end{figure*}

 \begin{figure*}
    \centering  
   \includegraphics[width=8cm,height=6.5cm]{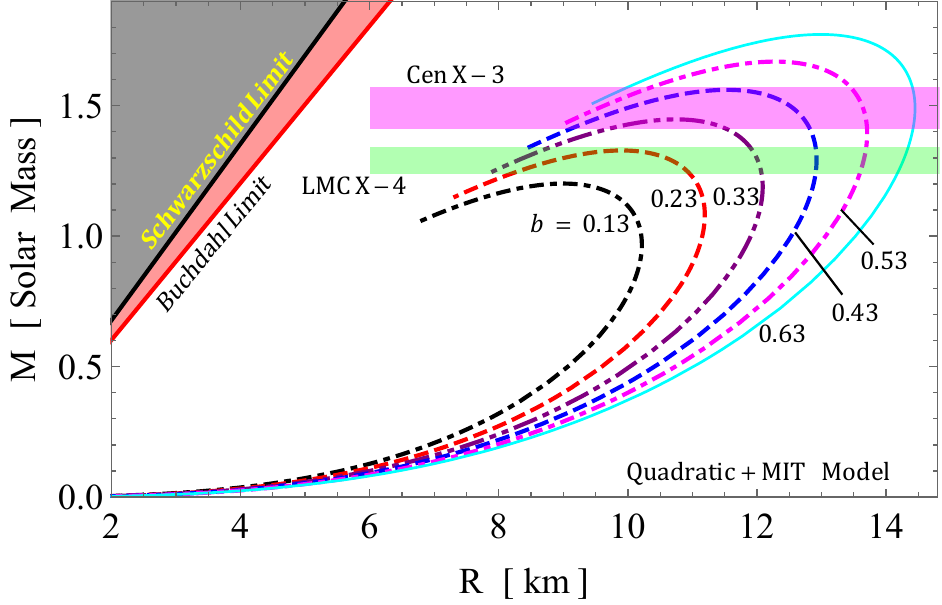}~~~~~~
   \includegraphics[width=8cm,height=6.5cm]{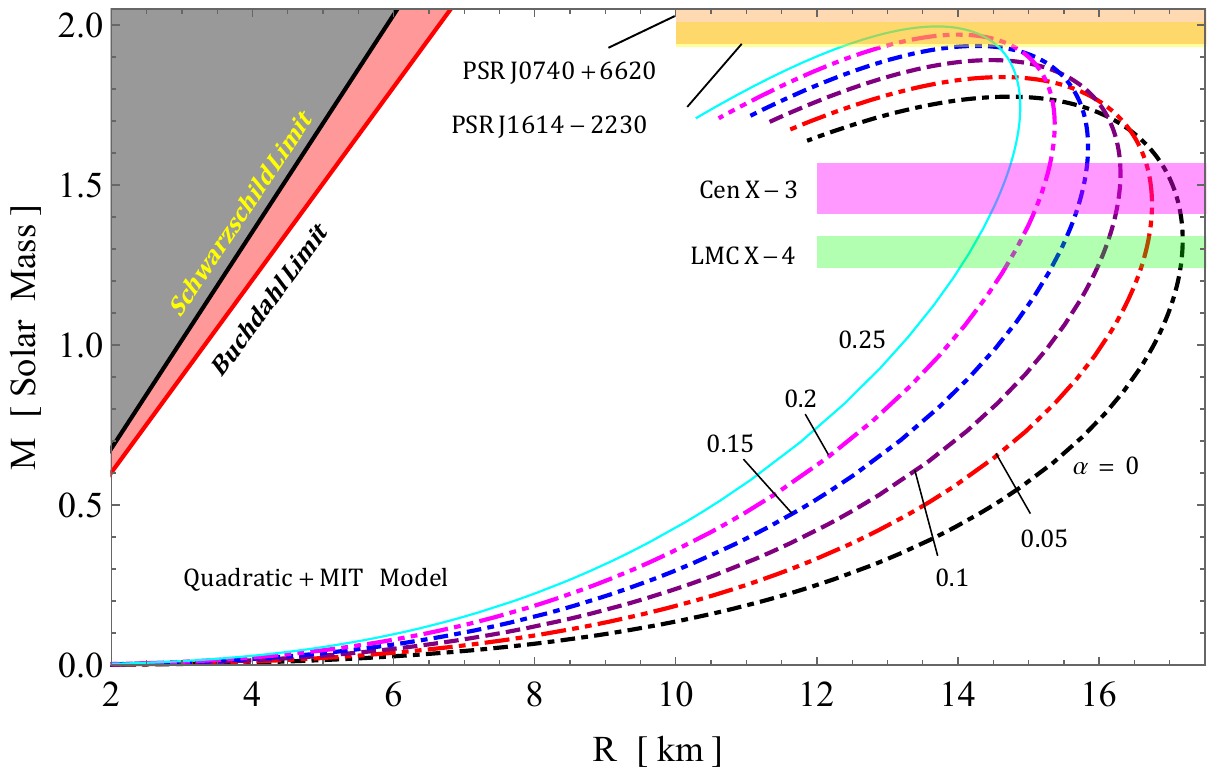}~~~ \caption{The above $M-R$ curve is plotted for describing the upper limit of mass-radius relationship in quadratic+MIT bag model for different $b$ with fixed $\alpha=0.2\,\text{km}^2$, $a=2\,\text{km}^2$  and $\mathcal{B}_g=60\,MeV/fm^3$ (left panel) while the right panel shows a quadratic+MIT bag model for different $\alpha$ with fixed $a=2\,\text{km}^2$, $b=1/3$  and $\mathcal{B}_g=60\,MeV/fm^3$ for the solution \ref{solB} ($\theta^1_1=p_r$).}
    \label{fig12a}
\end{figure*}

\subsection{Constraining upper limit of maximum mass for SS via M-R diagrams} \label{5.2}
 { In this section we investigate the effect of the decoupling parameter ($\alpha$), bag constant ($\mathcal{B}_g$) and EOS parameters on the mass-radius relations of various compact objects and compare our theoretical values to observational constraints. Based on the EOS
 \begin{eqnarray}
p_r &=& a\, \rho^{1+1/n}+b \rho + c, \label{EOSd}
\end{eqnarray} we classify the stellar fluid as 
\begin{itemize}
 \item[(1)] $a \neq 0, b \neq 0 \rightarrow$ quadratic EOS\\
 \item[(2)] $a \neq 0, b = 0 \rightarrow$  pure quadratic EOS\\
 \item[(3)] $a = 0, b \neq 0 \rightarrow$  linear EOS (MIT bag model)
\end{itemize}}

\subsubsection{ {Linear EOS with constancy in bag constant, fixed decoupling parameter and varying EOS parameter}} \label{5.2.1}

We have up to this point tested the physical viability of our solutions in describing self-gravitating stellar objects. We have shown, through tests based on regularity, causality, and stability that these solutions describe physically realizable stellar configurations. In this section, we constrain the free parameters in our solutions by using observational data of compact stars LMC X-4, Cen X-3, PSR J1614-2230, PSR J0740+6620, and the supposedly secondary component of GW190814. The robustness of our solutions allows us to predict the observed mass-radius relations of these stars. Starting off with solution A emanating from the $\rho=\theta^0_0$ sector, we refer to Table \ref{table1} and Fig. \ref{fig7a} (left panel). Here we focus on the MIT-bag models where we have fixed the Bag constant and decoupling parameter while varying the linear EOS parameter, $b$. { At the outset we must point out that a fixed linear EOS parameter $b$, leads to the prediction of a category of stars with similar trends in mass and radii and not individual stars.} From Table \ref{table1}, we observe that for small values of $b$, i.e., linear contributions from the energy density, do account for low-mass stars such as LMC X-4 and Cen X-3.  { Our model shows that by fixing the parameter $b$, we can obtain { masses of stars beyond 2 $M_{\odot}$ }}. For example, for $b = 0.38$, our model predicts the existence { of a class of stars befitting} of the secondary component of the GW190814 event with a mass range of 2.5-2.67 $M_{\odot}$ and a radius of 15.74$_{-0.21}^{+0.40}$. On the right panel, we see the trend in the M-R curves for the $p_r=\theta^1_1$ sector in the linear regime. A similar trend in the increase in the linear EOS parameter on the radii of compact stars is observed in this case. However, there is one notable difference, while the predicted radii increase as $b$ increases, the $p_r=\theta^1_1$ sector predicts smaller radii, i.e., more compact stellar models. For example, in the case of the GW190814 event, the secondary component has a radius of 14.02$_{-0.08}^{+0.09}$  km for the upper value of $b$, which is $10\%$ less than its $\rho=\theta^0_0$ counterpart.

\subsubsection{ {Linear EOS with constancy in bag constant, fixed EOS parameter and varying decoupling constant}}  \label{5.2.2}
We now look at Table \ref{table2} in conjunction with Figure \ref{fig8a} in which we have fixed the bag constant and the EOS parameter, $b$ while varying $\alpha$. From Table \ref{table2} and the left panel of Fig. \ref{fig8a}, we observe that an increase in the decoupling parameter results in a decrease in radii leading to more compact configurations. A similar trend is observed in the $p_r=\theta^1_1$ sector. We also observe that for the chosen values of the bag constant and EOS parameter, our solutions are able to predict a wider range of radii for all compact objects under investigation. A comparison of the left and right panels reveal that the allowable radii for the compact objects under investigation is more tightly bounded for the $\rho=\theta^0_0$ sector. The left panel shows that low mass stars such as  LMC X-4 (1.29 $\pm$ 0.05 $M_{\odot}$) and Cen X-3 (1.49 $\pm$ 0.08 $M_{\odot}$) exist for radii greater 10 km. For the  $p_r=\theta^1_1$, there is wider range of radii for low mass stars closer to 9 km. More noticeably is the prediction of the existence of higher-mass compact objects beyond  2 $M_{\odot}$. For small $\alpha$, there exist stars with masses closer to 3 $M_{\odot}$. For the vanishing of the decoupling parameter ($\alpha = 0$), stellar masses exceed 3 $M_{\odot}$. For the right panel, we note that the maximum mass is in the order of  3.5 $M_{\odot}$. In a study of Burgio et al.~\citep{burgio}, they questioned the existence of compact stars with small radii giving rise to the $GW170817/AT2017gfo$ signals. In one of their proposals, the so-called `twin-star' scenario they concluded that the merger involved the coalescence of a hadronic star and a quark matter star with radii in the range $10.7 \text{km} < R_{1.5} < 12 \text{km}$ with $R_{1.5}$ {($14.28-15.41$ km for $\theta^0_0=\rho$ and $9.70-14.18$ km for $\theta^1_1=p_r$)} being the radius of a 1.5 $M_{\odot}$ compact star. Figure \ref{fig8a} accounts for the existence of a myriad of observed compact objects including those falling into the `twin star' category scenario with the latter being {more suitable for $p_r=\theta^1_1$ sector}.

\subsubsection{ { Quadratic EOS with fixed bag constant, fixed decoupling constant with varying quadratic EOS parameter}}  \label{5.2.4}
Here we focus on Table \ref{table4}. and Figure \ref{fig10a}. It is clear in this scenario, both of our solutions do not predict very well the existence of compact objects beyond  2 $M_{\odot}$.  Even in the case of PSR J1614-2230 with a mass of 1.97$\pm$0.04 $M_{\odot}$, which is less than the 2.0 $M_{\odot}$, fails to exist. The lower mass stars have larger radii, thus indicating that these candidates have low densities compared to their linear EOS counterparts as displayed in Figure 8. In a study by  \citep{Astash01}, they showed that the observations of GW190814 can be accounted for within the framework of $f(R)$ gravity and the inclusion of rotation. They utilized three frameworks, i.e., classical GR, $f(R)$ gravity with $f(R) = R + \alpha R^2$, where $\alpha$ is quadratic curvature correction and finally $f(R) = R^{1+\epsilon}$ where $\epsilon$ is a measure of a small deviation from GR. Utilizing 
different EOS's, they showed that in the absence of rotation, the models of compact objects for both GR and $f(R)$ gravity with varying $\alpha$ lead to masses less than  2.5 $M_{\odot}$ and radii between 10 and 14 km.  With the inclusion of rotation, the $f(R)$ models can easily account for masses greater than 2.5 $M_{\odot}$ this accounting for the observed mass of the secondary component of GW190814. For observed rotational frequencies of neutron stars, $f(R)$ gravity predicts the existence of a 2.0 $M_{\odot}$ compact object. In the case of $f(R) = R^{1+\epsilon}$ framework, they obtained masses in the range of 2.5- 2.67 $M_{\odot}$ for $\epsilon$ in the range 0.005 and 0.008. Interestingly, the radii of these compact objects in $f(R)$ gravity with rotation can vary between 12 to 18 km. We have also obtained models of compact objects with large radii in this category of stars.

\subsubsection{ Quadratic EOS with fixed bag constant, fixed quadratic EOS parameter with constant versus varying  decoupling constant for mimicking of radial pressure, $p_r=\theta^1_1$} \label{5.2.6}

In this subsection, we pay particular attention to the M-R curves arising in the composite linear-quadratic EOS and the effect of the decoupling constant. Table \ref{table6}. shows that these models predict the existence of low-mass stars with an increase in the decoupling constant leading to higher-mass stars with larger radii very similar to the trends obtained by \citep{Astash1} in describing rotating neutron stars in $f(R) = R + \alpha R^2$ gravity. It is clear from Table \ref{table6} and Figure \ref{fig12a} that the mixed EOS model fails to predict masses which can account for the LIGO observation of approximately 2.6  $M_{\odot}$ of the secondary component of the binary coalescence GW190814. It appears that when the linear and quadratic EOS parameters are switched on simultaneously, the decoupling constant quenches any increase in the mass of the compact object. While these models fail to predict masses above 2.6 $M_{\odot}$, the quadratic EOS predicts the existence of well-known pulsars and neutron stars to a very good approximation. 

In comparing the models derived from the $\rho=\theta^0_0$ and $p_r=\theta^1_1$ sectors we compared and contrasted their predicting powers when it comes to low mass stars as well as extreme masses bordering on the mass of the lightest black hole. We have shown that the predicted stellar masses are sensitive to the decoupling constant, $\alpha$, the linear EOS parameter, $b$, the quadratic EOS parameter, $a$, and the bag constant. By varying these constants in particular sets, our models describe a family of compact objects with varying EOS's ranging from the simplest linear EOS (MIT bag model), through to the pure quadratic EOS and the more complex quadratic EOS. The tabulated data and plots reveal that the {quadratic EOS accounts for a wide spectrum of observed compact objects in the low-mass limit and stars which qualify as the secondary component of the GW190814 event.}  

\begin{table*}
\footnotesize
\centering
 \rotatebox{90}{%
   \begin{minipage}{1.1\textheight}
\caption{The predicted radii of compact stars LMC X-4, Cen X-3, PSR J1614-2230, PSR J0740+6620, and GW190814 for MIT-bag model (see Fig.\ref{fig7a}). }\label{table1}
 \scalebox{1}{\begin{tabular}{| *{12}{c|} }
\hline
{Objects} & {$\frac{M}{M_\odot}$}   & \multicolumn{5}{c|}{{Predicted $R$ km}} & \multicolumn{5}{c|}{{Predicted $R$ km}}  \\[0.15cm]
\cline{3-12}
&& \multicolumn{5}{c|}{ For solution $\rho=\theta^0_0$} & \multicolumn{5}{c|}{For solution $p_r=\theta^1_1$} \\[0.15cm]
\cline{3-12}
&  &  $b=0.18$ & $b=0.23$  & $b=0.28$  & $b=0.33$  &$b=0.38$  & $b=0.13$ & $b=0.18$ & $b=0.23$  & $b=0.28$  & $b=0.33$  \\[0.15cm] \hline
LMC X-4 \citep{star3}  &  1.29 $\pm$ 0.05  & $9.39_{-0.48}^{+0.24}$  &  $11.65_{-0.01}^{+0.02}$  &    $13.01_{-0.08}^{+0.06}$  &  $14.05_{-0.1}^{+0.1}$  &  $14.94_{-0.10}^{+0.12}$ & $8.23_{-0.04}^{+0.06}$ & $9.45_{-0.11}^{+0.05}$ & $10.32_{-0.09}^{+0.11}$ & $11.05_{-0.11}^{+0.10}$ & $11.67_{-0.13}^{+0.13}$  \\[0.15cm]
\hline
Cen X-3 \citep{star3} & 1.49$\pm$0.08  & -  &  $11.63_{-0.06}^{+0.13}$ & 13.25$_{-0.08}^{+0.05}$ & 14.44$_{-0.14}^{+0.12}$ & 15.43$_{-0.16}^{+0.14}$ & 8.45$_{-0.08}^{+0.02}$ &  $9.79_{-0.06}^{+0.09}$ & 10.72$_{-0.14}^{+0.13}$ & 11.55$_{-0.19}^{+0.16} $ &  $12.21_{-0.18}^{+0.18}$\\[0.15cm]
\hline
 PSR J1614-2230 \citep{star1} & 1.97$\pm$0.04  & -  & - & 12.89$_{-0.11}^{+0.17}$ & 14.81$_{-0.01}^{+0.01}$ & 16.11$_{-0.02}^{+0.01}$ & - &  $10.21_{-0.01}^{+0.02}$ & 11.44$_{-0.05}^{+0.05}$ & 12.39$_{-0.04}^{+0.07} $ &  $13.21_{-0.06}^{+0.05}$\\[0.15cm]
\hline
PSR J0740+6620 \citep{Cromartie} & $2.14^{+0.2}_{-0.17}$ & - & -  & - & 14.68$_{-0.57}^{+0.13}$ & 16.18$_{-0.10}^{+0.02}$ & - & 10.19$_{-0.53}^{+0.01}$ & 11.63$_{-0.21}^{+0.08}$ & 12.63$_{-0.27}^{+0.20}$ &  $13.47_{-0.36}^{+0.27}$  \\[0.15cm]
\hline
GW190814 \citep{wenbin} & 2.5-2.67 & - & -  & - & - & 15.74$_{-0.21}^{+0.40}$ & - & - & 11.64$_{-0.07}^{+0.13}$ & 13.03$_{-0.06}^{+0.02}$ & 14.02$_{-0.08}^{+0.09}$  \\[0.15cm]
\hline
\end{tabular}}\\\\

\caption{The predicted radii of compact stars LMC X-4, Cen X-3, PSR J1614-2230, PSR J0740+6620, and GW190814 for MIT-bag model (see Fig.\ref{fig8a}). }\label{table2}
 \scalebox{0.92}{\begin{tabular}{| *{13}{c|} }
\hline
{Objects} & {$\frac{M}{M_\odot}$}   & \multicolumn{5}{c|}{{Predicted $R$ km}} & \multicolumn{6}{c|}{{Predicted $R$ km}}  \\[0.15cm]
\cline{3-13}
&& \multicolumn{5}{c|}{ For solution $\rho=\theta^0_0$} & \multicolumn{6}{c|}{For solution $p_r=\theta^1_1$} \\[0.15cm]
\cline{3-13}
&  & $\alpha=0.0$ &  $\alpha=0.1$ &  $\alpha=0.2$ &  $\alpha=0.3$ &  $\alpha=0.4$ &  $\alpha=0.0$ &  $\alpha=0.1$ &  $\alpha=0.2$ &  $\alpha=0.3$ &  $\alpha=0.4$ & $\alpha=0.5$ \\[0.15cm] \hline
LMC X-4 \citep{star3}  &  1.29 $\pm$ 0.05  & $14.86_{-0.10}^{+0.09}$  &  $14.58_{-0.08}^{+0.12}$  &  $14.36_{-0.12}^{+0.10}$   &  $14.14_{-0.07}^{+0.07}$  &  $13.93_{-0.08}^{+0.06}$ & $13.64_{-0.15}^{+0.12}$ & $12.44_{-0.14}^{+0.13}$ & $11.71_{-0.18}^{+0.17}$ & $10.88_{-0.14}^{+0.12}$ & $10.02_{-0.13}^{+0.10}$  & $9.17_{-0.02}^{+0.14}$ \\[0.15cm]
\hline
Cen X-3 \citep{star3} & 1.49$\pm$0.08  & $15.14_{-0.14}^{+0.11}$  & $15.10_{-0.16}^{+0.09}$ & 14.82$_{-0.11}^{+0.10}$ & 14.54$_{-0.09}^{+0.08}$ & 14.26$_{-0.09}^{+0.08}$ & 14.18$_{-0.20}^{+0.16}$ & $13.19_{-0.18}^{+0.18}$ & 12.28$_{-0.21}^{+0.15}$ & 11.38$_{-0.10}^{+0.18} $ & $10.00_{-0.19}^{+0.18}$ & $9.67_{-0.19}^{+0.15}$\\[0.15cm]
\hline
 PSR J1614-2230 \citep{star1} & 1.97$\pm$0.04  & $16.42_{-0.02}^{+0.03}$  & $15.90_{-0.01}^{+0.01}$ & 15.39$_{-0.01}^{+0.01}$ & 14.92$_{-0.01}^{+0.01}$ & 14.40$_{-0.02}^{+0.04}$ & 15.11$_{-0.07}^{+0.05}$ & $14.16_{-0.04}^{+0.07}$ & 13.23$_{-0.05}^{+0.07}$ & 12.36$_{-0.05}^{+0.05} $ & $11.46_{-0.03}^{+0.05}$ & $10.58_{-0.05}^{+0.04}$ \\[0.15cm]
\hline
PSR J0740+6620 \citep{Cromartie} & $2.14^{+0.2}_{-0.17}$ & $16.79_{-0.14}^{+0.05}$ & $16.11_{-0.04}^{+0.01}$  & 15.48$_{-0.22}^{+0.07}$ & 14.82$_{-0.50}^{+0.17}$ & 13.96$_{-1.31}^{+0.81}$ & 15.37$_{-0.29}^{+0.23}$ & 14.43$_{-0.34}^{+0.27}$ & 13.51$_{-0.34}^{+0.28}$ & 13.60$_{-0.35}^{+0.26}$  & $11.74_{-0.33}^{+0.27}$ & $10.83_{-0.31}^{+0.24}$ \\[0.15cm]
\hline
GW190814 \citep{wenbin} & 2.5-2.67 & $17.42_{-0.02}^{+0.02}$ & $16.32_{-0.16}^{+0.16}$  & $14.86_{-0}^{+0.09}$ & - & - & 15.84$_{-0.07}^{+0.05}$ & 14.96$_{-0.08}^{+0.06}$ & 14.09$_{-0.10}^{+0.07}$ & 13.19$_{-0.08}^{+0.07}$ & $12.28_{-0.08}^{+0.07}$ & $11.31_{-0.08}^{+0.03}$ \\[0.15cm]
\hline
\end{tabular}}\\ \\

\caption{The predicted radii of compact stars LMC X-4, Cen X-3, PSR J1614-2230, PSR J0740+6620, and GW190814 for Fig.\ref{fig10a}. }\label{table4}
 \scalebox{0.99}{\begin{tabular}{| *{12}{c|} }
\hline
{Objects} & {$\frac{M}{M_\odot}$}   & \multicolumn{5}{c|}{{Predicted $R$ km}} & \multicolumn{5}{c|}{{Predicted $R$ km}}  \\[0.15cm]
\cline{3-12}
&& \multicolumn{5}{c|}{ for solution $\rho=\theta^0_0$} & \multicolumn{5}{c|}{for solution $p_r=\theta^1_1$} \\[0.15cm]
\cline{3-12}
&  &$a=3$ & $a=6$ & $a=9$ &$a=12$ & $a=15$ & $a=25$ & $a=30$ & $a=35$ & $a=40$ & $a=45$  \\ \hline
LMC X-4 \citep{star3}  &  1.29 $\pm$ 0.05  & $13.71_{-0.06}^{+0.04}$  &  $14.44_{-0.05}^{+0.07}$  &   $15.13_{-0.07}^{+0.07}$  &  $15.78_{-0.07}^{+0.08}$  &  $17.0_{-0.10}^{+0.08}$ & $13.11_{-0.06}^{+0.05}$ & $13.74_{-0.08}^{+0.09}$ & $14.31_{-0.11}^{+0.11}$ & $14.81_{-0.12}^{+0.10}$ & $15.25_{-0.15}^{+0.13}$ \\[0.15cm]
\hline
Cen X-3 \citep{star3} & 1.49$\pm$0.08  & $13.84_{-0.03}^{+0.01}$  & $14.63_{-0.04}^{+0.02}$ & 15.36$_{-0.06}^{+0.03}$ & 16.05$_{-0.09}^{+0.05}$ & 16.71$_{-0.12}^{+0.06}$ & 13.15$_{-0.15}^{+0.05}$ & $13.95_{-0.03}^{+0.04}$ & 14.62$_{-0.10}^{+0.05}$ & 15.21$_{-0.14}^{+0.10} $ & $15.72_{-0.15}^{+0.14}$\\[0.15cm]
\hline
 PSR J1614-2230 \citep{star1} & 1.97$\pm$0.04  & -  & $13.72_{-0.27}^{+0.34}$ & 14.88$_{-0.12}^{+0.17}$ & 15.80$_{-0.11}^{+0.09}$ & 16.62$_{-0.06}^{+0.06}$ & - & - & - & 14.19$_{-0}^{+0.5}$ & 15.75$_{-0.20}^{+0.13} $ \\[0.15cm]
\hline
PSR J0740+6620 \citep{Cromartie} & $2.14^{+0.2}_{-0.17}$ & - & -  & - & 14.99$_{-0.81}^{+0.91}$ & 16.11$_{-1.46}^{+0.58}$ & - & - & - & - & 15.78$_{-0.45}^{+0.87}$  \\[0.15cm]
\hline
GW190814 \citep{wenbin} & 2.5-2.67 & - & -  & - & - & - & - & - & - & - & -  \\[0.15cm]
\hline
\end{tabular}}\\\\

\end{minipage}}
\end{table*}

\begin{table}
\footnotesize
\centering
 \rotatebox{90}{%
   \begin{minipage}{1.1\textheight}
\caption{The predicted radii of compact stars LMC X-4, Cen X-3, PSR J1614-2230, PSR J0740+6620, and GW190814 for solution $p_r=\theta^1_1$ corresponding Fig.\ref{fig12a}.}\label{table6}
 \scalebox{0.99}{\begin{tabular}{| *{12}{c|} }
\hline
{Objects} & {$\frac{M}{M_\odot}$}   & \multicolumn{5}{c|}{{Predicted $R$ km}} & \multicolumn{5}{c|}{{Predicted $R$ km}}  \\[0.15cm]
\cline{3-12}
&& \multicolumn{5}{c|}{ Quadratic model+MIT bag model with different $b$} & \multicolumn{5}{c|}{Quadratic model+MIT bag model with different $\alpha$ } \\[0.15cm]
\cline{3-12}
&  & $b=0.23$ & $b=0.33$ & $b=0.43$ & $b=0.53$ & $b=0.63$ & $\alpha=0$ & $\alpha=0.05$ & $\alpha=0.1$ & $\alpha=0.15$ & $\alpha=0.20$  \\[0.15cm] \hline
LMC X-4 \citep{star3}  &  1.29 $\pm$ 0.05  & $10.67_{-0.68}^{+0.28}$  &  $12.01_{-0.02}^{+0.06}$  &   $12.92_{-0.01}^{+0.01}$  &  $13.65_{-0.06}^{+0.05}$  &  $14.26_{-0.09}^{+0.08}$ & $17.19_{-0.04}^{+0.01}$ & $16.66_{-0.06}^{+0.06}$ & $16.07_{-0.09}^{+0.08}$ & $15.43_{-0.09}^{+0.12}$ & $14.81_{-0.14}^{+0.12}$ \\[0.15cm]
\hline
Cen X-3 \citep{star3} & 1.49$\pm$0.08  & -  & - & 12.56$_{-0.87}^{+0.26}$ & 13.65$_{-0.18}^{+0.06}$ & 14.45$_{-0.03}^{+0.01}$ & 17.05$_{-0.17}^{+0.12}$ & $16.74_{-0.07}^{+0.02}$ & 16.29$_{-0.05}^{+0.02}$ & 15.77$_{-0.10}^{+0.05} $ & $15.21_{-0.14}^{+0.10}$\\[0.15cm]
\hline
 PSR J1614-2230 \citep{star1} & 1.97$\pm$0.04  & -  & - & - & - & - & - & - & - & - & 14.12$_{-0}^{+0.65} $ \\[0.15cm]
\hline
PSR J0740+6620 \citep{Cromartie} & $2.14^{+0.2}_{-0.17}$ & - & -  & - & - & - & - & - & - & - & -  \\[0.15cm]
\hline
GW190814 \citep{wenbin} & 2.5-2.67 & - & -  & - & - & - & - & - & - & - & -  \\[0.15cm]
\hline
\end{tabular}}
 \end{minipage}}
\end{table}
\section{Energy exchange between the fluid distributions for $\hat{T}_{\lowercase{ij}}$ and $\theta_{\lowercase{ij}}$ }  \label{sec6}

Let us now discuss about the energy exchange requirement in connection to the extended gravitational decoupling. Ovalle in his work~\citep{OvallePLB2019} showed that both the sources $\hat{T}_{ij}$ and $\theta_{ij}$ can be decoupled in a successful way as long as the exchange of energy is there in between them. Denoting $\mathcal{G}_{ij}$ as the equations of motion of the line element in $f(Q)$-gravity, i.e. Eq. (\ref{eq8}), can be considered as follows:
\begin{eqnarray}
  \mathcal{G}_{ij}=  \frac{2}{\sqrt{-g}}\nabla_k\left(\sqrt{-g}\,f_Q\,P^k_{\,\,\,\,i j}\right)+\frac{1}{2}g_{i j}f 
+f_Q\big(P_{i\,k l}\,Q_j^{\,\,\,\,k l}\nonumber\\-2\,Q_{k l i}\,P^{k l}_{\,\,\,\,\,j}\big) = T_{ij} = \hat{T}_{ij}+\alpha ~\theta_{ij}.~~~~~~~\label{eq94}
\end{eqnarray}

The conservation equation can be found by Bianchi identity $ \bigtriangledown^i\mathcal{G}_{ij}=0$, given by
\begin{eqnarray}
&& \hspace{-0.2cm}\frac{H^{\prime}}{2}\,({T}^1_{1}-{T}^0_{0})+({T}^1_{1})^{\prime}-\frac{2}{r}\,({T}^2_{2}-{T}^1_{1})=\frac{\alpha \eta^{\prime}}{2} ({T}^0_{0}-{T}^1_{1}).~~~~~~\label{eq95}
\end{eqnarray}

The above Eq.  (\ref{eq95}) can be provided in a more suitable form, which is
\begin{small}
\begin{eqnarray}
&& \hspace{-0.1cm}-\frac{H^{\prime}}{2}\,(\hat{T}^0_{0}-\hat{T}^1_{1})+(\hat{T}^1_{1})^{\prime}-\frac{2}{r}\,(\hat{T}^2_{2}-\hat{T}^1_{1})-\frac{\alpha \eta^{\prime}}{2} (\hat{T}^0_{0}-\hat{T}^1_{1})  \nonumber\\&& \hspace{-0.1cm} \frac{\alpha \Phi'}{2} \Big([T^\theta]^1_1-[T^\theta]^0_0\Big) +\alpha \Big([T^\theta]^{1}_{1}\Big)^{\prime}=\frac{2 \alpha}{r}\,\Big([T^\theta]^{2}_{2}-[T^\theta]^{1}_{1}\Big).~~~~~~\label{eq96}
\end{eqnarray} 
\end{small}
Now there is a notable point in the context of the TOV equation for $f(\mathcal{Q})$ that under the linear functional form it corresponds to the static and spherically symmetric line element of general relativity.  This ensures that $\mathcal{G}^{\{H,W\}}_{ij}$ for the metric (\ref{eq43}) should satisfy its corresponding Bianchi identity. Again, this also suggests that the energy-momentum tensor $\hat{T}_{ij}$ should be conserved with the spacetime geometry $\{H, W\}$ of Eq. (\ref{eq42}). Hence, one can provide   
\begin{eqnarray}
    \bigtriangledown^{\{H,W\}}_i \, \hat{T}^{i}_ {j}=0. \label{eq97} 
\end{eqnarray}

It is also instructive, in connection to Eq.(\ref{eq13}), that
\begin{eqnarray}
    \bigtriangledown_i \, \hat{T}^{i}_{ j} =  \bigtriangledown^{\{H,W\}}_i \, \hat{T}^{i}_{j}-\frac{\alpha \,\eta'}{2} (\hat{T}^0_0-{T}^1_1) \delta^1_\nu. \label{eq98} 
\end{eqnarray}

As a linear combination of the Einstein field equations (\ref{eq40})--(\ref{eq42}), we obtain from Eq. (\ref{eq97}) the following explicit form:
\begin{eqnarray}
&& \hspace{-0.5cm}-\frac{H^{\prime}}{2}\,(\hat{T}^0_{0}-\hat{T}^1_{1})+(\hat{T}^1_{1})^{\prime}-\frac{2}{r}\,(\hat{T}^2_{2}-\hat{T}^1_{1})=0.~~~~\label{eq99}
\end{eqnarray}

This at once indicates that the source $\hat{T}_{ij}$ can be decoupled from the system of equations (\ref{eq39})--(\ref{eq41}) in a well defined manner and eventually, based on the condition (\ref{eq96}), one may get from Eq. (\ref{eq97}) the following forms:
\begin{eqnarray}
   &&  \bigtriangledown_\epsilon \, \hat{T}^{i}_{j} =-\frac{\alpha\, \eta'}{2} (\hat{T}^0_0-\hat{T}^1_1) \delta^1_j, \label{eq100}
   \end{eqnarray}
   and 
   \begin{eqnarray}
    && \bigtriangledown_i \, \theta^{i}_{ j}= \frac{\alpha\, \eta'}{2} (\hat{T}^0_0-\hat{T}^1_1) \delta^1_j. \label{eq101}
\end{eqnarray}

At this juncture, it is to be informed that (i) in Eqs. (\ref{eq39})--(\ref{eq41}), the divergence has been calculated in connection to the deformed spacetime (\ref{eq16}), (ii) Eq. (\ref{eq101}) is nothing but a linear combination of ``quasi-Einstein'' field equations, i.e. (\ref{eq45})--(\ref{eq47}) under the platform $f(\mathcal{Q})$-gravity, and (iii) as long as there is an exchange of energy between the sources $\hat{T}_{ij}$ and $\theta_{ij}$ decoupling can be successfully performed. Following the works~\citep{Ener1,Ener2}, the energy exchange between the sources can be expressed as follows:
\begin{eqnarray}
    \Delta E=\frac{\eta^\prime }{2} \big(p_r + \rho\big). \label{eq102}
\end{eqnarray}

Therefore, $p_r$ and $\rho$ being two positive physical quantities, the above Eq. (\ref{eq102}) can help explore the following situations: (i) if $\eta^\prime>0$ then $\Delta E>0$ which implies $\bigtriangledown_i \, \theta^{i}_{ j}>0$, i.e. the new source $ \theta_{i j}$ supplies energy to the environment, and (ii) if $\eta^\prime<0$ then $\Delta E<0$ which implies $\bigtriangledown_i \, \theta^{i}_{ j}<0$, i.e. the new source $ \theta_{i j}$ extracts energy from the environment.

It is noted that temporal deformation is the same for both solutions, therefore the expressions of energy exchange will be the same for both cases but the amount of energy exchange will be different for both cases. Now inserting the expressions for seed pressure and density along with the temporal deformation function $\eta$, we find
\begin{small}
\begin{eqnarray}
&&\hspace{-0.15cm} \Delta E=\frac{r}{8} \Bigg[\frac{1}{4 \left(N r^2+1\right)^4}\Big\{\Big(2 \beta_1 (L-N) \left(N r^2+3\right)+\beta_2 \nonumber\\
&&\hspace{0.3cm} \times \left(N r^2+1\right)^2\Big)  \Big(2 a \beta_1 (L-N) \left(N r^2+3\right) +(a \beta_2 -2 b ) \nonumber\\
&&\hspace{0.3cm} \times \left(N r^2+1\right)^2\Big)\Big\}  +c-\frac{\beta_2}{2}-\frac{\beta_1 (L-N) \left(N r^2+3\right)}{\left(N r^2+1\right)^2}\Bigg] \nonumber\\
&&\hspace{0.3cm} \times \Bigg[\Big\{4 \beta_1 L^3 (3 a \beta_2 -6 a \beta_1 N -3 b-1)+L^2 \big\{\beta_2 (a \beta_2-2 b-2) \nonumber\\
&&\hspace{0.3cm} +8 \beta_1 N (1-2 a \beta_2+2 b)  +4 a \beta_1^2 N^2\big\}-2 L N  [\beta_2 (a \beta_2-2 b-2) \nonumber\\
&&\hspace{0.3cm} +2 \beta_1 N (1-a \beta_2+b)] +\beta_2 N^2 (a \beta_2-2 b-2)+36 a \beta_1^2 L^4 \nonumber\\
&&\hspace{0.3cm} +4 a_3 (L-N)^2\Big\}  \frac{1}{\beta_1 L (L-N) \left(L r^2+1\right)} \nonumber\\
&&\hspace{0.3cm} +\frac{N (\beta_2 (a \beta_2-2 b-2)+4 c)}{\beta_1 L}+\frac{1}{(N-L) \left(N r^2+1\right)} \nonumber\\
&&\hspace{0.3cm} \times \Big[4 N \Big(a \big\{9 \beta_1 L^2 +2 \beta_2 L-6 \beta_1 L N+\beta_1 N^2-2 \beta_2 N\big\} \nonumber\\
&&\hspace{0.3cm}-2 b (L-N)\Big)\Big]-\frac{16 a \beta_1 N (2 L-N)}{\left(N r^2+1\right)^2}-\frac{16 a \beta_1 N (L-N)}{\left(N r^2+1\right)^3}\Bigg],~~~~~~ \label{eq103}
\end{eqnarray}
\end{small}

\subsubsection{For solution \ref{solA} ($\rho=\theta^0_0$)} \label{6.0.1}
\begin{figure*}
    \centering
   \includegraphics[width=3.4cm,height=3.4cm]{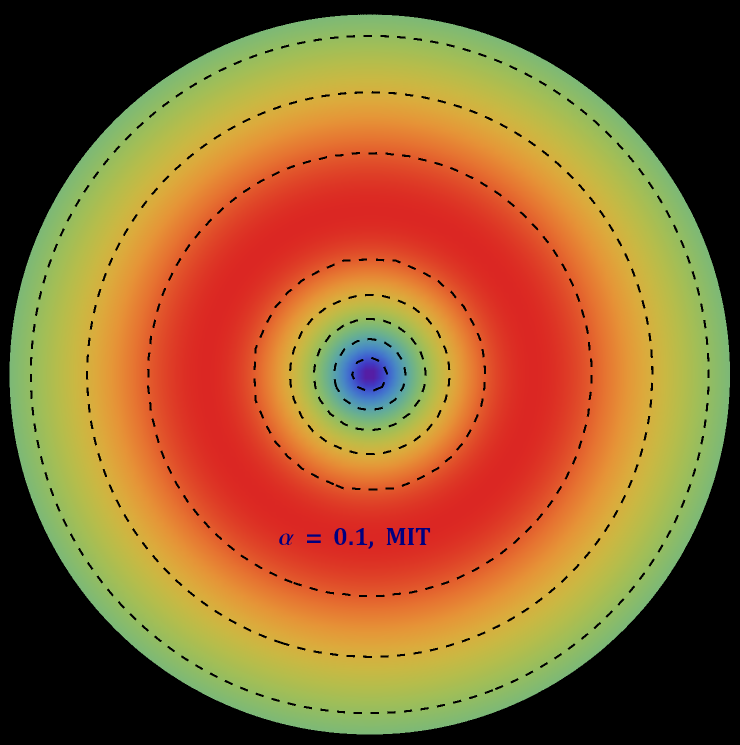}\includegraphics[width=1cm,height=3.4cm]{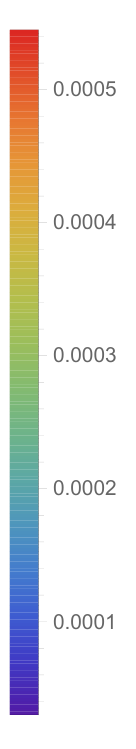} \includegraphics[width=3.4cm,height=3.4cm]{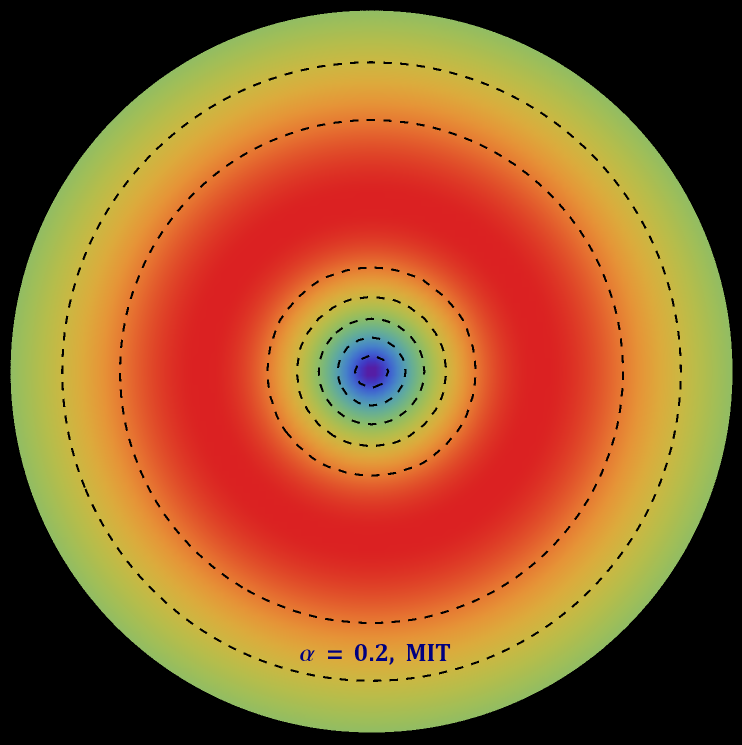}\includegraphics[width=1cm,height=3.4cm]{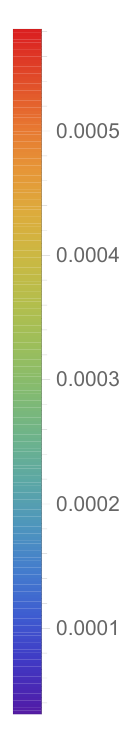}
      \includegraphics[width=3.4cm,height=3.4cm]{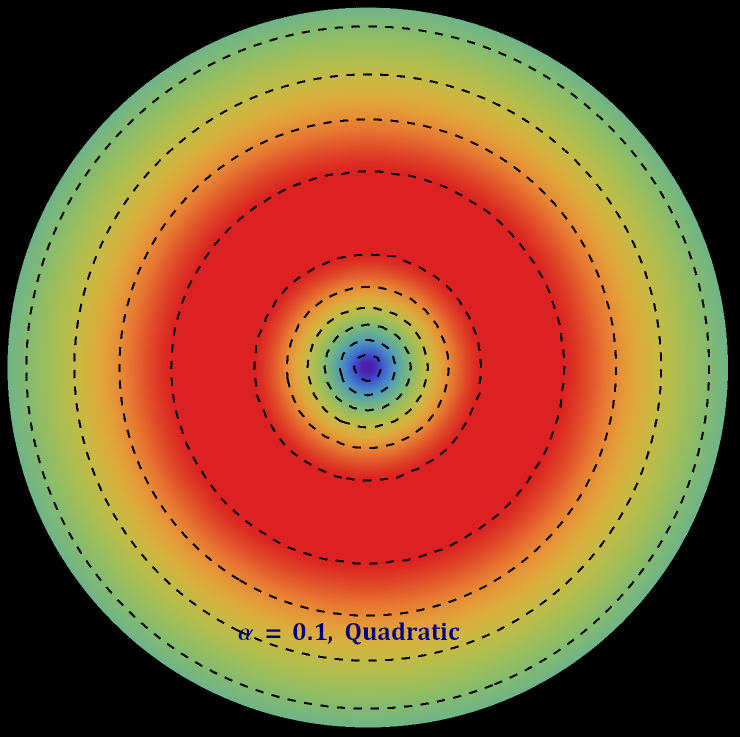}\includegraphics[width=1cm,height=3.4cm]{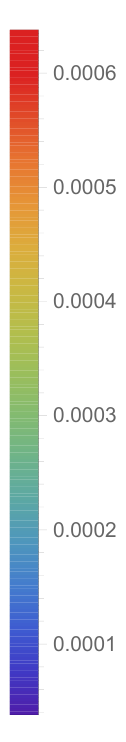}\includegraphics[width=3.4cm,height=3.4cm]{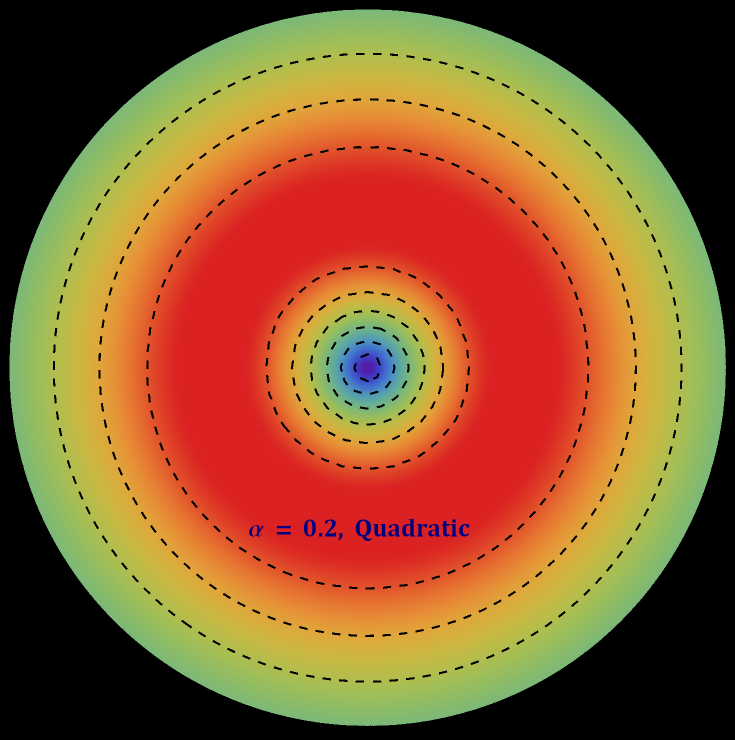}\includegraphics[width=1cm,height=3.4cm]{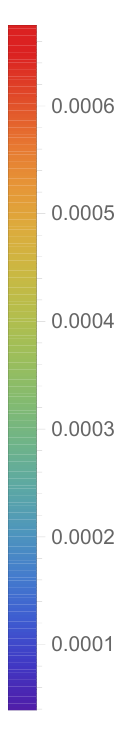}
    \caption{The flow of $energy-exchange$ between the fluid distributions for MIT Bag model by taking $a=0$ and $\mathcal{B}_g=60\,MeV/fm^3$ (first two panels) and quadratic model by taking $a=5\,\text{km}^2,~b=1/3$ and $\mathcal{B}_g=60\,MeV/fm^3$ (right two panels) with two different values of $\alpha$ for solution \ref{solA}($\rho=\theta^0_0)$). The same values of constants are employed here as used in Fig.\ref{fig1}.}
     \label{fig13}
\end{figure*}

In this Section, we discuss the amount of energy exchange ($\Delta E$) between the generic fluid $\theta_{ij}$ and anisotropic fluid $\hat{T}_{ij}$ for solution \ref{solA}. To see this distribution, we plotted Figure \ref{fig13} to show the energy exchange between the relativistic fluids via the density plots. The first left figure is plotted in the context of the pure  MIT bag model by taking bag constant value $\mathcal{B}_g=60~MeV/fm^3$ and decoupling constant $\alpha=0.1$. We observe that the $\Delta E$  is positive and minimum at the center while starts increasing towards the boundary and attaining its maximum value within the star rather than the boundary. The maximum value of $\Delta E$ is $0.00054~\text{km}^{-3}$. Now we move the second figure from the left which is plotted for the pure MIT bag model by taking $\mathcal{B}_g=60~MeV/fm^3$ with decoupling constant $\alpha=0.2$.  We observe that the same situation occurs as happened before but the amount of energy exchange increases which is  $\Delta E_{max}\approx 0.00058~\text{km}^{-3}$.

The third and fourth panels show the distribution of energy for the quadratic model with decoupling constant values $\alpha=0.1$ and  $\alpha=0.2$, respectively for the same value of bag constant $\mathcal{B}_g=60~MeV/fm^3$. The pattern of energy change within the star for the quadratic model is similar to the MIT bag model but $|\Delta E|_{qua}>|\Delta E|_{MIT}$. The maximum value of $\Delta E$ is $0.00063~\text{km}^{-3}$ and $0.00067~\text{km}^{-3}$ at $\alpha=0.1$ and $\alpha=0.2$, respectively.
 
 Finally, we conclude that the interaction between both fluids increases significantly when moving towards the boundary and reaches its maximum value within the star, not at the boundary in all cases for the solution \ref{solA} ($\rho=\theta^0_0$). Also, $\Delta E$ is positive throughout in the radial direction as well as its magnitude of maximum value increases when the decoupling constant $\alpha$ increases. Furthermore, the generic source $\theta_{ij}$ gives more energy to the environment in the presence of the quadratic EOS as compared to MIT bag model EOS.
 
 \begin{figure*}
    \centering
  \includegraphics[width=3.4cm,height=3.4cm]{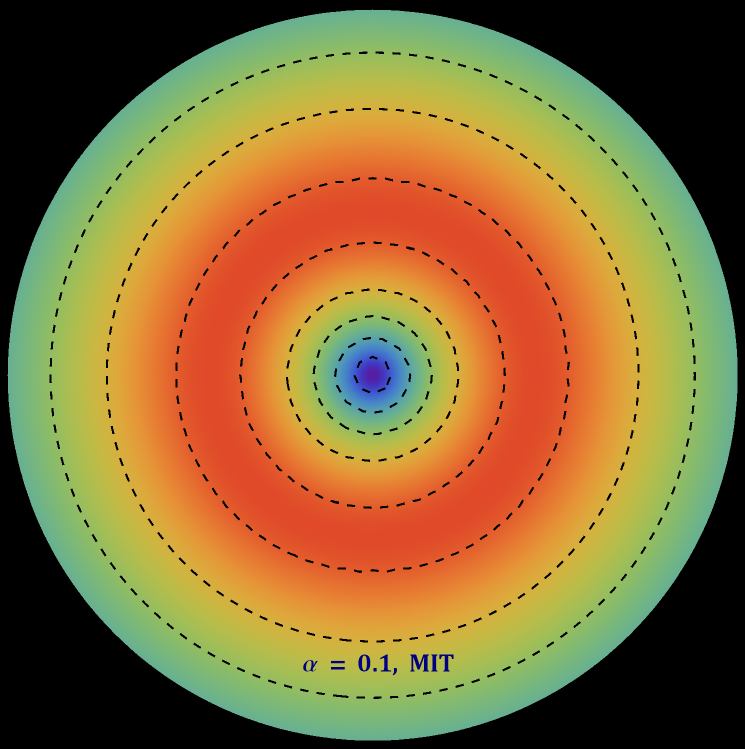}\includegraphics[width=1cm,height=3.4cm]{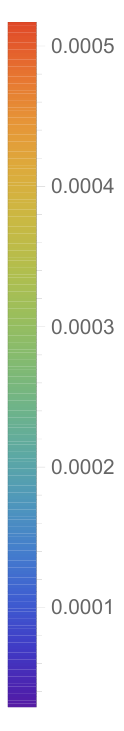} \includegraphics[width=3.4cm,height=3.4cm]{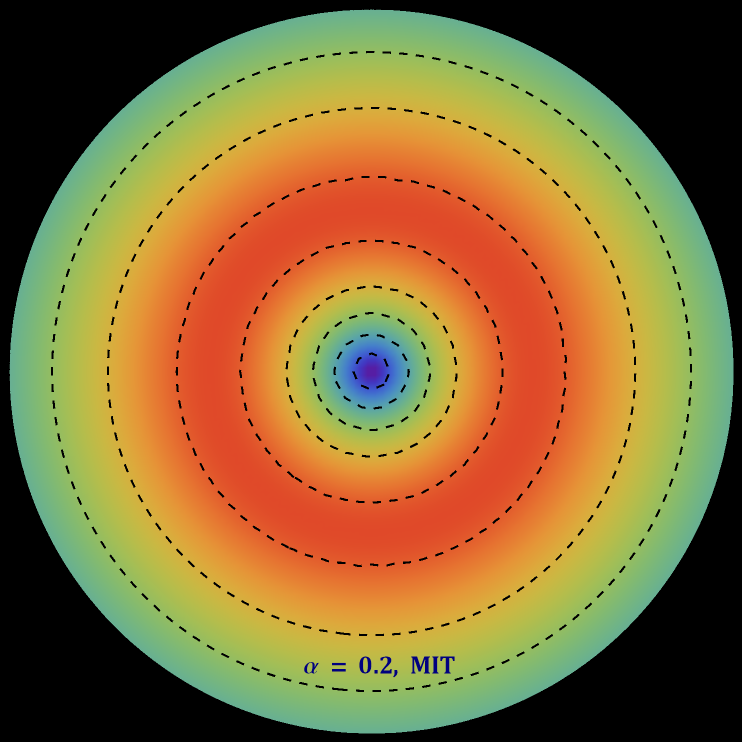}\,\includegraphics[width=1cm,height=3.4cm]{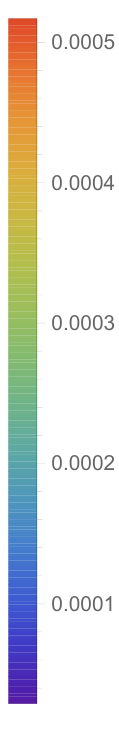}
      \includegraphics[width=3.4cm,height=3.4cm]{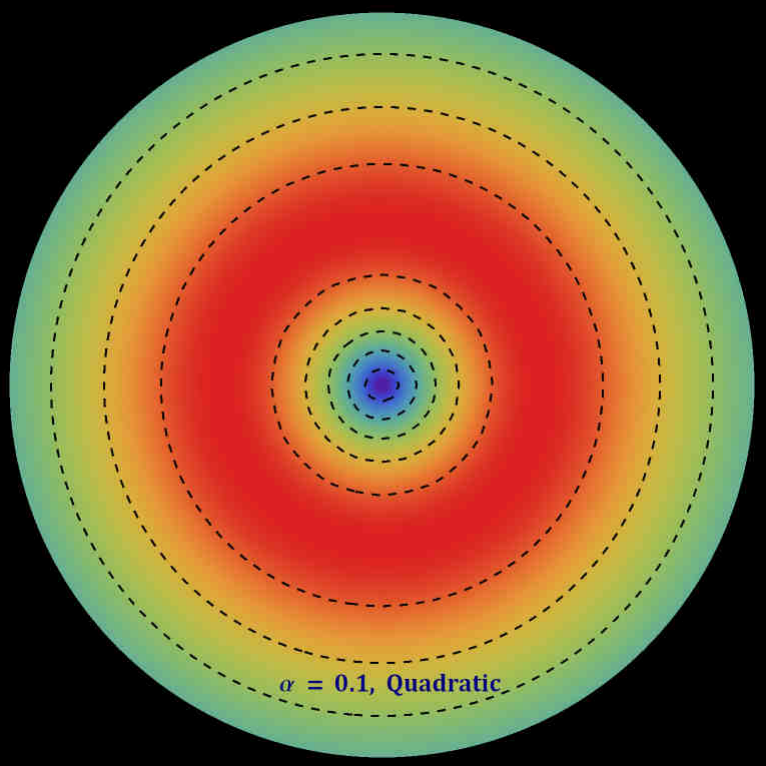}\includegraphics[width=1cm,height=3.4cm]{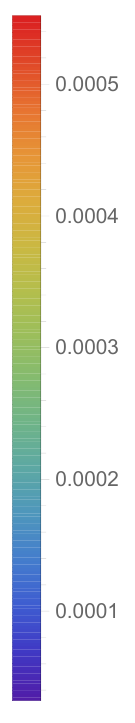}\includegraphics[width=3.4cm,height=3.4cm]{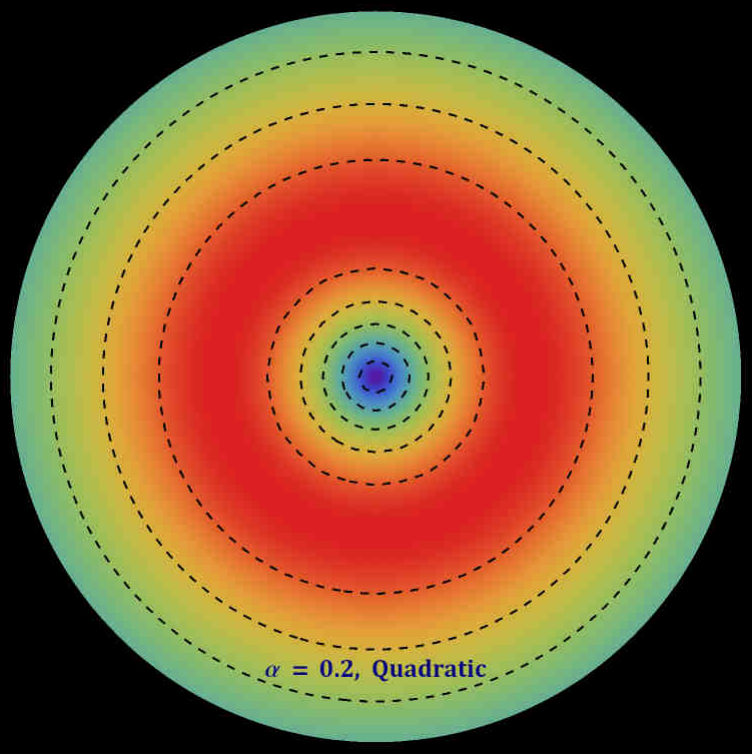}\includegraphics[width=1cm,height=3.4cm]{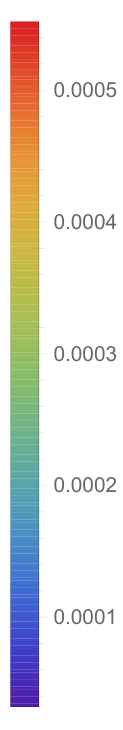}
    \caption{The flow of $energy-exchange$ between the fluid distributions for MIT Bag model by taking $a=0$ and $\mathcal{B}_g=60\,MeV/fm^3$ (first two panels) and quadratic model by taking $a=2\,\text{km}^2,~b=1/3$ and $\mathcal{B}_g=60\,MeV/fm^3$ (right two panels) with two different values of $\alpha$ for solution \ref{solB}($p_r=\theta^1_1)$). The same values of constants are employed here as used in Fig.\ref{fig4}.}
    \label{fig14}
\end{figure*}

\subsubsection{For solution \ref{solB} ($p_r=\theta^1_1$)} \label{6.0.2}

This section contains the analysis of energy exchange ($\Delta E$) distributions between the generic fluid $\theta_{ij}$ and anisotropic fluid $\hat{T}_{ij}$ for solution \ref{solB}. For this purpose, the density plot for Fig. \ref{fig14} reveals the flow of energy between the relativistic fluids.  The first two left figures are plotted in the context of the pure MIT bag model for decoupling constant $\alpha=0.1$ and $\alpha=0.2$, respectively using the bag constant value $\mathcal{B}_g=60~MeV/fm^3$. We observe that the $\Delta E$  is positive and minimum at the center while it starts increasing when moving towards the boundary but the maximum value of $\Delta E$ is achieved within the star, not at the boundary. On the other hand, we observe one interesting point here that no impact of the gravitational decoupling on the energy exchange fluid distributions is noticed under the mimic-to-pressures constraint approach. Therefore, the same maximum value of $\Delta E \approx 0.0005~\text{km}^{-3}$ is observed for both values of $\alpha=0.1$ and $0.2$. From the third and fourth panels, we show the distribution of energy for the quadratic model with decoupling constant values $\alpha=0.1$ and  $\alpha=0.2$, respectively for the same value of bag constant $\mathcal{B}_g=60~MeV/fm^3$. We find that the behavior of energy exchange within the star for the quadratic model is similar to the MIT bag model but $|\Delta E|_{qua}>|\Delta E|_{MIT}$ as happened in solution \ref{solA}. The maximum value of $\Delta E$ is $0.00054~\text{km}^{-3}$ at $\alpha=0.1$ and $\alpha=0.2$, both.

For the above solution, we finally can conclude that the interaction between both fluids also increases significantly when moving towards the boundary and reaches its maximum value within the star, not at the boundary, in all scenarios for the solution \ref{solB} ($p_r=\theta^1_1$) as well as $\Delta E$ is positive throughout in the radial direction. Furthermore, the magnitude of the maximum value for $\Delta E$ is independent of decoupling constant $\alpha$ i.e. $\Delta E$ remains the same for each $\alpha$. In fact here also, the generic source $\theta_{ij}$ gives more energy to the environment in the presence of the quadratic EOS as compared to MIT bag model EOS.

\begin{figure}
    \centering
 \includegraphics[width=8.3cm,height=6.7cm]{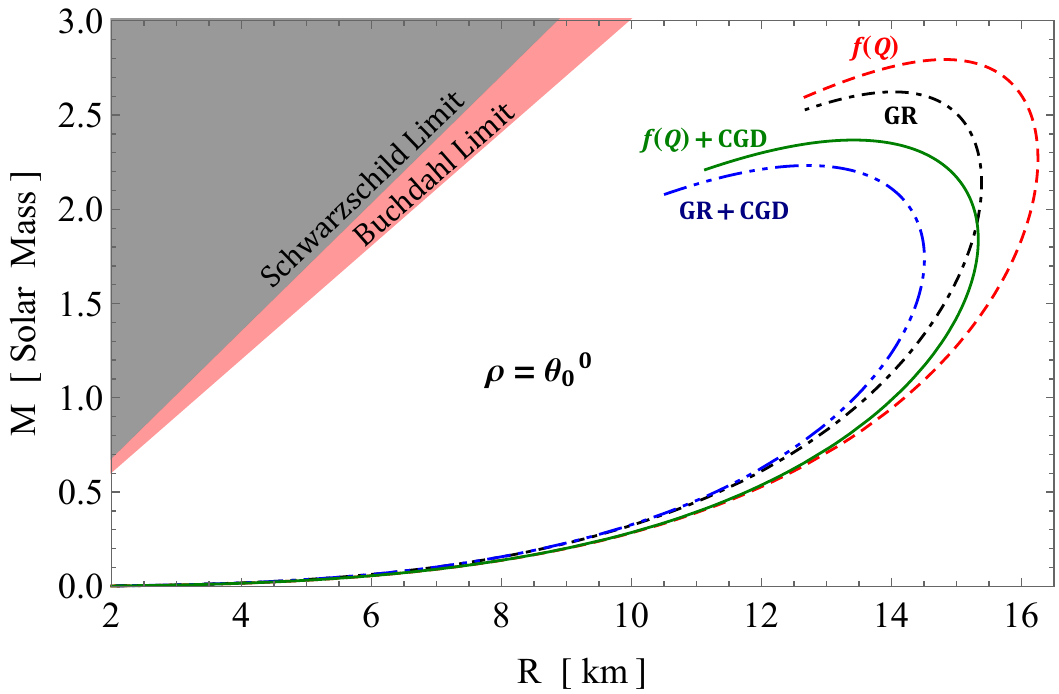}\\  \includegraphics[width=8.3cm,height=6.7cm]{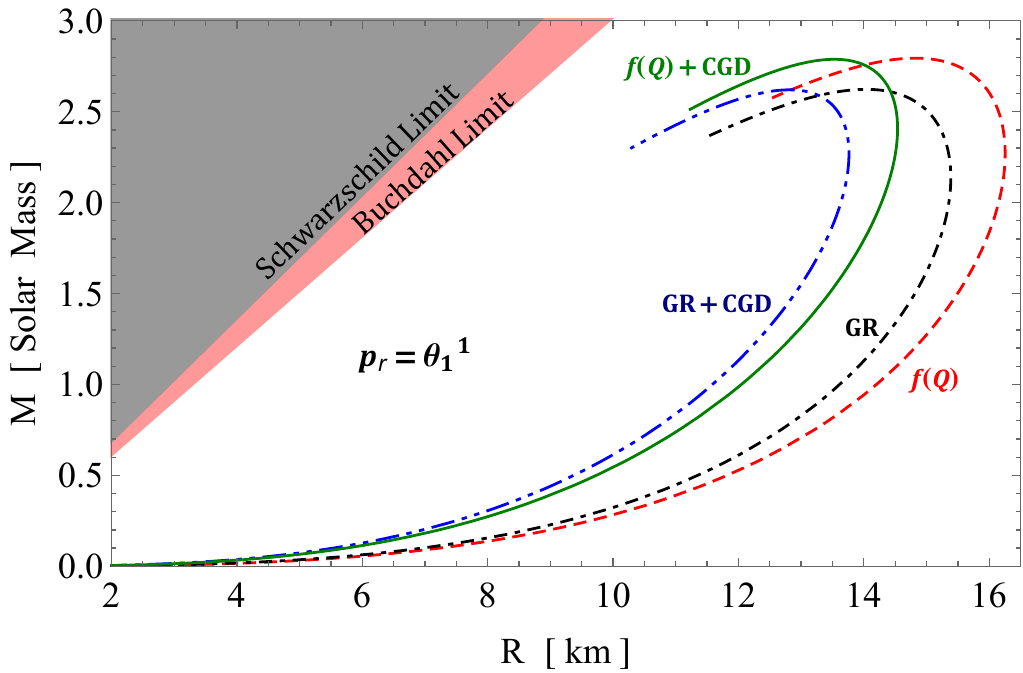}
    \caption{Comparison of GR, GR+CGD, $f(Q)$ and $f(Q)+$CGD for both the solutions with the same value of constants employed as in Fig. \ref{fig4}.}
    \label{fig20}
\end{figure}

{\section{Comparative study of model's arising in GR, GR+CGD, $\lowercase{f}(\mathcal{Q})$ and $\lowercase{f}(\mathcal{Q})$+CGD gravity}  \label{sec7}
In this section we highlight the key differences arising in general relativity (GR) and  $f(\mathcal{Q})$ gravity with and without CGD via our modeling framework. To this end, we draw the reader's attention to Fig. \ref{fig20}. For the first case ($\rho=\theta_0^0$), we have plotted the M-R curves for pure GR ~[$\beta_1=1,~\alpha=0$] and pure $f(\mathcal{Q})$~ [$\beta_1=1.1,~\alpha=0$] in Fig. \ref{fig20} (top). We note that for the same EOS, the $f(\mathcal{Q})-$gravity can generate higher $M_{max}$ than GR iff $\beta_1>1$, otherwise the reverse will happen. However, when CGD turns on i.e. $\alpha \neq 0$, the corresponding EOS gets softer resulting in lower $M_{max}$ for both GR and $f(\mathcal{Q})$ gravity. Therefore, when the $tt-$ component of the CGD-induced stress-tensor $\theta_i^j$ mimics the density, the $M_{max}$ is lowered due to softening EOS. This can be ascribed to the effective density i.e. $\epsilon=(1+\alpha)\rho$ increasing with the CGD strength leading to a denser interior, which will eventually trigger many exotic processes such as hyperon and kaon productions. On the other hand, the second solution where $\theta_1^1$ mimics pressure leads to the lowering of the effective pressure $P_r=(1-\alpha)p_r$. In view of the EOS within the core, lower radial pressure supports lower interior density which may suppress any exotic phase transitions. This makes the EOS stiffer when the CGD strength increases leading to higher $M_{max}$ (see Fig. \ref{fig20} bottom) and compactness parameter. Furthermore, it can also be observed that both $f(\mathcal{Q})$ models yield roughly the same masses and radii when CGD is turned off. However, when CGD turns on the first case immediately decrease the $M_{max}$ and corresponding radius while the second solution yields a slightly higher maximum mass with smaller  radius. In the GR limit, both solutions yield the same $M-R$ curves as the seed solution is unaffected by $\beta$.}

\section{Concluding Remarks and Astrophysical Implications} \label{sec8}

To make some concluding remarks on the present work and its outcomes, let us arrange the steps point-wise that what was our motivation and how we have carried out that in a successful way:

(i) Inspired by the recent gravitational event, i.e. GW190814 which brings to light the coalescence of a 23 $ M_{\odot}$ black hole with a yet-to-be-determined secondary component, we try to model compact objects under the framework of $f(\mathcal{Q})$ gravity theory along with the method of gravitational decoupling.

(ii) We assume a suitable quadratic equation of state (EOS) for the interior matter distribution of a compact star, especially a neutron/quark star, which in the appropriate limit reduces to the MIT bag model. 

(iii) The concerned field equations under $f(\mathcal{Q})$ gravity are subjected to gravitational decoupling bifurcates into the $\rho=\theta^0_0$ and $p_r=\theta^1_1$ sectors leading two distinct classes of solutions. 

(iv) Both families of solutions are subjected to rigorous tests qualifying them to describe a class of compact stellar objects which include neutron stars, strange stars, and the possible progenitor of the secondary component of GW190814.

(v) Using observational data of mass-radius relations for compact objects, e.g. LMC X-4, Cen X-3, PSR J1614-2230, and PSR J0740+6620, we show that it is possible to generate stellar masses and radii beyond 2.5 $ M_{\odot}$. 

(vi) The outcomes of the present work reveal that the quadratic EOS is versatile enough to account for a range of low-mass stars as well as typical stellar candidates describing the secondary component of GW190814. 

Based on the above-mentioned steps, the obtained results have been already classified with detailed discussions in several tables and figures. Let us now put some salient features of the outcomes as follows:

1. {\bf Graphical plots for the basic physical parameters}: Distribution of the energy density (Figs. \ref{fig1} and \ref{fig4}), the radial pressure (Figs.\ref{fig2} and \ref{fig5}), and the anisotropy (Figs. \ref{fig3} and \ref{fig6}) are exhibited for different values of the model parameters. Discussions in a length have been done already in \ref{5.1.1} in connection to the solution \ref{solA} ($\rho=\theta^0_0$) and in \ref{5.1.2} in connection to the solution \ref{solB} ($p_r=\theta^1_1$) respectively. We note that all the features are satisfactory as far as physical attributes are concerned.

2. {\bf Constraining upper limit of maximum mass via the graphical plots for M-R diagrams}: For different choices of the linear EOS with constancy in bag constant, fixed decoupling parameter, and varying EOS parameter we have shown results in Tables \ref{table1}-\ref{table6} and figures \ref{fig7a}-\ref{fig12a} which are as follows.\\
In Table \ref{table1} and Fig. \ref{fig7a}, we observe linear contributions from the energy density which do account for low-mass stars such as LMC X-4 and Cen X-3. \\
From Table  \ref{table2} and Fig. \ref{fig8a} (left panel), we observe that an increase in the decoupling parameter results in a decrease in radii leading to more compact configurations. The left panel also shows that low mass stars such as  LMC X-4 (1.29 $\pm$ 0.05 $M_{\odot}$) and Cen X-3 (1.49 $\pm$ 0.08 $M_{\odot}$) do exist for radii greater 10 km. \\

 Figure 8 predicts a wide range of masses and radii applicable to observed compact objects, including the 'twin star' scenario, the latter of which is favoured in the $p_r=\theta^1_1$ sector. Interestingly, the radii of these compact objects in $f(R)$ gravity with rotation can vary between 12 to 18 km.\\
We ascertain from Table \ref{table6} and Fig. \ref{fig12a} that the quadratic EOS model fails to predict masses which can account for the LIGO observation of approximately 2.6  $M_{\odot}$ of the secondary component of the binary coalescence GW190814. It appears that when the linear and quadratic EOS parameters are switched on simultaneously, the decoupling constant quenches any increase in the mass of the compact object. While these models fail to predict masses above 2.6 $M_{\odot}$, the quadratic EOS predicts the existence of well-known pulsars and neutron stars to a very good approximation. The tabulated data and plots reveal that the quadratic EOS which has the linear limit successfully predicts the existence of low-mass stars as well as neutron stars with masses beyond 2 $M_{\odot}$. 

3. {\bf Graphical plots for the energy exchange between the fluid distributions}: We would now discuss The most important aspect of our findings, ie., the energy exchange arising from the extended gravitational decoupling. It is noted that temporal deformation is the same for both the solutions and hence the expressions of energy exchange will be the same for both cases but the amount of energy exchange will be different for both cases. \\
(i) \textbf{For solution \ref{solA} ($\rho=\theta^0_0$):} In Fig. \ref{fig13} we have exhibited the energy exchange between the relativistic fluids via the density plots. The first left figure is plotted in the context of the pure MIT bag model where we observe that the $\Delta E$  is positive and minimum at the center while starts increasing towards the boundary and attaining its maximum value within the star rather than the boundary. Now we move the second figure from the left which is plotted for the pure MIT bag model and one can note that the same situation occurs as happened before but the amount of energy exchange increases which is  $\Delta E_{max}\approx 0.00058~\text{km}^{-3}$.\\
The third and fourth panels show the distribution of energy for the quadratic model where the pattern of energy change within the star for the quadratic model is similar to the MIT bag model but $|\Delta E|_{qua}>|\Delta E|_{MIT}$. The maximum value of $\Delta E$ is $0.00063~\text{km}^{-3}$ and $0.00067~\text{km}^{-3}$ at $\alpha=0.1$ and $\alpha=0.2$, respectively.

 Based on the above observations, one may conclude that the interaction between both fluids increases significantly when moving towards the boundary and reaches its maximum value within the star, not at the boundary in all cases for the solution \ref{solA} ($\rho=\theta^0_0$). Also, $\Delta E$ is positive throughout in the radial direction as well as its magnitude of maximum value increases when the decoupling constant $\alpha$ increases. Furthermore, the generic source $\theta_{ij}$ gives more energy to the environment in the presence of the quadratic EOS as compared to MIT bag model EOS.
 
(ii) {\bf For solution \ref{solB} ($p_r=\theta^1_1$)}: The density plot for Fig. \ref{fig14} has been shown to observe the flow of energy exchange between the relativistic fluids. The first two left figures are plotted in the context of the pure MIT bag model where we observe that the $\Delta E$  is positive and minimum at the center while it starts increasing when moving towards the boundary but the maximum value of $\Delta E$ is achieved within the star, not at the boundary. On the other hand, we observe one interesting point here that no impact of the gravitational decoupling on the energy exchange fluid distributions is noticed under the mimic-to-pressures constraint approach. From the third and fourth panels, we show the distribution of energy for the quadratic model and find that the behavior of energy exchange within the star for the quadratic model is similar to the MIT bag model but $|\Delta E|_{qua}>|\Delta E|_{MIT}$ as happened in solution \ref{solA}. 

Therefore, we can conclude that the interaction between both fluids also increases significantly when moving towards the boundary and reaches its maximum value within the star, not at the boundary in all the situations for the solution \ref{solB} ($p_r=\theta^1_1$) as well as $\Delta E$ is positive throughout in the radial direction. Furthermore, the magnitude of the maximum value for $\Delta E$ is independent of decoupling constant $\alpha$, i.e. $\Delta E$ remains the same for each $\alpha$. In fact here also, the generic source $\theta_{ij}$ gives more energy to the environment in the presence of the quadratic EOS as compared to MIT bag model EOS.

The overall findings of the models presented reveal that it is possible to put suitable constraints on the upper limit of the mass-radius relation of the secondary component of GW190814 and other self-bound strange star configurations under gravitational decoupling in $f(\mathcal{Q})$-gravity theory which may provide the observational signature of the objects in a significant way.

 \section*{Acknowledgement}
The author SKM is thankful for continuous support and encouragement from the administration of the University of Nizwa to carry out the research work. G. Mustafa is very thankful to Prof. Gao Xianlong from the Department of Physics, Zhejiang Normal University, for his kind support and help during this research. Further, G. Mustafa acknowledges Grant No. ZC304022919 to support his Postdoctoral Fellowship at Zhejiang Normal University. KNS and SR are also thankful to the authorities of the Inter-University Centre for Astronomy and Astrophysics, Pune, India for providing the research facilities where SR specifically expresses thanks to ICARD of IUCAA at GLA University. The authors are also thankful to Prof. Jorge Ovalle for his help in the coding of the contour diagrams. 

\section*{DATA AVAILABILITY:}
No new data were generated or analyzed in support of this research.

\section*{Appendix}
\begin{small}
\begin{eqnarray}
   &&\hspace{-0.15cm} \theta_{11}(r)=\frac{1}{4}  \Bigg[\frac{1}{\beta_1 L (L-N) \left(L r^2+1\right)}\Big(-4 \beta_1 L^3 (1-3 a \beta_2 \nonumber\\&& \hspace{0.6cm} +6 a \beta_1 N+3 b)  +L^2 \big(\beta_2 (a \beta_2-2 b-2)-8 \beta_1 N (2 a \beta_2\nonumber\\&& \hspace{0.6cm} -2 b-1)+4 a \beta_1^2 N^2\big) -2 L N (\beta_2 (a \beta_2-2 b-2)+2 \beta_1 \nonumber\\&& \hspace{0.6cm} \times N (-a \beta_2+b+1)) +\beta_2 N^2 (a \beta_2 -2 b-2)+36 a \beta_1^2 \nonumber\\&& \hspace{0.6cm} \times L^4+4 c (L-N)^2\Big)+\frac{N (\beta_2 (a \beta_2-2 b-2) +4 c)}{\beta_1 L}  \nonumber\\&& \hspace{0.6cm} +\frac{4 N \big(a \big(9 \beta_1 L^2+2 \beta_2 L-6 \beta_1 L N+\beta_1 N^2-2 \beta_2 N\big)\big)}{(N-L) \big(N r^2+1\big)} \nonumber\\&& \hspace{0.6cm}  -\frac{16 a \beta_1 N (2 L-N)-2 b (N r^2+1)}{\big(N r^2+1\big)^2} -\frac{16 a \beta_1 N (L-N)}{\big(N r^2+1\big)^3}\Bigg],\,\nonumber
   \end{eqnarray}
   \begin{eqnarray}
 &&\hspace{-0.15cm}   \theta_{22}(r)=-\frac{1}{4 \beta_1 \big(L r^2+1\big)^2 \big(N r^2+1\big)^4}\Big[-a \beta_2^2-4 \beta_1 L^2 \nonumber\\&& \hspace{0.6cm} \times \big(9 a \beta_1+N^4 r^{10} (1-a \beta_2+b) +2 N^3 r^8  (-5 a \beta_2 \nonumber\\&& \hspace{0.6cm} +3 a \beta_1 N+5 b+2)+N^2 r^6  (-20 a \beta_2+53 a \beta_1 N  +20 b \nonumber\\&& \hspace{0.6cm} +6)+N r^4 (-14 a \beta_2+101 a \beta_1 N +14 b+4)+r^2 (-3 a \nonumber\\&& \hspace{0.6cm} \times (\beta_2+3 \beta_1 N)+3 b+1)\big)+L \big\{4 \beta_1 \big(N r^2+1\big)^2  \big[-3 a \beta_2 \nonumber\\&& \hspace{0.6cm} +N^3 r^6 (1-a \beta_2)+3 N^2 r^4 (1-3 a \beta_2)-3 N r^2 (a \beta_2  -1) \nonumber\\&& \hspace{0.6cm}+b \big(N^3 r^6+9 N^2 r^4+3 N r^2+3\big)+1\big] +\beta_2 r^2 \big(N r^2-1\big) \nonumber\\&&\hspace{0.5cm} \times \big(N r^2+1\big)^4 (-a \beta_2+2 b+2)+4 a \beta_1^2 N \big(3 N^4 r^8+25 N^3 \nonumber\\&& \hspace{0.6cm} \times r^6+31  N^2 r^4-45 N r^2+18\big)\big\}+12 a \beta_1^2 L^3 r^2 \big(N^3 r^6 \nonumber\\&& \hspace{0.6cm} +9 N^2 r^4+19 N r^2 +3\big)-3 a \beta_2^2 N^5 r^{10}+4 a \beta_1 \beta_2 N^5 r^8 \nonumber\\&& \hspace{0.6cm} +4 a \beta_1^2 N^5 r^6 -13 a \beta_2^2 N^4 r^8 +8 a \beta_1 \beta_2 N^4 r^6+52 a \beta_1^2 \nonumber\\&& \hspace{0.6cm} \times N^4 r^4-22 a \beta_2^2 N^3 r^6+16 a \beta_1 \beta_2 N^3 r^4  +108 a \beta_1^2 N^3 r^2 \nonumber\\&& \hspace{0.6cm} -36 a \beta_1^2 N^2-18 a \beta_2^2 N^2 r^4+24 a \beta_1 \beta_2 N^2 r^2  +12 a \beta_1 \beta_2 N \nonumber\\&& \hspace{0.6cm} -7 a \beta_2^2 N r^2+2 b \beta_2+6 b \beta_2 N^5 r^{10} -4 \beta_1 b N^5 r^8  +26 b \beta_2 N^4 \nonumber\\&& \hspace{0.6cm} \times r^8-8 \beta_1 b N^4 r^6+44 b \beta_2 N^3 r^6-16 \beta_1 b N^3 r^4+36 b \beta_2 \nonumber\\&& \hspace{0.6cm} \times N^2  r^4-24 \beta_1 b N^2 r^2 -12 \beta_1 b N+14 b \beta_2 N r^2+2 \beta_2-4 c \nonumber\\&&\hspace{0.6cm} \times \big(N r^2+1\big)^4 \big(L r^2 \big(N r^2-1\big)+3 N r^2+1\big)+6 \beta_2 N^5 r^{10} \nonumber\\&&\hspace{0.6cm} -4 \beta_1  N^5 r^8+26 \beta_2 N^4 r^8-16 \beta_1 N^4 r^6+44 \beta_2 N^3 r^6-24 \beta_1 \nonumber\\&&\hspace{0.6cm} \times N^3 r^4  +36 \beta_2 N^2 r^4-16 \beta_1 N^2 r^2 -4 \beta_1 N+14 \beta_2 N r^2\Big]. \nonumber
\end{eqnarray}
\begin{eqnarray}
   && \hspace{-0.1cm} \theta_{23} (r)=\theta^2_{11}(r)  r^2    \left(N r^2+1\right) \big[6 \beta_1  \big\{k L r^2+k+\big(\alpha ^2 +2 \alpha +3\big) \nonumber\\&& \hspace{0.6cm} \times  L r^2 -\alpha ^2 N r^2 -2 \alpha  N r^2 -N r^2+2\big\}  +(\alpha +1)^2 \beta_2 r^2 \nonumber\\&& \hspace{0.6cm} \times \left(N r^2+1\right)\big]+4 \theta_{11}(r)  \big(3 \beta_1 \big((\alpha +2)  L r^2  \left(N r^2+2\right) \nonumber\\&& \hspace{0.6cm} -(\alpha +1)-N^2 r^4-2 (\alpha +1) N r^2+1\big)+(\alpha +1) \beta_2 \nonumber\\&& \hspace{0.6cm} \times \left(N r^3+r\right)^2\big) +4 \big(\beta_2+6 \beta_1 L+\beta_2 N^2 r^4 -6 \beta_1 N \nonumber\\&& \hspace{0.6cm} +2 \beta_2 N r^2\big),\nonumber\\
   &&\hspace{-0.2cm}    \Psi_{21}(r)= 4 \beta_1 \big(N r^2+1\big)^2 \big[N r^2 \big(\alpha -3 \alpha  a \beta_2+\alpha  L r^2 (a \beta_2-1) \nonumber\\&&\hspace{0.3cm} +a \beta_2  \big(L r^2-3\big)+2\big) +3 \alpha  a \beta_2 L r^2+3 a \beta_2 L r^2-(\alpha +1) N^2 r^4 \nonumber\\&&\hspace{0.3cm} \times (a \beta_2-1)-(\alpha +1) b r^2 (L-N) \big(N r^2+3\big)-\alpha  L r^2+1\big] \nonumber\\&&\hspace{0.3cm} +(\alpha +1) \beta_2 r^2 \big(N r^2+1\big)^4 (a \beta_2-2 b-2)+4 (\alpha +1) a \beta_1^2 r^2 \nonumber\\&&\hspace{0.3cm} \times (L-N)^2 \big(N r^2+3\big)^2+4 (\alpha +1) c r^2 \big(N r^2+1\big)^4, \nonumber
\end{eqnarray}
\begin{eqnarray}
  &&\hspace{-0.2cm}  \Psi_{22} (r)=+N \big(36 a \beta_1+r^2 (3-14 a \beta_2)\big)+2 b \big(N r^2+1\big)^2 \big(N r^2\nonumber\\&&\hspace{0.2cm}+3\big) +1\big] +4 a \beta_1^2 L^2 \big(N r^2+3\big)^2+a \beta_2^2 N^4 r^8-4 a \beta_1   \beta_2 N^4 r^6 \nonumber\\&&\hspace{0.2cm} +4 a \beta_1^2 N^4 r^4 \  +4 a \beta_2^2 N^3 r^6-20 a \beta_1 \beta_2 N^3 r^4+24 a \beta_1^2 N^3 r^2 \nonumber\\&&\hspace{0.2cm} +36 a \beta_1^2 N^2+6 a \beta_2^2  N^2 r^4-28 a \beta_1 \beta_2 N^2 r^2-12 a \beta_1 \beta_2 N \nonumber\\&&\hspace{0.2cm}  +4 a \beta_2^2 N r^2 -2 b \beta_2-2 b  \beta_2 N^4 r^8+4 \beta_1 b N^4 r^6-8 b \beta_2 N^3 r^6 \nonumber\\&&\hspace{0.2cm}  +20 \beta_1 b N^3 r^4-12 b \beta_2 N^2 r^4  +28 \beta_1 b N^2 r^2+12 \beta_1 b N-8 b \beta_2 \nonumber\\&&\hspace{0.2cm}  \times N r^2-\beta_2+4 c \big(N r^2+1\big)^4-\beta_2  N^3 r^6(1+4N r^2)+2 \beta_1 N^4 r^6 \nonumber\\&&\hspace{0.2cm}+6 \beta_1 N^3 r^4-6 \beta_2 N^2 r^4+6 \beta_1 N^2 r^2 +2 \beta_1 N-4 \beta_2 N r^2, \nonumber
\end{eqnarray}
\begin{eqnarray}
 &&\hspace{-0.1cm}\Psi_{00} (r)=   \big[2 a \beta_2^2 -\beta_1 L \big(N^2 \big(r^4 (3-6 a \beta_2)  +8 a \beta_1 r^2\big) -14 a \beta_2 \nonumber\\&&\hspace{0.3cm} +N \big(24 a \beta_1 +r^2  (6-20 a \beta_2)\big)+2 b \big(3 N^2 r^4+10 N r^2+7\big)\nonumber\\&&\hspace{0.3cm}+3\big)+4 a \beta_1^2 L^2  \big(N r^2+3\big)  +2 a N^3( \beta_2^2 r^6-3 \beta_1 \beta_2 r^4+2 \beta_1^2 r^2) \nonumber\\&&\hspace{0.3cm} +12 a  \beta_1^2 N^2+6 a \beta_2^2 N^2 r^4-20 a \beta_1 \beta_2 N^2 r^2-14 a \beta_1 \beta_2 N  \nonumber\\&&\hspace{0.3cm} +6 a \beta_2^2 N r^2-4 b \beta_2-4 b \beta_2 N^3 r^6+6 \beta_1 b N^3 r^4-12 b \beta_2 N^2 r^4 \nonumber\\&&\hspace{0.3cm} +20 \beta_1 b N^2 r^2 +14 \beta_1 b N-12 b \beta_2 N r^2  -2 \beta_2+8 c \big(N r^2+1\big)^3 \nonumber\\&&\hspace{0.3cm} -2 \beta_2 N^3 r^6 +3 \beta_1 N^3 r^4-6 \beta_2 N^2 r^4+6 \beta_1 N^2 r^2+3 \beta_1 N -6 \beta_2 \nonumber\\&&\hspace{0.3cm} \times  N r^2\big] \big[4 \beta_1  \big(N r^2+1\big)^2 \big(N r^2 \big(\alpha -3 \alpha  a \beta_2+\alpha  L r^2 (a \beta_2-1) \nonumber\\&&\hspace{0.3cm} +a \beta_2 \big(L r^2 -3\big)+2\big)+3 \alpha  a \beta_2 L r^2  + 3 a \beta_2 L r^2  - (\alpha +1) N r^4 \nonumber\\&&\hspace{0.3cm} \times (a \beta_2-1)  -(\alpha +1) b r^2 (L-N) \big(N r^2+3\big)-\alpha  L r^2+1\big)  \nonumber\\&&\hspace{0.3cm} +(\alpha +1) \beta_2 r^2  \big(N r^2+1\big)^4    (a \beta_2-2 b-2)+4 (\alpha +1) a \beta_1^2 r^2 \nonumber\\&&\hspace{0.3cm} \times (L-N)^2   \big(N r^2+3\big)^2+4 (\alpha +1) c r^2 \big(N r^2+1\big)^4\big] \nonumber
\end{eqnarray}
\end{small}
\begin{small}
\begin{eqnarray}
&&\hspace{-0.2cm} f_1(r)=\frac{1}{\left(N r^2+1\right)^3 \left(\beta_1+\beta_1 L r^2\right)}\Big[a \beta_2^2-4 \beta_1 L \big(-3 a \beta_2+N^3 \nonumber\\&&\hspace{0.2cm} \times \big(r^6 (1-a \beta_2)+2 a \beta_1 r^4\big)+N^2  \big(r^4 (3-5 a \beta_2)+12 a \beta_1 r^2\big)\nonumber\\&&\hspace{0.2cm}+N  \big(18 a \beta_1+r^2 (3-7 a \beta_2)\big)+b \big(N r^2+1\big)^2 \big(N r^2+3\big)+1\big) \nonumber\\&&\hspace{0.2cm} +4 a \beta_1^2  L^2 \big(N r^2+3\big)^2+a \beta_2^2 N^4 r^8-4 a \beta_1 \beta_2 N^4 r^6 +4 a \beta_1^2 N^4 r^4 \nonumber\\&&\hspace{0.2cm} +4 a \beta_2^2  N^3 r^6-20 a \beta_1 \beta_2 N^3 r^4  +24 a \beta_1^2 N^3 r^2+36 a \beta_1^2 N^2 \nonumber\\&&\hspace{0.2cm} +6 a \beta_2^2 N^2 r^4 -28 a \beta_1 \beta_2 N^2 r^2-12 a \beta_1 \beta_2 N+4 a \beta_2^2 N r^2-2 b \beta_2  \nonumber\\&&\hspace{0.2cm} -2 b \beta_2 N^4 r^8  +4 \beta_1 b N^4 r^6-8 b \beta_2 N^3 r^6+20 \beta_1 b N^3 r^4-12 b \beta_2 \nonumber\\&&\hspace{0.2cm} \times N^2 r^4+28 \beta_1 b  N^2 r^2+12 \beta_1 b N  -8 b \beta_2 N r^2-2 \beta_2+4 b \nonumber\\&&\hspace{0.2cm} \times \big(N r^2+1\big)^4-2 \beta_2 N^4 r^8 +4 \beta_1 N^4 r^6-8 \beta_2 N^3 r^6+12 \beta_1 N^3 r^4 \nonumber\\&&\hspace{0.2cm} -12 \beta_2 N^2 r^4+12 \beta_1 N^2 r^2 \nonumber\\&&\hspace{0.2cm} +4 \beta_1 N-8 \beta_2 N r^2\Big],\nonumber
\end{eqnarray}
\begin{eqnarray}
  &&\hspace{-0.2cm}  f_2(r)=2 \Big[\Big\{(N-L) \Big(a N^4 \beta_2^2 r^8-2 b N^4 \beta_2 r^8-2 N^4 \beta_2 r^8+4 a N^3 \nonumber\\&&\hspace{0.2cm} \times  \beta_2^2 r^6  -8 b N^3 \beta_2 r^6-8 N^3 \beta_2 r^6-4 a N^4 \beta_1^2 r^4+6 a N^2 \beta_2^2 r^4 \nonumber\\&&\hspace{0.2cm} -8 b N^3 \beta_1 r^4 -12 b N^2 \beta_2 r^4-12 N^2 \beta_2 r^4+8 a N^3 \beta_1 \beta_2 r^4-40 a \nonumber\\&&\hspace{0.2cm} \times N^3 \beta_1^2 r^2+4 a N \beta_2^2 r^2-16 b N^2 \beta_1 r^2-8 b N \beta_2 r^2-8 N \beta_2 r^2\nonumber\\&&\hspace{0.2cm} +16 a N^2 \beta_1 \beta_2 r^2  +4 b \big(N r^2+1\big)^4-84 a N^2 \beta_1^2+4 a L^2 \big(2 N^3 r^6 \nonumber\\&&\hspace{0.2cm} 
 +17 N^2 r^4+36 N r^2 +9\big) \beta_1^2+a \beta_2^2-8 b N \beta_1-2 b \beta_2+8 a N \beta_1 \nonumber\\&&\hspace{0.2cm} \times \beta_2-2 \beta_2-4 L \beta_1   \big[\big((1-a \beta_2) r^8+2 a \beta_1 r^6\big) N^4+4 r^4 \nonumber\\&&\hspace{0.2cm} \times \big((1-2 a \beta_2) r^2+4 a \beta_1\big) N^3 +\big((6-16 a \beta_2) r^4+26 a \beta_1 r^2\big) N^2\nonumber\\&&\hspace{0.2cm} -4 \big((3 a \beta_2-1) r^2+3 a \beta_1\big) N  +b \big(N r^2+1\big)^2 \big(N^2 r^4+6 N r^2\nonumber\\&&\hspace{0.2cm} +3\big)-3 a \beta_2+1\big]\Big)\Big\}\Big] r^2,\nonumber
  \end{eqnarray}
  
\begin{eqnarray}
&&\hspace{-0.2cm} f_3(r)=a N^4 \beta_2^2 r^8-2 b N^4 \beta_2 r^8-2 N^4 \beta_2 r^8+4 a N^3 \beta_2^2 r^6 \nonumber\\&&\hspace{0.2cm} +4 b N^4 \beta_1 r^6+4 N^4 \beta_1 r^6-8 b N^3 \beta_2 r^6-8 N^3 \beta_2 r^6-4 a N^4 \beta_1 \nonumber\\&&\hspace{0.2cm} \times \beta_2 r^6+4 a N^4 \beta_1^2 r^4+6 a N^2 \beta_2^2 r^4+20 b N^3 \beta_1 r^4+12 N^3 \beta_1 r^4 \nonumber\\&&\hspace{0.2cm} -12 b N^2 \beta_2 r^4-12 N^2 \beta_2 r^4-20 a N^3 \beta_1 \beta_2 r^4+24 a N^3 \beta_1^2 r^2 \nonumber\\&&\hspace{0.2cm} +4 a N \beta_2^2 r^2+28 b N^2 \beta_1 r^2+12 N^2 \beta_1 r^2-8 b N \beta_2 r^2-8 N \beta_2 r^2 \nonumber\\&&\hspace{0.2cm} -28 a N^2 \beta_1 \beta_2 r^2+4 b \big(N r^2+1\big)^4+36 a N^2 \beta_1^2+4 a L^2 \nonumber\\&&\hspace{0.2cm} \times \big(N r^2+3\big)^2 \beta_1^2+a \beta_2^2+12 b N \beta_1+4 N \beta_1-2 b \beta_2-12 a N \beta_1 \beta_2\nonumber\\&&\hspace{0.2cm} -2 \beta_2-4 L \beta_1 \big[\big\{(1-a \beta_2) r^6+2 a \beta_1 r^4\big\} N^3+\big[(3-5 a \beta_2) r^4 \nonumber\\&&\hspace{0.2cm} +12 a \beta_1 r^2\big] N^2+\big((3-7 a \beta_2) r^2+18 a \beta_1\big) N+b \big(N r^2+1\big)^2 \nonumber\\&&\hspace{0.2cm} \times \big(N r^2+3\big)-3 a \beta_2+1\big],\nonumber
\end{eqnarray}

\begin{eqnarray}
&&\hspace{-0.2cm} f_4(r)=-2 N \big(L r^2+1\big) \big[4 c r^2 (\alpha +1) (N r^2+1)^4+r^2 (\alpha +1) \beta_2 \nonumber\\&&\hspace{0.2cm} \times  (a \beta_2-2 b-2) (N r^2+1)^4 +4 \beta_1 \big(N^2 (\alpha +1) (1-a \beta_2-1) r^4 \nonumber\\&&\hspace{0.15cm} -L \alpha  r^2  -b (L-N) (N r^2+3) (\alpha +1) r^2+3 a L \beta_2 r^2 +3 a L \alpha  \beta_2 r^2 \nonumber\\&&\hspace{0.2cm} +N  [L \alpha  (a \beta_2-1) r^2+\alpha +a (L r^2-3) \beta_2-3 a \alpha  \beta_2+2] r^2+1\big) \nonumber\\&&\hspace{0.2cm} \times (N r^2+1)^2  +4 a (L-N)^2 r^2 (N r^2+3)^2 (\alpha +1) \beta_1^2\big] f_5(r) \nonumber\\&&\hspace{0.2cm} +2 L(N r^2+1) \big[4 c r^2 (\alpha +1) (N r^2+1)^4+r^2 (\alpha +1) \beta_2 (-2 b   \nonumber\\&&\hspace{0.2cm} +a \beta_2-2)(N r^2+1)^4+4 \beta_1 \big(-N^2 (\alpha +1) (a \beta_2-1) r^4 \nonumber\\&&\hspace{0.2cm}-L \alpha  r^2-b (L-N) (N r^2+3) (\alpha +1) r^2  +3 a L \beta_2 r^2+3 a L \alpha \nonumber\\&&\hspace{0.2cm} \times \beta_2 r^2+N [L \alpha  (a \beta_2-1) r^2+\alpha +a (L r^2-3) \beta_2-3 a \alpha  \beta_2+2] \nonumber\\&&\hspace{0.2cm} \times r^2 +1\big) (N r^2+1)^2 +4 a (L-N)^2 r^2 (N r^2+3)^2 (\alpha +1) \beta_1^2\big] \nonumber\\&&\hspace{0.2cm} \times f_6(r)+f_{44}(r), \nonumber
\end{eqnarray}

\begin{eqnarray}
    && \hspace{-0.2cm} f_4(r)=-2 \big(L r^2+1\big)  \big(N r^2+1\big) \big[a N^4 \beta_2^2 r^8-2 b N^4 \beta_2 r^8-N^4 \nonumber\\&&\hspace{0.2cm} \times \beta_2 r^8+4 a N^3 \beta_2^2 r^6+4 b N^4 \beta_1 r^6+2 N^4 \beta_1 r^6-8 b N^3 \beta_2 r^6  \nonumber\\&&\hspace{0.2cm} -4 N^3 \beta_2 r^6-4 a N^4 \beta_1 \beta_2 r^6+4 a N^4 \beta_1^2 r^4+6 a N^2 \beta_2^2 r^4+20 b \nonumber\\&&\hspace{0.2cm} \times N^3 \beta_1 r^4+6 N^3 \beta_1 r^4-12 b N^2 \beta_2 r^4  -6 N^2 \beta_2 r^4-20 a N^3 \nonumber\\&&\hspace{0.2cm} \times  \beta_1 \beta_2 r^4+24 a N^3 \beta_1^2 r^2+4 a N \beta_2^2 r^2+28 b N^2 \beta_1 r^2+6 N^2 \beta_1 r^2 \nonumber\\&&\hspace{0.2cm} -8 b N \beta_2 r^2-4 N \beta_2 r^2  -28 a N^2 \beta_1 \beta_2 r^2+4 c (N r^2+1)^4+36 a \nonumber\\&&\hspace{0.2cm} \times N^2 \beta_1^2+4 a L^2 (N r^2+3)^2 \beta_1^2+a \beta_2^2+12 b N \beta_1+2 N \beta_1-2 b \beta_2 \nonumber\\&&\hspace{0.2cm} -12 a N \beta_1 \beta_2-\beta_2-2 L \beta_1 \big(\{(1-2 a \beta_2) r^6+4 a \beta_1 r^4\} N^3 \nonumber\\&&\hspace{0.2cm} +[(3-10 a \beta_2) r^4+24 a \beta_1 r^2] N^2+\big((3 -14 a \beta_2) r^2 +36 a \beta_1\big) N \nonumber\\&&\hspace{0.2cm} +2 b (N r^2+1)^2 (N r^2+3)-6 a \beta_2+1\big)\big] f_7(r)+3 \big(L r^2+1\big)\nonumber\\&&\hspace{0.2cm} \times \big(N r^2+1\big) \Big[4 c r^2 (\alpha +1) \big(N r^2+1\big)^4+r^2 (\alpha +1) \beta_2  (-2 b \nonumber\\&&\hspace{0.2cm} +a \beta_2-2) \big(N r^2+1\big)^4+4 \beta_1 \big(N^2 (\alpha +1) (1-a \beta_2) r^4-L \alpha  r^2 \nonumber\\&&\hspace{0.2cm}-b (L-N) \big(N r^2+3\big)  (\alpha +1) r^2+3 a L \beta_2 r^2+3 a L \alpha  \beta_2 r^2+N \nonumber\\&&\hspace{0.2cm} \times \big(L \alpha  (a \beta_2-1) r^2+\alpha +a \big(L r^2-3\big) \beta_2-3 a \alpha  \beta_2+2\big) r^2+1\big) \nonumber\\&&\hspace{0.2cm} \times  \big(N r^2+1\big)^2+4 a (L-N)^2 r^2 \big(N r^2+3\big)^2 (\alpha +1) \beta_1^2\Big] f_8(r)\nonumber
\end{eqnarray}
\begin{eqnarray}
&&\hspace{-0.15cm} f_5(r)=  \Big[a N^4 \beta_2^2 r^8-2 b N^4 \beta_2 r^8-N^4 \beta_2 r^8+4 a N^3 \beta_2^2 r^6  \nonumber\\&&\hspace{0.2cm} +4 b N^4 \beta_1 r^6  +2 N^4 \beta_1 r^6-8 b N^3 \beta_2 r^6-4 N^3 \beta_2 r^6 \nonumber\\&&\hspace{0.2cm} -4 a N^4 \beta_1 \beta_2 r^6+4 a N^4 \beta_1^2 r^4+6 a N^2 \beta_2^2 r^4  +20 b N^3 \beta_1 r^4 \nonumber\\&&\hspace{0.2cm} +6 N^3 \beta_1 r^4-12 b N^2 \beta_2 r^4-6 N^2 \beta_2 r^4-20 a N^3 \beta_1 \beta_2 r^4 \nonumber\\&&\hspace{0.2cm} +24 a N^3 \beta_1^2 r^2+4 a N \beta_2^2 r^2  +28 b N^2 \beta_1 r^2+6 N^2 \beta_1 r^2-8 b N \nonumber\\&&\hspace{0.2cm} \times \beta_2 r^2-4 N \beta_2 r^2-28 a N^2 \beta_1 \beta_2 r^2 +4 c \big(N r^2+1\big)^4+36 a N^2 \nonumber\\&&\hspace{0.2cm} \times\beta_1^2  +4 a L^2 \big(N r^2+3\big)^2 \beta_1^2+a \beta_2^2+12 b N \beta_1+2 N \beta_1-2 b \beta_2 \nonumber\\&&\hspace{0.2cm} -12 a N \beta_1 \beta_2-\beta_2-2 L \beta_1 \big(\big((1-2 a \beta_2) r^6  +4 a \beta_1 r^4\big) N^3 \nonumber\\&&\hspace{0.2cm} +\big((3-10 a \beta_2) r^4+24 a \beta_1 r^2\big) N^2  +\big((3-14 a \beta_2) r^2  +36 a \beta_1\big) \nonumber\\&&\hspace{0.2cm} \times N  +2 b \big(N r^2+1\big)^2 \big(N r^2+3\big)-6 a \beta_2+1\big)\Big] r^2, \nonumber
\end{eqnarray}

\begin{eqnarray}
  &&\hspace{-0.15cm}  f_6(r)=\Big[a N^4 \beta_2^2 r^8-2 b N^4 \beta_2 r^8-N^4 \beta_2 r^8+4 a N^3 \beta_2^2 r^6+4 b N^4 \nonumber\\&&\hspace{0.2cm} \times \beta_1 r^6   +2 N^4 \beta_1 r^6-8 b N^3 \beta_2 r^6-4 N^3 \beta_2 r^6-4 a N^4 \beta_1 \beta_2 r^6 \nonumber\\&&\hspace{0.2cm} +4 a N^4 \beta_1^2 r^4+6 a N^2 \beta_2^2 r^4+20 b N^3 \beta_1 r^4+6 N^3 \beta_1 r^4 -12 b N^2 \nonumber\\&&\hspace{0.2cm} \times \beta_2 r^4 -6 N^2 \beta_2 r^4-20 a N^3 \beta_1 \beta_2 r^4+24 a N^3 \beta_1^2 r^2+4 a N \beta_2^2 r^2 \nonumber\\&&\hspace{0.2cm} +28 b N^2 \beta_1 r^2+6 N^2 \beta_1 r^2 -8 b N \beta_2 r^2-4 N \beta_2 r^2-28 a N^2 \beta_1 \nonumber\\&&\hspace{0.2cm}  \times \beta_2 r^2 +4 c \big(N r^2+1\big)^4+36 a N^2 \beta_1^2+4 a L^2 \big(N r^2+3\big)^2 \beta_1^2 \nonumber\\&&\hspace{0.2cm}+a \beta_2^2 +12 b  N \beta_1+2 N \beta_1-2 b \beta_2-12 a N \beta_1 \beta_2-\beta_2-2 L \beta_1 \nonumber\\&&\hspace{0.2cm} \times \big(\big((1-2 a \beta_2)   r^6+4 a \beta_1 r^4\big) N^3+\big((3-10 a \beta_2) r^4 +24 a \beta_1 r^2\big) \nonumber\\&&\hspace{0.2cm} \times N^2+\big((3 -14 a \beta_2) r^2+36 a \beta_1\big) N+2 b \big(N r^2+1\big)^2 \big(N r^2+3\big) \nonumber\\&&\hspace{0.2cm} -6 a \beta_2+1\big)\Big] r^2, \nonumber
\end{eqnarray}
\begin{eqnarray}
&&\hspace{-0.15cm}  f_7(r)=\Big[4 c r (\alpha +1) \big(N r^2+1\big)^4+r (\alpha +1)   \beta_2 (-2 b+a \beta_2-2) \nonumber\\
&&\hspace{0.2cm} \times \big(N r^2+1\big)^4+4 r \beta_1 \big(-2 N^2 (\alpha +1) (a \beta_2-1) r^2-b (L-N) \nonumber\\
&&\hspace{0.2cm} \times \big(2 N r^2+3\big) (\alpha +1) +N \big(2 L \alpha  (a \beta_2-1) r^2+\alpha +a \big(2 L r^2-3\big) \nonumber\\
&&\hspace{0.2cm} \times \beta_2-3 a \alpha  \beta_2+2\big)+L (3 a \beta_2+\alpha  (3 a \beta_2-1))\big) \big(N r^2+1\big)^2  \nonumber\\
&&\hspace{0.2cm} +8 N r \beta_1 \big(-N^2 (\alpha +1) (a \beta_2-1) r^4-L \alpha  r^2-b (L-N) \nonumber\\
&&\hspace{0.2cm} \times \big(N r^2+3\big) (\alpha +1) r^2+3 a L \beta_2 r^2+3 a L \alpha  \beta_2 r^2  +N \big(L \alpha  (a \beta_2 \nonumber\\
&&\hspace{0.2cm} -1) r^2+\alpha +a \big(L r^2-3\big) \beta_2-3 a \alpha  \beta_2+2\big) r^2+1\big) \big(N r^2+1\big) \nonumber\\
&&\hspace{0.2cm} +4 a (L-N)^2 r \big(N r^2+3\big)^2   (\alpha +1) \beta_1^2+8 a (L-N)^2 N r^3 \nonumber\\
&&\hspace{0.2cm} \times \big(N r^2+3\big) (\alpha +1) \beta_1^2+16 c N \big(N r^3+r\big)^3  (\alpha +1)+4 N\nonumber\\
&&\hspace{0.2cm} \times \big(N r^3+r\big)^3  (\alpha +1) \beta_2 (-2 b+a \beta_2-2)\Big] r, \nonumber
\end{eqnarray}
\begin{eqnarray}
&&\hspace{-0.15cm}    f_8(r)=\Big[a N^4 \beta_2^2 r^8-2 b N^4 \beta_2 r^8-N^4 \beta_2 r^8+4 a N^3 \beta_2^2 r^6  +4 b N^4\nonumber\\
&&\hspace{0.2cm} \times \beta_1 r^6 +2 N^4 \beta_1 r^6-8 b N^3 \beta_2 r^6-4 N^3 \beta_2 r^6-4 a N^4 \beta_1 \beta_2 r^6 \nonumber\\
&&\hspace{0.2cm} +4 a N^4  \beta_1^2 r^4+6 a N^2 \beta_2^2 r^4+20 b N^3 \beta_1 r^4+6 N^3 \beta_1 r^4-12 b N^2 \nonumber\\
&&\hspace{0.2cm} \times \beta_2 r^4  -6 N^2 \beta_2 r^4-20 a N^3 \beta_1 \beta_2 r^4+24 a N^3 \beta_1^2 r^2+4 a N \beta_2^2 r^2\nonumber\\
&&\hspace{0.2cm} +28 b  N^2 \beta_1 r^2  +6 N^2 \beta_1 r^2-8 b N \beta_2 r^2-4 N \beta_2 r^2-28 a N^2 \beta_1 \beta_2 r^2 \nonumber\\
&&\hspace{0.2cm} +4 c \big(N r^2+1\big)^4+36 a N^2 \beta_1^2  +4 a L^2 \big(N r^2+3\big)^2 \beta_1^2 +a \beta_2^2 \nonumber\\
&&\hspace{0.2cm} +12 b N \beta_1+2 N \beta_1-2 b \beta_2-12 a N \beta_1 \beta_2-\beta_2-2 L \beta_1 \nonumber\\
&&\hspace{0.2cm} \times \big\{\big((1-2 a \beta_2) r^6+4 a \beta_1 r^4\big) N^3+\big((3-10 a \beta_2) \times r^4+24 a \nonumber\\
&&\hspace{0.2cm} \times\beta_1 r^2\big) N^2+\big((3-14 a \beta_2) r^2+36 a \beta_1\big) N+2 b \big(N r^2+1\big)^2 \nonumber\\
&&\hspace{0.2cm} \times \big(N r^2+3\big)-6 a \beta_2+1\big\}\Big]. \nonumber
\end{eqnarray}
\end{small}


\begin{thebibliography}{99}

\bibitem[\protect\citeauthoryear{Abott et al.}{2016}]{waves1}  Abbott B. P. et. al, 2016, PhRvL,  116, 6

\bibitem[\protect\citeauthoryear{Abott et al.}{2017}]{waves2}  Abbott B. P. et. al, 2017, ApJL.,  848:L12

\bibitem[\protect\citeauthoryear{Abott et al.}{2020}]{waves3}  Abbott R., et al., 2020a, arXiv e-prints, p. arXiv:2006.12611

\bibitem[\protect\citeauthoryear{Abramowicz}{1983}]{EOS1} Abramowicz M.A., 1983, Acta Astron. 33, 313 

\bibitem[\protect\citeauthoryear{Astashenok et al.}{2020}]{Astash01}  Astashenok A. V., Capozziello, S., Odintsov, S. D. \& Oikonomou, V. K., 2020, Phys. Lett. B, 811, 135910 

\bibitem[\protect\citeauthoryear{Astashenok et al.}{2021}]{Astash1}  Astashenok A. V., Capozziello, S., Odintsov, S. D. \& Oikonomou, V. K., 2021, Phys. Lett. B, 816, 136222 

\bibitem[\protect\citeauthoryear{Azam \& Mardan}{2017}]{EOS12} Azam M., Mardan S.A., 2017, Eur. Phys. J. C, 77, 113

\bibitem[\protect\citeauthoryear{Azam \& Mardan}{2017}]{EOS14} Azam M., Mardan S.A., 2017, JCAP, 01, 040 

\bibitem[\protect\citeauthoryear{Azam et al.}{2016}]{EOS13} Azam M., Gigot G., Witt\'e I., Schatz B., 2016, Eur. Phys. J. C, 76, 315

\bibitem[\protect\citeauthoryear{Azam et al.}{2015}]{EOS10} Azam M., Mardan S.A., Rehman M.A., 2015, Astrophys. Space Sci., 359, 14

\bibitem[\protect\citeauthoryear{Bekenstein}{1960}]{EOS8} Bekenstein J.D., 1960, Phys. Rev. D, 4, 2185

\bibitem[\protect\citeauthoryear{Boehmer et al.}{2011}]{Boehmer:2011gw} B{\"o}hmer, C.~G., Mussa, A. \& Tamanini, N., 2011, Classical Quant. Grav. 28, 245020 

\bibitem[\protect\citeauthoryear{Capano et al.}{2020}]{bau} Capano C.D. et al., 2020, Nat. Astron., 2020, 4, 625

\bibitem[\protect\citeauthoryear{Cosenza et al.}{1981}]{EOS2} Cosenza M., Herrera L., Esculpi M.,  Witten L., 1981, J. Math. Phys., 22, 118

\bibitem[\protect\citeauthoryear{Chavanis}{2014a}]{EOS1c} Chavanis P.H., 2014, Eur. Phys. J. Plus, 129, 38

\bibitem[\protect\citeauthoryear{Chavanis}{2014b}]{EOS2c} Chavanis P.H., 2014, Eur. Phys. J. Plus, 129, 222

\bibitem[\protect\citeauthoryear{Contreras \& Stuchlik}{2022}]{Ener2}  Contreras E., Stuchlik Z., 2022 Eur. Phys. J. C  82, 365 

\bibitem[\protect\citeauthoryear{Contreras \& Fuenmayor}{2021}]{Contreras} Contreras E., Fuenmayor E., 2021, Phys. Rev. D 103, 124065

\bibitem[\protect\citeauthoryear{Cromartie et al.}{2019}]{Cromartie} Cromartie H. T., Fonseca E., Ransom S. M., et al., 2020, Nat. Astron. 4, 72

\bibitem[\protect\citeauthoryear{Darmois}{1927}]{Darmois1927} Darmois G. J., 1927, Memorial des Sciences Mathematiques, 25, 1

\bibitem[\protect\citeauthoryear{Demorest et al.}{2010}]{star1} Demorest P., Pennucci T., Ransom S. et al., 2010, Nature, 467, 1081

\bibitem[\protect\citeauthoryear{Dimakis et al.}{2022}]{Dimakis22} Dimakis, N., Paliathanasis, A., Roumeliotis, M. , \& Christodoulakis, T., 2021, Phys. Rev. D 106, 043509

\bibitem[\protect\citeauthoryear{Douchin \& Haensel}{2001}]{Douchin} Douchin, F. and Haensel, P., 2001, A \& A, {\bf 380}, 151

\bibitem[\protect\citeauthoryear{Fattoyev et al.}{2020}]{fatt} Fattoyev F. J., Horowitz C. J., Piekarewicz J. and Reed, B., 2020, Phys. Rev. C, 102, 065805

\bibitem[\protect\citeauthoryear{Ferreira et al.}{2022}]{ferr1} Ferreira J., Barreiro T., Mimoso J., Nunes N. J.,  2022, Phys. Rev. D, 105, 123531

\bibitem[\protect\citeauthoryear{Ferraro \& Fiorini}{2011}]{rm7} Ferraro, R., Fiorini, F., 2011, Phys. Rev. D 84, 083518

\bibitem[\protect\citeauthoryear{Freitas \& Goncalves}{2014}]{EOS3c} Freitas R. C., Goncalves S. V. B., 2014, Eur. Phys. J. C, 74, 3217

\bibitem[\protect\citeauthoryear{Dev \& Gleiser}{2002}]{Gleiser2002} Dev K., Gleiser M, 2002, Gen. Relativ. Gravit., 34, 1793

\bibitem[\protect\citeauthoryear{Godzieba et al.}{2022}]{god1}  Godzieba D. A., Radice D., Bernuzzi S., 2021, ApJ, 908:122

\bibitem[\protect\citeauthoryear{Haensel \& Potekhin}{2004}]{Haensel2004} Haensel, P. and Potekhin, A. Y., 2004, A \& A, 498, 191 

\bibitem[\protect\citeauthoryear{Hansraj et al.}{2022}]{hansraj1}  Hansraj S., Govender M., Moodly L., Singh Ksh. Newton, 2022, Phys. Rev. D, 105, 044030 

\bibitem[\protect\citeauthoryear{Herrera \& Barreto}{2004}]{EOS3} Herrera L.,  Barreto W., 2004, Gen. Relativ. Gravit., 36, 127 

\bibitem[\protect\citeauthoryear{Herrera \& Barreto}{2013}]{EOS5} Herrera L.,  Barreto W., 2013, Phys. Rev. D, 88, 084022

\bibitem[\protect\citeauthoryear{Herrera \& Santos}{1997}]{Herrera1997} Herrera L., Santos N. O., 1997, Phys. Rept., 286, 53

\bibitem[\protect\citeauthoryear{Herrera et al.}{2004}]{EOS4} Herrera L. et al., 2004, Phys. Rev. D, 69, 084026

\bibitem[\protect\citeauthoryear{Herrera et al.}{2014}]{EOS6} Herrera L., Di Prisco A., Barreto W., Ospino J., 2014, Gen. Relativ. Gravit., 46, 1827

\bibitem[\protect\citeauthoryear{Herrera et al.}{2016}]{EOS7} Herrera L., Fuenmayor E., Leon P., 2016, Phys. Rev.D, 93, 024247

\bibitem[\protect\citeauthoryear{Hererra}{2020}]{Hererra2020} Herrera L., 2020, Phys. Rev. D 101, 104024

\bibitem[\protect\citeauthoryear{Israel}{1966}]{Israel1966} Israel W., 1966, Nuo. Cim. B, 44, 1

\bibitem[\protect\citeauthoryear{Jones}{1975}]{Jones1975} Jones P. B., 1975, Astrophys. Space Sci., 33, 215

\bibitem[\protect\citeauthoryear{ Liebling \& Palenzuela}{2012}]{Liebling2012} Liebling S. L., Palenzuela C., 2012, Living Rev. Rel., 15, 6

\bibitem[\protect\citeauthoryear{Lu et al.}{2021}]{lu1} Lu W., Beniamini P., Bonnerot C.,  2021, MNRAS, 500, 1817 

 \bibitem[\protect\citeauthoryear{Beniamini \& Bonnerot}{2021}]{wenbin} Lu W., Beniamini P., Bonnerot C., 2021, MNRAS, 500, 1817  

\bibitem[\protect\citeauthoryear{Mandal et al.}{2022}]{Mandal1} Mandal S., Mustafa G., Hassan Z., Sahoo P. K., 2022, PDU, 35, 100934

\bibitem[\protect\citeauthoryear{Maurya et al.}{2020}]{sunil5d1} Maurya, S. K., Errehymy A., Singh K. N., Tello-Ortiz F., Daoud M., 2020, Phys. Dark Univ. 30, 100640 

\bibitem[\protect\citeauthoryear{Maurya et al.}{2021}]{sunil5d2} Maurya S. K., Pradhan A., Tello-Ortiz F., Banerjee A., Nag R., 2021, Eur. Phys. J. C, 81, 848 

\bibitem[\protect\citeauthoryear{Maurya et al.}{2022}]{sunil5d3} Maurya S. K., Singh K. N., Govender M., Hansraj S., 2022, Astrophys. J. 925, 208

\bibitem[\protect\citeauthoryear{Maurya et al.}{2023}]{sunil5d4}  Maurya S. K., Singh K. N., Govender M., Ray S., 2023, MNRAS, 519, 4303

\bibitem[\protect\citeauthoryear{Maurya et al.}{2022}]{sunil5d5} Maurya S. K., Singh K. N., Lohakare S. V., Mishra B., 2022, Forts. der Phys. 70, 2200061

\bibitem[\protect\citeauthoryear{Mohanty et al.}{2023}]{mohanty1} Mohanty, S. R., Ghosh, S., Routaray, P., Das, H. C. and Kumar, B., arXiv preprint arXiv:2305.15724 

\bibitem[\protect\citeauthoryear{Naeem et al.}{2021}]{EOS11} Naeem R., Azam M., Abbas G., Nazar H., 2021, New Astron., 89, 101651 

\bibitem[\protect\citeauthoryear{Oppenheimer \& Volkoff}{1939}]{OV1939} Oppenheimer J., Volkoff G.M., 1939, Phys. Rev. 55, 374

\bibitem[\protect\citeauthoryear{Ovalle}{2017}]{OvallePRD2017} Ovalle J., 2017, Phys. Rev. D, 95, 104019

\bibitem[\protect\citeauthoryear{Ovalle}{2019}]{OvallePLB2019} Ovalle J., 2019, Phys. Lett. B, 788, 213

\bibitem[\protect\citeauthoryear{Ovalle et al.}{2022}]{Ener1} Ovalle J., Contreras E., Stuchlik Z., 2022  Eur. Phys. J. C 82, 211 

\bibitem[\protect\citeauthoryear{Pandharipande \& Ravenhall}{1989}]{Pandhar} Pandharipande, V. R. and Ravenhall, D. G. 1989, Hot Nuclear Matter in Nuclear Matter and Heavy Ion Collisions, NATO ADS Ser., {\bf B205}, ed. Soyeur, M., Flocard, H.,  Tamain, B. and Porneuf, M. (Dordrecht: Reidel), 103

\bibitem[\protect\citeauthoryear{Potekhin et al.}{2013}]{Potekhin} Potekhin, A. Y., Fantina, A. F., Chamel, N., Pearson, J. M. and Goriely, S., 2013, A \& A, {\bf 560} A48

\bibitem[\protect\citeauthoryear{Kippenhahn \& Weigert}{1990}]{Kippen1990} Kippenhahn R., Weigert A., Stellar Structure and Evolution (Springer-Verlag, Berlin, 1990)

\bibitem[\protect\citeauthoryear{Radice et al.}{2018}]{radice}  Radice D., Perego A., Zappa F., Bernuzzi S. 2018, ApJL, 852, L29

\bibitem[\protect\citeauthoryear{Burgio et al.}{2018}]{burgio}  Burgio, G. F., et al., 2018, ApJ, 860, 139 

\bibitem[\protect\citeauthoryear{Rawls et al.}{2011}]{star3} Rawls M. L., Orosz J. A., McClintock J. E., Torres M. A. P., Bailyn C. D., Buxton M. M., 2011, ApJ, 730, 25

\bibitem[\protect\citeauthoryear{Ruderman}{1972}]{Ruderman1972} Ruderman M. A., 1972, Annu. Rev. Astron. Astrophys., 10, 427

\bibitem[\protect\citeauthoryear{Ruiz et al.}{2018}]{ruiz}  Ruiz M., Shapiro S. L., Tsokaros A. 2018, PhRvD, 97, 021501

\bibitem[\protect\citeauthoryear{Sharif et al.}{2020a}]{sharif1} Sharif M., Majid A., 2020a, Astrophys Space Sci., 365, 42 

\bibitem[\protect\citeauthoryear{Sharif et al.}{2020b}]{sharif2} Sharif M., S. Saba S., 2020b, Int. J. Mod. Phys. D, 29, 2050041 

\bibitem[\protect\citeauthoryear{Sharif et al.}{2021}]{sharif3} Sharif M., Aslam M., 2021, Eur. Phys. J. C, 81, 641 

\bibitem[\protect\citeauthoryear{Sawyer}{1972}]{Sawyer1972} Sawyer R. F., 1972, Phys. Rev. Lett., 29, 382

\bibitem[\protect\citeauthoryear{Sokolov}{1980}]{Sokolov1980} Sokolov A. I., 1980, JETP, 79, 1137

\bibitem[\protect\citeauthoryear{Takisa \& Maharaj}{2013}]{EOS9} Takisa P.M., Maharaj S.D., 2013, Astrophys. Space Sci., 45, 1951

\bibitem[\protect\citeauthoryear{Tangphati et al.}{2022}]{tang1} Tangphati T, Karar I, Pradhan A., Banerjee A, 2022, Eur. Phys. J., 82, 57

\bibitem[\protect\citeauthoryear{Tangphati et al.}{2021}]{tang2} Tangphati T., Pradhan A., Errehym A., Banerjee A., 2021, Phys. Lett. B., 819, 136423  

\bibitem[\protect\citeauthoryear{Tews et al.}{2021}]{tews} Tews I, Pang P. T. H., Dietrich T., Coughlin M. W., Antier S., Bulla M., Heinzel J., Issa, L., 2021, ApJ, 908, L1

\bibitem[\protect\citeauthoryear{Tolman}{1939}]{Tolman1939} Tolman R. C., 1939, Phys., Rev. 55, 364

\bibitem[\protect\citeauthoryear{Wang et al.}{2022}]{Wang} Wang W., Chen,H., Katsuragawa T., 2022, Phys. Rev. D, 105, 024060

\bibitem[\protect\citeauthoryear{Weber}{1999}]{Weber1999} Weber F., Pulsars as Astrophysical Observatories for Nuclear and Particle Physics (IOP Publishing, Bristol, 1999)
 
\bibitem[\protect\citeauthoryear{Zhao}{2022}]{Zhao} Zhao D., 2022, Eur. Phys. J. C, 82, 303

\end{thebibliography}
\end{document}